\documentclass[a4paper,12pt,twoside]{article}

\RequirePackage[OT1]{fontenc}
\RequirePackage{amsthm,amsmath,amssymb}
\RequirePackage{natbib}

\usepackage{dsfont}
\usepackage{xcolor}
\usepackage{hyperref}

\DeclareMathAlphabet\mathbfcal{OMS}{cmsy}{b}{n}

\usepackage{tikz,soul}                     
\usetikzlibrary{arrows.meta}     
\usepackage[ruled,norelsize,vlined]{algorithm2e}
\usepackage[noend]{algpseudocode}      
\usepackage{dsfont}                 
\usepackage{gensymb}             
\usepackage{multirow}  
\usepackage{makecell}
\usepackage{mathrsfs}

\newcommand{\X}{\textbf{X}}
\newcommand{\x}{\textbf{x}}
\newcommand{\bDelta}{\boldsymbol{\Delta}}
\newcommand{\bSigma}{\boldsymbol{\Sigma}}

\newcommand{\bphi}{\boldsymbol{\phi}}
\newcommand{\btheta}{\boldsymbol{\theta}}
\newcommand{\Z}{\mathbfcal{Z}}
\newcommand{\W}{\mathbfcal{W}}

\newcommand{\spartaco}{\textsc{SpaRTaCo}}

\newcommand{\bTheta}{\boldsymbol{\Theta}}
\newcommand{\review}[1]{{\color{black}#1}}
\newcommand{\rowepsilon}{\varepsilon^{\mathrm{rows}}_k}
\newcommand{\colepsilon}{\varepsilon^{\mathrm{cols}}_r}

\DeclareMathOperator*{\argmax}{arg\,max}

\usepackage{fancyhdr}
\title{\bf Co-clustering of Spatially Resolved Transcriptomic Data\vspace{.5cm}}

\author{{A. Sottosanti and D. Risso}
\\[4ex]
{\small University of Padova, Department of Statistical Sciences,}
\\[.3ex]
{\small via Cesare Battisti 241-243, Padova, Italy}
}

\date{
Address for correspondence: \texttt{andrea.sottosanti@unipd.it}}

\begin{document}
\maketitle
\begin{abstract}
	Spatial transcriptomics is a groundbreaking technology that allows the measurement of the activity of thousands of genes in a tissue sample and maps where the activity occurs. 
	This technology has enabled the study of the spatial variation of the genes across the tissue. Comprehending gene functions and interactions in different areas of the tissue is of great scientific interest, as it might lead to a deeper understanding of several key biological mechanisms, such as cell-cell communication or tumor-microenvironment interaction.
	To do so, one can group cells of the same type and genes that exhibit similar expression patterns.
	However, adequate statistical tools that exploit the previously unavailable spatial information to more coherently group cells and genes are still lacking.

	In this work, we introduce \spartaco, a new statistical model that clusters the spatial expression profiles of the genes according to a partition of the tissue. 
	This is accomplished by performing a co-clustering, i.e., inferring the latent block structure of the data and inducing two types of clustering: of the genes, using their expression across the tissue, and of the image areas, using the gene expression in the \textit{spots} where the RNA is collected.
	Our proposed methodology is validated with a series of simulation experiments and its usefulness in responding to specific biological questions is illustrated with an application to a human brain tissue sample processed with the 10X-Visium protocol.
\end{abstract}

\section{Introduction}
\subsection{The rise of spatial transcriptomics}
\label{subsec:spatial_transcriptomics_intro}

In the last few years, we have witnessed a dramatic improvement in the efficiency of DNA sequencing technologies that ultimately gave rise to new advanced protocols for single-cell RNA sequencing (scRNA-seq) and, more recently, spatial transcriptomics. In particular, spatial transcriptomics has been chosen as \textit{method of the year 2020} \citep{methodoftheyear}. With respect to scRNA-seq, spatial transcriptomic platforms are able to provide, in addition to the abundance, the locations of thousands of genes in a tissue sample. 

\citet{Righelli_etal.2021} classify spatial transcriptomic protocols into \emph{molecule-based} and \emph{spot-based} methods. \review{Among molecule-based methods, seqFISH \citep{Lubeck_etal.2014} and similar methods, such as MERFISH \citep{Chen_etal.2015}, are capable of providing the spatial expression of thousands of transcripts at a sub-cellular level, but the setup necessary to perform this kind of spatial experiments is often complex and expensive to recreate.}
\review{Spot-based methods, such as Slide-seq \citep{Rodriques_etal.2019} or the 10X Genomics \emph{Visium} platform \citep{Rao_etal.2020}, have substantially lower resolution than seqFISH, but allow scientists to measure close to the whole transcriptome of (small pools of) cells across a tissue in a relatively easy manner.}

\begin{figure}[th!]
\centering
\includegraphics[width=0.6\linewidth]{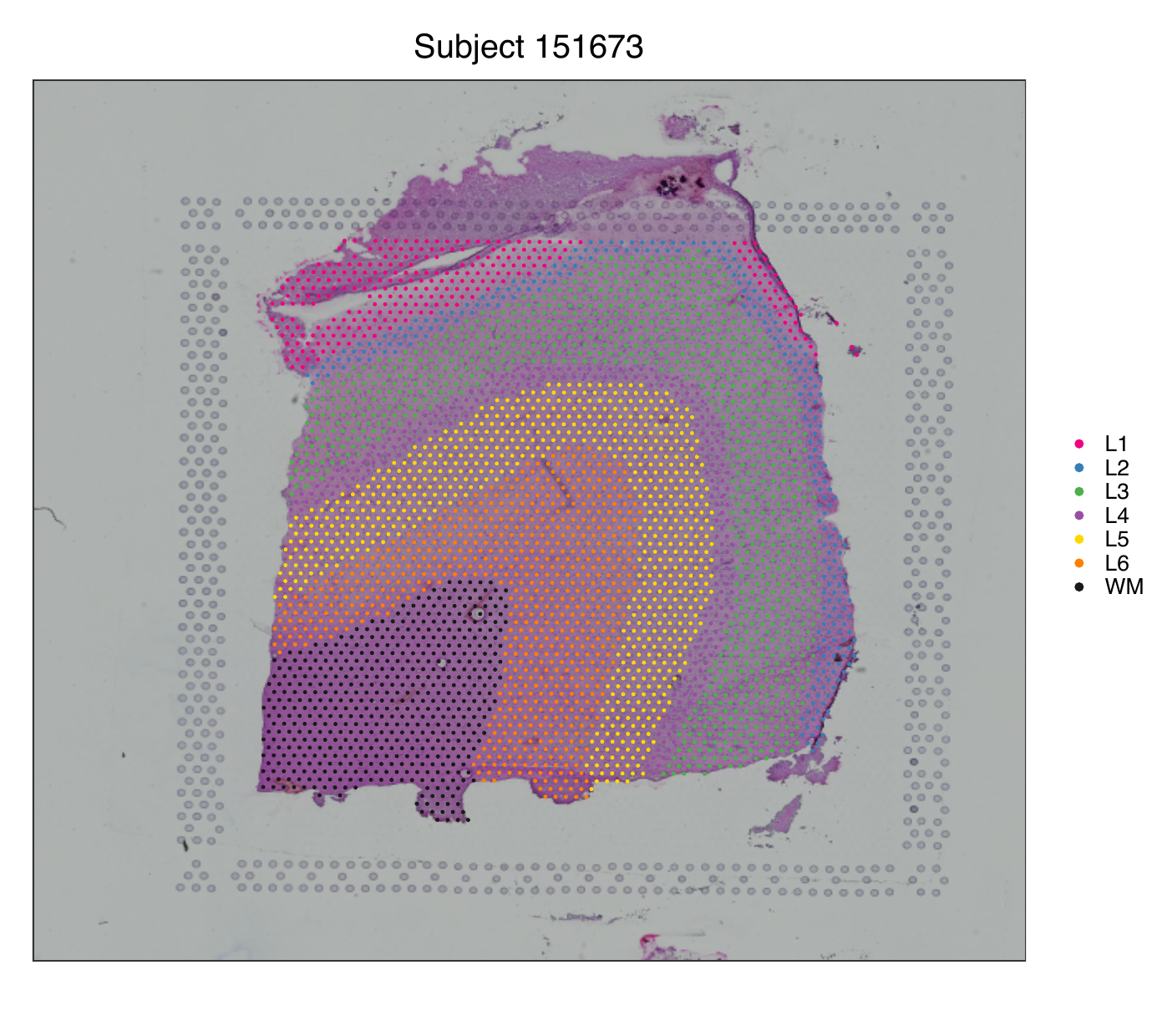}
\caption{Tissue sample of LIBD human dorsolateral prefrontal cortex (DLPFC) processed with Visium platform and stored in the \textnormal{\texttt{R}} package \textnormal{\texttt{spatialLIBD}}. The dots represent the spots over the chip surface. Different colors denote a manual annotation of the areas performed by \citet{Maynard_etal.2021}: they recognize a White Matter (\textnormal{WM}) stratum in the bottom-left part of the image, and 6 Layers (from \textnormal{L6} to \textnormal{L1}) moving toward the top-right.}
\label{figure:spatialLIBD_intro}
\end{figure}

\review{Briefly, in the Visium platform, the data collection process is performed by placing a slice of the tissue of interest over a grid of spots, so
that every spot contains a few neighboring cells.} The gene expression of each spot is then characterized, resulting in a dataset made of tens of thousands of genes for each spot,
together with the spatial location of the spots. 
Figure \ref{figure:spatialLIBD_intro} shows an example of human dorsolateral prefrontal cortex (DLPFC) processed with Visium at the Lieber Institute for Brain Development \citep{Maynard_etal.2021}. The colored dots denote a manual annotation of the spots performed by \cite{Maynard_etal.2021}. The dataset is available in the \texttt{R} package \texttt{spatialLIBD} \citep{Pardo_etal.2021}.

The rise of spatial transcriptomics has motivated the development of new statistical methods
that handle the identification of \emph{spatially expressed} (s.e.) genes, i.e., genes with spatial patterns of
expression variation across the tissue.
Specific inferential procedures for detecting such kind of
genes, such as SpatialDE \citep{Svensson_etal:2018} and Trendsceek \citep{Edsgard_etal.2018}, have been proposed only in the last years. These methods are widely
computationally efficient, but sometimes they reach discordant inferential conclusions, and
additionally they fail to account for the correlation of the genes. The very recent algorithm by
\cite{Sun_etal:2020}, called SPARK, has addressed some of the limitations of the earlier methods.
However, the additional information brought by the new spatial transcriptomic platforms has
raised several questions, both on the biological and the statistical side: detecting the s.e. genes is thus
not the end of the analysis but just its beginning.
In this article, we want to focus on three specific research questions, i.e., to determine:
\begin{enumerate}
\item[\emph{i.)}]  \review{
	the clustering of the areas of the tissue sample according to the spatial variation of the genes};
\item[\emph{ii.)}] the existence of clusters of genes which are s.e. only in some of the areas discovered from \emph{i.)};
\item[\emph{iii.)}] the highly variable genes in the areas discovered from \emph{i.)} net of any spatial effect. 

\end{enumerate}
Research question \emph{i.)} is fundamental for the analysis of tissue samples because it is the starting point for successive downstream analyses.  The recent GIOTTO \citep{Dries_etal.2021} and BayesSpace \citep{Zhao_etal.2021} methods are unsupervised clustering algorithms for spot-based spatial transcriptomics, \review{designed for inferring the cell types making up a tissue. They perform a clustering based on the principle that neighboring spots are likely to be annotated with the same label, without exploiting the information carried by s.e. genes. Thus, these methods respond to a substantially different research question than \emph{i.)}. } 

Research question \emph{ii.)} is of great scientific interest, but, to the best of our knowledge, has not been tackled yet. Discovering that some genes are s.e. only in some areas of the tissue would play a core role in comprehending some fundamental biological mechanisms, and ultimately discovering new ones. Even the very recent SPARK method for detecting s.e. genes is not designed to state if  the spatial expression activity of a gene is restricted to specific areas of the tissue. With the existing statistical tools, one can approach this issue with a two-step analysis, first clustering the image using BayesSpace or GIOTTO, and then applying SPARK to each of the discovered clusters. However, such heuristic procedure has some \review{severe} limitations. 
First, repeating the tests in each of the image cluster requires to control for multiple testing, e.g., by controlling the False Discovery Rate \citep{Benjamini_Hochberg.1995}. Second, even after the s.e. genes are isolated, an additional clustering of the genes is necessary to perform specific downstream analyses \citep{Svensson_etal:2018, Sun_etal:2020}. Last, if indeed there are clusters of genes, such information should be accounted for in the first step of the procedure, when the image is clustered. \review{However, this is something that cannot be accomplished with BayesSpace or GIOTTO.}

Finally, research question \emph{iii.)} has the goal of determining which genes are active in each of the image cluster. Thanks to the spatial mapping of the spots, it will be possible to separate the presence of spatial effects from the total variation of each gene, providing a more accurate list of highly variable genes.




\subsection{A co-clustering perspective}
\label{subsec:intro_coclustering_perspective}
In this article, we consider the problem of modelling and clustering gene expression profiles in  a tissue sample processed with a spot-based spatial transcriptomic method, such as 10X Visium, and measured over a set of spatially located sites. 

In the remainder of the article, we use ``spots'' to denote the spots in the tissue from which RNA is extracted and ``genes'' to denote the variables measured in each spot, using a terminology typical of the Visium platform. However, the method presented here is more general and can be applied to any spatial transcriptomic technology and, more broadly, to any dataset for which the rows or the columns are measured in some observational sites with known coordinates.

We tackle the research questions outlined above as a single, two-directional clustering problem: 
of the genes, using spots as variables, and of the spots, using genes as variables. This kind of  procedure is known in the literature as \emph{co-clustering} (or \emph{block-clustering}, \citealp{Bouveyron_etal.2019}) and denotes the act of clustering both the rows and the columns of a data matrix, which, in this way, is partitioned into rectangular, non-overlapping sub-matrices called \emph{co-clusters} (or \emph{blocks}).

\cite{Bouveyron_etal.2019} distinguish between \emph{deterministic} and \emph{model-based} co-clustering approaches. Model-based methods are designed to simultaneously perform the clustering and reconstruct the probabilistic generative mechanism of the data. 
The model-based co-clustering literature is centered around the Latent Block Model \citep[LBM;][]{Govaert_Nadif.2013}, an extension of the standard mixture modelling approach when both rows and columns of a data matrix are deemed to come from some underlying clusters. Thanks to the ease of interpretation and to the raise of new advanced computational methods, the LBM has been extensively explored as a tool for modelling continuous \citep[Chapter 5]{Govaert_Nadif.2013}, categorical \citep{Keribin_etal.2015}, count \citep{Govaert_Nadif.2010}, binary \citep{Govaert_Nadif.2008} and recently even functional data \citep{Bouveyron_etal.2018,Casa_etal.2021}. In addition, both frequentist \citep{Govaert_Nadif.2008,Bouveyron_etal.2018} and Bayesian \citep{Wyse_Friel.2012,Keribin_etal.2015} approaches have been proposed for fitting these models.
The conditional independence assumption of LBM states that the observations within the same co-cluster are independent.  
Surely, this hypothesis is computationally attractive, yet it is incompatible with the high correlation levels shown by gene expression data \citep{Efron.2009}.

\cite{Tan_Witten.2014} overcome the conditional independence assumption proposing a co-clustering model based on the matrix variate Gaussian distribution \citep{Gupta_Nagar.2018}, which accounts for the dependency across the rows and the columns in a block with two non-diagonal covariance matrices. Their model represents a first attempt to extend k-means-type algorithms for co-clustering to the case where the data entries in a block are not independent. The estimation of the needed covariance matrices is challenging; a challenge that can be overcome with the aid of a penalization term, such as the LASSO \citep{Witten_Tibshirani.2009}, to avoid singularity problems. However, with spatial data, it is natural to leverage the spatial dependencies observed in the data to aid the covariance matrix estimation. 

Here, we propose \spartaco~(SPAtially Resolved TrAnscriptomics CO-clustering), a novel co-clustering technique designed for discovering the hidden block structure of spatial transcriptomic data. Since the spots in which gene expression is measured are spatially located on a grid, our model 
expresses the correlation across transcripts in different spots as a function of their distances.
As a consequence, differently from the rest of the co-clustering models proposed in the literature, \spartaco~divides the data matrix into blocks based on the estimated means, variances, and spatial covariances. 
In addition, we use gene-specific random effects to account for the remaining covariance  not explained by the spatial structure.

Although the published literature is not always clear about the distinction between \emph{co-clustering} and \emph{biclustering}, in accordance with the recent works of \cite{Moran_etal.2021} and \cite{Murua_Quintana.2021} here we adopt the following terminology: both co-clustering and biclustering are families of techniques used to group the rows and the columns of a data matrix. However, in biclustering the groups formed, called \emph{biclusters}, can take any possible shape, while co-clustering is limited to rectangular, non-overlapping blocks. In addition, biclustering algorithms do not necessarily allocate all the data entries into one of the existent biclusters, and so some entries can be left unassigned. Although biclustering methods are more flexible, the main advantage of co-clustering is that the returned blocks are often easier to interpret both from a statistical and practical perspective.

\subsection{Outline}
The rest of the manuscript is structured as follows. Section 2 illustrates the \spartaco~modelling approach and reviews some competing co-clustering models, highlighting the similarities and the differences with our proposal. Section 3 discusses some identifiability issues, illustrates our classification-stochastic EM (CS-EM) algorithm for parameter estimation, \review{proposes a measure to quantify the clustering uncertainty}, and derives a model selection criterion based on the \emph{integrated completed log-likelihood} \citep{Biernacki_etal.2000}. Section 4 proposes five simulated spatial experiments of growing complexity with whom we compare \spartaco~with other co-clustering models. Section 5 shows how our proposal allows to answer our three research questions using the human brain tissue sample displayed in Figure \ref{figure:spatialLIBD_intro}. The manuscript is concluded by some considerations of the possible future extensions.

\section{The statistical model}
\label{sec:the_statistical_model}


Let $\X=(x_{ij})_{1\leq i\leq n, 1\leq j \leq p}$ be the $n\times p$ matrix of a spatial experiment processed by a spot-based spatial transcriptomic platform, i.e, containing the expression of $n$ genes over a grid of $p$ spots on the chip surface. 
The spatial location of the spot $j$ over the chip surface is known through its spatial coordinates $\mathbf{s}_j = (s_{jx},s_{jy})$; we name as $\mathbf{S}=(\mathbf{s}_j)_{1\leq j \leq p}$ the $p\times 2$ matrix containing the  coordinates of the $p$ spots. From this point, we assume that the data entries in $\X$ have been properly pre-processed, and so $x_{ij}\in \mathbb{R}$ for any $i$ and $j$ \review{(see Section \ref{sec:Application})}. 


\subsection{Model formulation}
\label{subsec:coclustering_model}

We assume there exist $K$ clusters of rows of $\X$, and $R$ clusters of columns of $\X$, forming a latent structure of $KR$ blocks. The vectors of random variables $\mathbfcal{Z} = (\mathcal{Z}_i)_{1 \leq i \leq n}$ and $\mathbfcal{W}=(\mathcal{W}_j)_{1 \leq j \leq p}$ denote to which cluster the rows and the columns belong, respectively. Thus, $\mathcal{C}_k = \{i = 1,\dots,n:\mathcal{Z}_i = k\}$ is the $k$-th row cluster, with $k = 1,\dots,K$, and $\mathcal{D}_r = \{j = 1,\dots,p:\mathcal{W}_j = r\}$ is the $r$-th column cluster, with $r = 1,\dots,R$. The cluster dimensions are $n_k = |\mathcal{C}_k|$ and $p_r=|\mathcal{D}_r|$. The notation used to refer to subsets of $\X$ is the following:  $\X^{kr}=(x_{ij})_{i\in\mathcal{C}_k,j\in\mathcal{D}_r}$ is the $kr$-th co-cluster (block), $\X^{k.}=(x_{ij})_{i\in\mathcal{C}_k,1\leq j \leq p}$ is the $n_k\times p$ matrix formed by all the rows in $\mathcal{C}_k$, and $\X^{.r}=(x_{ij})_{1\leq i \leq n, j\in\mathcal{D}_r}$ is the $n\times p_r$ matrix formed by all the columns in $\mathcal{D}_r$.
When it comes to access the elements of a block, we use the notation $\X^{kr} = (x^{kr}_{ij})_{1\leq i \leq n_k, 1\leq j \leq p_r}$. So, the $i$-th row vector and the $j$-th column vector of  $\X^{kr}$ are respectively $\x^{kr}_{i.}=(x^{kr}_{ij})_{1\leq j \leq p_r}$ and $\x^{kr}_{.j}=(x^{kr}_{ij})_{1\leq i \leq n_k}$.

The vector $\x^{kr}_{i.}$ contains the expression of the $i$-th gene in the cluster $\mathcal{C}_k$ across the $p_r$ spots in the cluster $\mathcal{D}_r$. We model $\x^{kr}_{i.}$  as
\begin{equation}
\label{formula:model_intrablock}
\x^{kr}_{i.}= \mu_{kr}\mathbf{1}_{p_r}+\sigma_{kr,i}\boldsymbol{\epsilon}^{kr}_{i.},\hspace{.7cm}\boldsymbol{\epsilon}^{kr}_{i.}\sim \mathcal{N}_{p_r}(\mathbf{0},\bDelta_{kr}),
\end{equation}
\begin{equation}
\label{formula:covariancematrix}
\bDelta_{kr} = \tau_{kr}\mathbfcal{K}(\mathbf{S}^r;\bphi_r)+\xi_{kr}\mathds{I}_{p_r},
\end{equation}
where $\mu_{kr}$ is a scalar mean parameter, $\mathbf{1}_{p_r}$ is a vector of ones, 
$\sigma^2_{kr,i}$ is a gene-specific variance, 
and $\bDelta_{kr}$ is the covariance matrix of the columns. Following \cite{Svensson_etal:2018} and \cite{Sun_etal:2020}, Formula \eqref{formula:covariancematrix} expresses $\bDelta_{kr}$ as a linear combination of two matrix terms: $\mathds{I}_{p_r}$ is a diagonal matrix of order $p_r$, $\mathbfcal{K}(\mathbf{S}^r;\bphi_r) = \left(\textit{k}(||\mathbf{s}^r_j-\mathbf{s}^r_{j'}||;\bphi_r)\right)_{1\leq j,j' \leq p_r}$ is the spatial covariance matrix, where ${\textit{k}}(\cdot;\bphi_r)$ is an \emph{isotropic} spatial covariance function \citep{Cressie:2015} 
parametrized by a vector $\bphi_r$, and $\mathbf{S}^r=(\mathbf{s}_j)_{j\in\mathcal{D}_r}$ is the sub-matrix of $\mathbf{S}$ containing the spots in $\mathcal{D}_r$. The term isotropic denotes that the covariance between two points $j,j'\in \mathcal{D}_r$ depends just on the distance between their two sites, $||\mathbf{s}^{r}_j-\mathbf{s}^{r}_{j'}||$. The positive parameters $\tau_{kr}$ and $\xi_{kr}$ in Formula \eqref{formula:covariancematrix} handle the linear combination between $\mathbfcal{K}$ and  $\mathds{I}_{p_r}$: the former measures the spatial dependence of the data, the latter is the so-called \emph{nugget effect}, a residual variance.  

According to Section 2.4 of \citet{Cressie:2015}, to select an adequate spatial covariance kernel for the data, one can explore the empirical spatial dependency through the \emph{variogram} and then select a kernel from a vast list of proposals (see for example \citealp{Rasmussen_Williams:2006}). However, under our model, this strategy would  be unfeasible because only the columns within the same cluster are spatially dependent, so the selection of the spatial covariance kernel should be performed simultaneously with the clustering of the data. 
As a compromise, \spartaco~considers the same covariance model $k(\cdot;\bphi_r)$ for every column cluster $\mathcal{D}_r$; the only difference among the  kernels of the clusters is the value of the model parameters $\bphi_r$. 


The scale parameters $\sigma^2_{kr,i}$ in \eqref{formula:model_intrablock}  aim to capture the variability left unexplained by the spatial covariance model \eqref{formula:covariancematrix}, and \review{possibly the extra source of variability due to the dependency across genes.} In the longitudinal data framework, \citet{Delacruz-masia_Marshall:2006} and \cite{Anderlucci_Viroli.2015} consider a random effect model  to account for the systematic dependency across subjects in the same group of study. We follow the same approach and we assume that every $\sigma^2_{kr,i}$ is a realization of an Inverse Gamma distribution $\mathcal{IG}(\alpha_{kr},\beta_{kr})$, where $\alpha_{kr}$ and $\beta_{kr}$ denote the shape and the rate, respectively. 
The Inverse Gamma is chosen for its conjugacy with the Gaussian distribution and allows to derive the marginal probability density of $\x^{kr}_{i.}$, that is 
\begin{equation}
\label{formula:marginal_model}
f(\x^{kr}_{i.};\btheta_{kr},\bphi_r) = \frac{1}{\sqrt{(2\pi)^{p_r} \mathrm{det}(\bDelta_{kr})}}\frac{\Gamma(\alpha^*_{kr,i})}{\Gamma(\alpha_{kr})}\frac{\beta_{kr}^{\alpha_{kr}}}{{\beta^*_{kr,i}}^{\alpha^*_{kr,i}}},
\end{equation}
\sloppy
where $\mathrm{det}(\cdot)$ denotes the matrix determinant, $\alpha^*_{kr,i}=p_r/2+\alpha_{kr}$ and $\beta^*_{kr,i} = (\x^{kr}_{i.}-\mu_{kr}\mathbf{1}_{p_r})^T\bDelta^{-1}_{kr}(\x^{kr}_{i.}-\mu_{kr}\mathbf{1}_{p_r})/2+\beta_{kr}$. 
Note that this formulation \review{corresponds to the probabilistic model $\x^{kr}_{i.}\sim t_{2\alpha_{kr}}({\mu}_{kr}\mathbf{1}_{p_r},\alpha^{-1}_{kr}\beta_{kr}\bDelta_{kr})$ and } is similar to that employed to \textit{shrink} the gene variances in the popular \textit{limma} model \citep{smyth2004linear}. The set of parameters $\btheta_{kr} = \{\mu_{kr},\tau_{kr}, \xi_{kr},\alpha_{kr},\beta_{kr}\}$ is specific of the data into the $(k,r)$-th co-cluster, while $\bphi_r$ is a parameter that is descriptive of the entire $r$-th column cluster. 

The model in Formula \eqref{formula:model_intrablock} can be rephrased with a probability distribution over the entire $kr$-th block, $\X^{kr}|\bSigma_{kr}\sim \mathcal{MVN}(\mu_{kr}\mathbf{1}_{n_k\times p_r}, \boldsymbol{\Sigma}_{kr},\boldsymbol{\Delta}_{kr})$, where $\mathcal{MVN}$ denotes the matrix-variate normal distribution and $\boldsymbol{\Sigma}_{kr} = \mathrm{diag}(\sigma^2_{kr,1},\dots,\sigma^2_{kr,n_k})$ is the (diagonal) covariance matrix of the genes. A consequence of the matrix-variate normal model is that every row, column and sub-matrix of $\X^{kr}$ is Gaussian \citep{Gupta_Nagar.2018}. For instance, the following model formulation is equivalent to Formula \eqref{formula:model_intrablock}:
\begin{equation*}
\x^{kr}_{.j}|\bSigma_{kr}\sim \mathcal{N}_{n_k}\left\{\mu_{kr}\mathbf{1}_{n_k},(\tau_{kr}+\xi_{kr})\bSigma_{kr}\right\},\hspace{.8cm}\mathrm{Cov}(\x^{kr}_{.j},\x^{kr}_{.j'})=\tau_{kr}k(||\mathbf{s}^r_j-\mathbf{s}^r_{j'}||;\bphi_r)\bSigma_{kr},
\end{equation*}
with $j,j'\in \mathcal{D}_r$.

\begin{figure}[th!]
\centering
\includegraphics[width=0.6\linewidth]{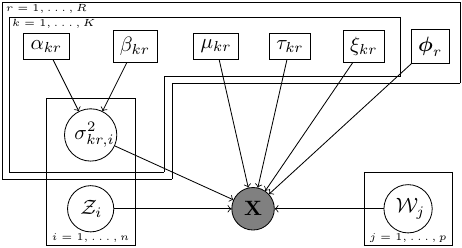}
\caption{DAG of the SpaRTaCo co-clustering model. Grey circle denotes the data, white circles are the latent random variables, and white rectangles are the model parameters.}
\label{figure:graph_model}
\end{figure}

Last, the clustering labels $\mathbfcal{Z}$ and $\mathbfcal{W}$ are  unknown independent random variables. Figure \ref{figure:graph_model} represents the relations across the elements of the model with a DAG.

\subsection{A comparison with other co-clustering models}
\label{subsec:comparison}
We review in this section some advanced co-clustering techniques that have some similarities with our proposal. The goal is to highlight, starting from the existing literature, how \spartaco~has been designed specifically for detecting  and clustering data based on their spatial covariance in some groups of observational sites. With respect to the distinction between deterministic and model-based co-clustering techniques we already discussed in Section \ref{subsec:intro_coclustering_perspective},  we choose to compare \spartaco~only with model-based techniques because they offer a clear advantage in the interpretation of the results. Some of the methods that we review here are named as biclustering models, but in practice they segment the data matrix into rectangular blocks. 





\emph{Sparse Biclustering} (\textsc{sparseBC}, \citealp{Tan_Witten.2014}) extends the \emph{k-means} algorithm to the co-clustering framework. The model corresponds to a probabilistic assumption on the block of the type $\X^{kr}\sim \mathcal{MVN}(\mu_{kr}\mathbf{1}_{n_k \times p_r},\mathds{I}_{n_k},\xi\mathds{I}_{p_r})$, where $\xi$ is an unknown scale parameter. In \textsc{sparseBC}, the  estimation of $\mu_{kr}$, for any $k$ and $r$, is regulated by a LASSO penalization. We thus distinguish the sparse estimation from the case of null penalization (\textsc{BC}). 

\emph{Matrix-Variate Normal Biclustering} (\textsc{MVNb}, \citealp{Tan_Witten.2014}) extends \textsc{sparseBC} by taking a probabilistic model on the blocks of the type $\X^{kr}\sim \mathcal{MVN}(\mu_{kr}\mathbf{1}_{n_k \times p_r},\bSigma^{\textsc{MVNb}}_k, \bDelta^{\textsc{MVNb}}_r)$, where both $\bSigma^{\textsc{MVNb}}_k$ and $\bDelta^{\textsc{MVNb}}_r$ are non-diagonal covariance matrices with respectively ${n_k (n_k+1)}/{2}$ and ${p_r (p_r+1)}/{2}$ free parameters. Together with the LASSO penalization on the centroids, handled by a parameter $\lambda$, the authors deploy a graphical LASSO penalization \citep{Witten_Tibshirani.2009} to practically solve the singularity problems in the estimate of $\bSigma^{\textsc{MVNb}}_{k}$ and $\bDelta^{\textsc{MVNb}}_{r}$. The penalization parameters involved are denoted by $\rho_{\bSigma}$ and $\rho_{\bDelta}$. With respect to the \textsc{MVNb}, \spartaco~has specific row and column covariance matrices $\bSigma_{kr}$ and $\bDelta_{kr}$ for each block, whose structure is described in Section \ref{subsec:coclustering_model}. The total number of free parameter, $K R|\boldsymbol{\theta}_{kr}|+R|\bphi_r|$, does not grow either with $n$ or $p$. As a direct consequence, the parameter estimation of \spartaco, conditioning on the clustering labels $\Z$ and $\W$, remains much less computationally prohibitive than the one of the \textsc{MVNb}, specially when the sample size becomes considerably large. 

\emph{Latent Block Model} 
is a vast class of statistical models that can be seen as an extension of the mixture model  for co-clustering problems. The model for continuous data (\citealp[Chapter 5]{Govaert_Nadif.2013}) can be written using the Matrix Variate Normal representation as $\X^{kr}\sim \mathcal{MVN}(\mu_{kr}\mathbf{1}_{n_k \times p_r},\mathds{I}_{n_k},\xi_{kr}\mathds{I}_{p_r})$ and so it is based on the assumption that the data entries in a block are independent given the clustering labels (conditional independence). The intra-block model is thus a special case of \spartaco~when $\bSigma_{kr} = \mathds{I}$ and $\tau_{kr}=0$, for all $k$ and $r$. However, the \textsc{LBM} is more general on the probabilitistic assumptions over the clustering variables. In fact, it assumes $\mathrm{Pr}(\mathcal{Z}_i = k)=\pi_k$ and  $\mathrm{Pr}(\mathcal{W}_j = r)=\rho_r$, where $(\pi_1,\dots,\pi_K)$ and $(\rho_1,\dots,\rho_R)$ are probability vectors such that $\sum_{k=1}^{K}\pi_k = \sum_{r=1}^{R}\rho_r = 1$, while \spartaco~implicitly assumes that $\mathrm{Pr}(\mathcal{Z}_i = k)=1/K$ and $\mathrm{Pr}(\mathcal{W}_j = r)=1/R$ for any $k$ and $r$.

Supplementary Figure 1 \citep{Sottosanti_Risso.2022_supplementary} gives a summary of the relations across \spartaco~and the co-clustering models discussed in this section.

\section{Inference}

\subsection{Identifiability}
\label{subsec:identifiability}
The model as expressed in Formula \eqref{formula:model_intrablock} is not identifiable in the covariance term: in fact, for any $a> 0$, $\sigma^2_{kr,i}\cdot\bDelta_{kr} = a\sigma^2_{kr,i}\cdot \bDelta_{kr}/a=\tilde{\sigma}^2_{kr,i}\cdot\tilde{\bDelta}_{kr}$. This issue generates in practice an infinite number of solutions for the parameter estimate. 

A typical workaround to get unique parameter estimates consists in setting the value of some covariance parameters. In our model, this would mean taking  $\sigma^2_{kr,i}=c$, for one $i$ in $\{1,\dots,n_k\}$, using an arbitrary positive constant $c$. Incidentally, this is equivalent to constraint $\mathrm{tr}(\bSigma_{kr})$, the trace of the matrix $\bSigma_{kr}$ \citep{Allen_Tibshinari.2010, Caponera_etal.2017}. However, we discard this solution as, under our model, the rows of the data matrix are involved into a clustering procedure. Thus, it is not possible to define which $i$ in a cluster should take the constraint. 


The solution we adopt for our model puts the identification constraint on $\bDelta_{kr}$ \citep{Anderlucci_Viroli.2015}. Since $\mathrm{tr}(\bDelta_{kr}) = p_r(\tau_{kr}+\xi_{kr})$, we constraint the quantity  $\tau_{kr}+\xi_{kr} = c_{\bDelta}$, where $c_{\bDelta}$ is an arbitrary positive constant. Such constraint has a notable practical consequence: in fact, once the estimate $\hat{\tau}_{kr}$ is determined within the constrained domain $(0,c_{\bDelta})$, then $\hat{\xi}_{kr}$ is simply taken by difference as $\hat{\xi}_{kr} = c_{\bDelta}-\hat{\tau}_{kr}$.
Hence, we can only interpret $\hat{\tau}_{kr}$ and $\hat{\xi}_{kr}$ in relation to each other and not in absolute terms. According to \cite{Svensson_etal:2018}, in our applications (Sections \ref{sec:Simulations} and \ref{sec:Application}) we will consider the quantity  $\tau_{kr}/\xi_{kr}$ that we called \emph{spatial signal-to-noise ratio}. This ratio is easily interpretable because it represents the amount of spatial expression of the genes in a cluster with respect to the nugget effect.

\subsection{Model estimation}
\label{subsec:model_estimation}

To estimate \spartaco, we propose an approach based on the maximization of the \emph{classification log-likelihood}, that is
\begin{equation}
\label{formula:classification_loglikelihood}
\log \mathcal{L}(\boldsymbol{\Theta},\mathbfcal{Z},\mathbfcal{W}) = \sum_{i = 1}^{n}\sum_{k = 1}^{K}\mathds{1}(\mathcal{Z}_i = k)\left\{\sum_{r = 1}^{R}\log f(\x^{.r}_{i.};\btheta_{kr},\bphi_r)\right\},
\end{equation}
where $\boldsymbol{\Theta}=\bigcup_r \left\{\bigcup_k \btheta_{kr},\bphi_r\right\}$, $\x^{.r}_{i.}$ is the $i$-th row of the matrix $\X^{.r}$ and $f(\cdot;\cdot)$ is given in Formula \eqref{formula:marginal_model}. Notice that the correlation across the columns does not allow to write the $\mathbfcal{W}$ explicitly. This issue does not concern the $\mathbfcal{Z}$, because the rows are independent.

Chapter 2 of \cite{Bouveyron_etal.2019} makes a clear distinction between the classification and the \emph{complete log-likelihood} (the latter includes an additional part related to the distribution of the clustering labels). However, since \spartaco~implicitly assumes that $\mathrm{Pr}(\mathcal{Z}_i = k)=1/K$ and $\mathrm{Pr}(\mathcal{W}_j = r)=1/R$ for any $k$ and $r$, then there is no practical difference between  classification and  complete log-likelihood.

The classification log-likelihood can be maximized with a \emph{classification EM} algorithm (CEM, \citealp{Celeux_Govaert.1992}), a modification of the standard EM  which allocates the observations into the clusters during the estimation procedure. The CEM  is an iterative algorithm which alternates between a classification step (CE Step), where the estimates of ${\Z}$ and ${\W}$ are updated, and a maximization step (M Step), which updates the parameter estimates of ${\boldsymbol{\Theta}}$. The benefits brought by such algorithm are particularly visible when complex models as the LBM are employed, because the joint conditional distribution $p(\Z,\W|\X;{\bTheta})$ is not directly available \citep{Govaert_Nadif.2013}. 

Under \spartaco, a direct update of ${\W}$ through a CE step is unfeasible due to the correlation across the columns, and so the estimation algorithm requires some modifications. This issue was already discussed by \cite{Tan_Witten.2014} for their \textsc{MVNb} model; however, their solution consists in an heuristic estimation algorithm with no guarantees of convergence. We propose to perform a stochastic allocation (SE step), where the column clustering configuration ${\W}$ is sampled from 
a Markov 
chain whose limit distribution is the conditional distribution 
$p(\W|{\Z},\X;{\bTheta})$. This step can be performed using the Metropolis-Hastings algorithm. 
A stochastic version of the EM algorithm was previously employed also for estimating the LBM by \cite{Keribin_etal.2015}, \cite{Bouveyron_etal.2018} and \cite{Casa_etal.2021}. Because of the alternation of a classification move, a stochastic allocation move and a maximization move, we name our algorithm \emph{classification-stochastic EM} (CS-EM).
We denote with $({\boldsymbol{\Theta}},{\Z},{\W})^{(t-1)}$ the estimate of the model parameters and of the clustering labels at iteration $t-1$. At step $t$, the algorithm executes the following steps:
\begin{itemize}
\item \textbf{CE Step}: keeping fixed $({\W},{\boldsymbol{\Theta}})^{(t-1)}$, update the row clustering labels with the following rule:
\begin{equation*}
\label{formula:row_clustering_allocation_rule}
{\mathcal{Z}}^{(t)}_i = \argmax_{k=1,\dots,K} \frac{\prod_{r=1}^{R}f\left(\x^{.r}_{i.};\btheta^{(t-1)}_{kr},\bphi^{(t-1)}_r\right)}{\sum_{k'=1}^{K} \left\{\prod_{r=1}^{R}f\left(\x^{.r}_{i.};\btheta^{(t-1)}_{k'r},\bphi^{(t-1)}_r\right) \right\}  },\hspace{1cm}i=1,\dots,n.
\end{equation*}

\item \textbf{SE Step}: keeping fixed ${\mathbfcal{Z}}^{(t)}$ and ${\boldsymbol{\Theta}}^{(t-1)}$, generate a candidate  clustering configuration $\W^*$ by randomly changing some elements from the starting configuration  $\W^{(t-1)}$. 
Let $m$ be the number of elements of $\W^{(t-1)}$ that we attempt to change: $m$ can be either fixed or randomly drawn from a discrete distribution. 
To formulate $\W^*$, we exploit two moves.

\vspace{.5cm}
\noindent
\textbf{(M1)} Two clustering labels $g_1\sim\mathcal{U}(\{1,\dots,R\})$ and $g_2\sim\mathcal{U}(\{1,\dots,R\}\setminus \{g_1\})$ are drawn.
The candidate configuration $\W^*$ is made by selecting $m$ observations from $\W^{(t-1)}$ at random with label $g_1$ and changing their label to $g_2$. 
The quantity
$$
\dfrac{q(\W^{(t-1)}|\W^*)}{q(\W^*| \W^{(t-1)})}=\dfrac{p_{g_1}!p_{g_2}!}{(p_{g_1}-m)!(p_{g_2}+m)!}
$$ 
is the ratio of transition probabilities employed by the Metropolis-Hastings algorithm to evaluate $\W^*$, where $q(\W^*| \W^{(t-1)})$ and $q(\W^{(t-1)}| \W^*)$  are respectively the probabilities of passing from configuration $\W^{(t-1)}$ to $\W^*$ and \emph{vice-versa}. This move almost coincides with the (M2) move of \cite{Nobile_Fearnside.2007}. 

\vspace{.5cm}
\noindent
\textbf{(M2)} For $h=1,\dots,m$, the clustering labels $g_{1h}\sim\mathcal{U}(\{1,\dots,R\})$ and $g_{2h}\sim\mathcal{U}(\{1,\dots,R\}\setminus \{g_{1h}\})$ are drawn.
Let $b_{lr} = \sum_{h=1}^{m}\mathds{1}(g_{lh}=r)$, for $l=1,2$ and $r = 1,\dots,R$. Then the candidate configuration $\W^*$ is made by changing the labels of $b_{1r}$ observations selected at random from the group $r$, when $b_{1r}>0$,  to $g_{2\kappa(r)}$, where $\kappa(r)=\{h=1,\dots,m: g_{1h}=r\}$. The ratio of transition probabilities is
$$
\dfrac{q(\W^{(t-1)}|\W^*)}{q(\W^*| \W^{(t-1)})}=\prod_{r:b_{2r}>0}\dfrac{b_{2r}!(p_r-b_{1r})!}{(p_r-b_{1r}+b_{2r})!}{\bigg /}\prod_{r:b_{1r}>0}\dfrac{b_{1r}!(p_r-b_{1r})!}{p_r!}.
$$

\vspace{.5cm}
\noindent
The choice between (M1) and (M2) is random. The candidate configuration $\W^*$ is accepted with probability $\min\{1, A\}$, where $A$ is the following Metropolis-Hastings ratio:
$$
A = \dfrac{\mathcal{L}(\boldsymbol{\Theta}^{(t-1)},\mathbfcal{Z}^{(t)},\mathbfcal{W}^{*})}{\mathcal{L}(\boldsymbol{\Theta}^{(t-1)},\mathbfcal{Z}^{(t)},\mathbfcal{W}^{(t-1)})} \dfrac{q(\W^{(t-1)}|\W^*)}{q(\W^*| \W^{(t-1)})}.
$$

Within the same iteration $t$, the SE Step can be run for an arbitrary large number of times to accelerate the exploration of the space of clustering configurations and so the convergence of the estimation algorithm to a stationary point. From our experience, we suggest to repeat the SE Step for at least 100 times per iteration.

\item \textbf{M Step}: using the rows in $\mathcal{C}^{(t)}_k$ and the columns in $\mathcal{D}^{(t)}_r$, update the parameter estimates $\btheta^{(t)}_{kr}$ and $\bphi^{(t)}_{r}$. The derivative of the log-likelihood with respect to $(\btheta_{kr},\bphi_r)$ does not lead to closed solutions for updating the model parameters, and for this reason a numerical optimizer must be applied. We exploit the \textsc{L-BFGS-B} algorithm of \cite{Byrd_etal.1995} implemented in the \texttt{stats} library of the \texttt{R} computing language, which allows constrained optimization; this aspect is particularly useful to estimate ${\tau}_{kr}$ under the identifiability constraint described in Section \ref{subsec:identifiability}.
\end{itemize}
\review{Following \cite{Tan_Witten.2014}, our implementation of the estimation algorithm alternates each allocation step, either the CE Step and the SE Step, with an M Step.}
As pointed by \cite{Keribin_etal.2015}, the SE Step is not
guaranteed to increase the classification log-likelihood at each iteration, but it generates
an irreducible Markov chain with a unique stationary distribution
which is expected to be concentrated around the maximum likelihood
parameter estimate. 
The estimation algorithm must be run for a \review{sufficiently} large number of iterations. \review{We additionally implemented a convergence criterion that stops the algorithm if the increment of the classification log-likelihood is smaller than a certain threshold for a given number of iterations in a row.} The final estimates of $(\hat\bTheta,\hat\Z,\hat\W)$ are the values obtained at the iteration from which \eqref{formula:classification_loglikelihood} is maximum.

\review{Notice that the criterion to form the co-clusters that \spartaco~uses  has also a geometrical interpretation; in fact, in the same way that \emph{k-means} minimizes the Euclidean distance between the observations and the centroids, \spartaco~minimizes the Mahalanobis distance of the observations from the block centroids, embedding the spatial structure of the data into the covariance matrix. Therefore, even when the data do not fully respect the probabilistic assumptions, the model is still valid, as a distance-based clustering algorithm.}

\review{
\subsection{Measuring the clustering uncertainty}
\label{subsec:cluster_evaluation}
The proposed estimation procedure should be run multiple times  from different starting points to check if the algorithm encounters some local maxima. In addition, the parallel runs can be used to quantify the uncertainty of the estimated co-clustering structure. In fact, if the analyzed data carry large evidence in favor of a unique clustering configuration, then the parallel runs will return approximately the same row and column clusters. If instead the clustering structure of the data that \spartaco~searches for is not evident, then the multiple runs of the algorithm will tend to discover different but equally likely solutions. 

Let us suppose to run the CS-EM algorithm $S$ times on the same dataset: $(\hat\bTheta^{(s)},\hat\Z^{(s)},\hat\W^{(s)})$ is the solution to the parameter estimate returned by the $s$-th run, for $s = 1,\dots,S$, and $\ell^{(s)} = \log\mathcal{L}(\hat{\bTheta}^{(s)}, \hat{\boldsymbol{\mathcal{Z}}}^{(s)}, \hat{\boldsymbol{\mathcal{W}}}^{(s)})$. 
In addition, let $s^* = \argmax_s \ell^{(s)}$: since the co-clustering structure $(\hat\Z^{(s^*)},\hat\W^{(s^*)})$ has found the largest evidence across the $S$ runs on the current data, it is the final estimate returned by the algorithm. The co-clustering uncertainty can be thought of as a function of the distances between the final estimate, $(\hat\Z^{(s^*)},\hat\W^{(s^*)})$, and the other estimates of lower evidence, $(\hat\Z^{(s)},\hat\W^{(s)})$, 
for $s\neq s^*$. Let $\mathbfcal{I}_k = \{\mathds{1}(\hat{\mathcal{Z}}_i^{(s^*)}=k)\}_{1\leq i \leq n}$  be the binary vector denoting which rows belong to the $k$-th row cluster given by the run  $s^*$, for $k=1,\dots,K$, and $\mathbfcal{I}_{h_s(k)} = [\mathds{1}\{\hat{\mathcal{Z}}_i^{(s)}=h_s(k)]_{1\leq i \leq n}$ be the binary vector denoting which observations belong to the cluster $h_s(k)$ given by the $s$-th run, where $h_s(k) = \argmax_{h = 1,\dots,K}\sum_{i=1}^{n}\mathds{1}(\mathcal{Z}^{(s^*)}_i = k,\mathcal{Z}^{(s)}_i = h)$, and $s\neq s^*$. In addition, let us consider the weights $\omega_s = 1/(\ell^{(s^*)}-\ell^{(s)})$.
The uncertainty of the row cluster $k$ is measured as
\begin{equation}
\label{formula:clustering_uncertainty}
\rowepsilon = \dfrac{\sum_{s \neq s^*}\omega_s\mathrm{CER}(\mathbfcal{I}_k, \mathbfcal{I}_{h_s(k)})}{\sum_{s \neq s^*}\omega_s},
\end{equation}
where $\mathrm{CER}(\cdot,\cdot)$ denotes the \emph{clustering error rate} (\citealp{witten_tibshirani.2010}), an index that measures the disagreement between a reference and an estimated clustering configuration: the closer is CER to 0, the larger is the agreement between the true and the estimated clusters. The  $\{\omega_s\}_{s\neq s^*}$ give a large weight to the CER between $\mathbfcal{I}_k$ and $\mathbfcal{I}_{h_s(k)}$ when $\ell^{(s^*)}-\ell^{(s)}$ is small, and \emph{vice-versa}. The reason is intuitively that, if both $\omega_s$  and $\mathrm{CER}(\mathcal{I}_k,\mathcal{I}_{h_s(k)})$ are large, then there are two considerably different clustering configurations that yield approximately the same log-likelihood value. Thus, the clustering structure of the data is uncertain. If instead $\omega_s$ is small, the difference between $\hat{\Z}^{(s^*)}$ and $\hat{\Z}^{(s)}$ is in practice irrelevant, because the evidence arising from the data clearly leans in favor of  $\hat{\Z}^{(s^*)}$.

Formula \eqref{formula:clustering_uncertainty} can be applied also for computing the uncertainties of the column clusters $(\varepsilon^{\mathrm{cols}}_1,\dots,\varepsilon^{\mathrm{cols}}_R)$, just replacing $\hat{\Z}^{(s)}$ with $\hat{\W}^{(s)}$. The uncertainty measure introduced here can be interpreted similarly to the CER index: the closer are $\rowepsilon$ and $\colepsilon$ to 0, the larger is the evidence of a unique co-clustering structure of the data.
}

\subsection{Model selection}
\label{subsec:model_selection}   
\spartaco~can be run with different spatial covariance models $k(\cdot;\cdot)$ and with different combinations of $K$ and $R$. We consider the problem of selecting the best model for the data, both in terms of the number of clusters and the spatial covariance function, using an information criterion. The most common criteria, the AIC and the BIC, cannot be derived under Model \eqref{formula:model_intrablock} because the  likelihood of the data $p(\X;\boldsymbol{\Theta})$, marginalized with respect to the latent variables $\Z$ and $\W$, is not available in closed form.

In this work, we propose to guide the model selection using the \emph{integrated completed log-likelihood} (ICL, \citealp{Biernacki_etal.2000}). The ICL is a well-established criterion for selecting the number of clusters \citep{Bouveyron_etal.2019} which has become popular in the co-clustering framework for selecting the size of LBM \citep{Keribin_etal.2015,Bouveyron_etal.2018,Casa_etal.2021}. Under Model \eqref{formula:model_intrablock}-\eqref{formula:covariancematrix}, its expression is
\begin{equation}
\label{formula:ICL}
\mathrm{ICL} = \log\mathcal{L}(\hat{\boldsymbol{\Theta}},\hat{\Z},\hat{\W})-n\log K -p\log R - \frac{4KR+\mathrm{dim}(\bphi)R}{2}\log np,
\end{equation}
where $\mathrm{dim}(\bphi)$ is the dimension of the parameter vector $\bphi_r$, which does not depend on $r$. The derivation of \eqref{formula:ICL} is described more in details in Supplementary Section 1. Operatively, the best model from a list of candidates corresponds to the one with the largest value of  \eqref{formula:ICL}.

In the presence of mixed effects, \cite{Delattre_etal.2014} argue that the actual sample size is not trivial to define, and thus the classical information criteria need to be modified. In particular, they derive an alternative formulation of the BIC which includes a  term that depends only on the parameters involved with the random effects. However, their model specification assumes that the marginal distribution of the data with the random parameters integrated out cannot be derived in closed form. Although the presence of the random variances $\sigma^2_{kr,i}$ makes \spartaco~a random effect model, the integration of $\sigma^2_{kr,i}$ from the density function of $\x^{kr}_{i.}|\sigma^2_{kr,i}$ leads to the marginal density \eqref{formula:marginal_model}. For this reason, we do not implement any modification based on the random effects into our information criterion \eqref{formula:ICL}.

\section{Simulation studies}
\label{sec:Simulations}
\subsection{Simulation model}
\label{subsec:simulation_model}

\begin{figure}[th!]
\centering
\includegraphics[width=0.45\linewidth]{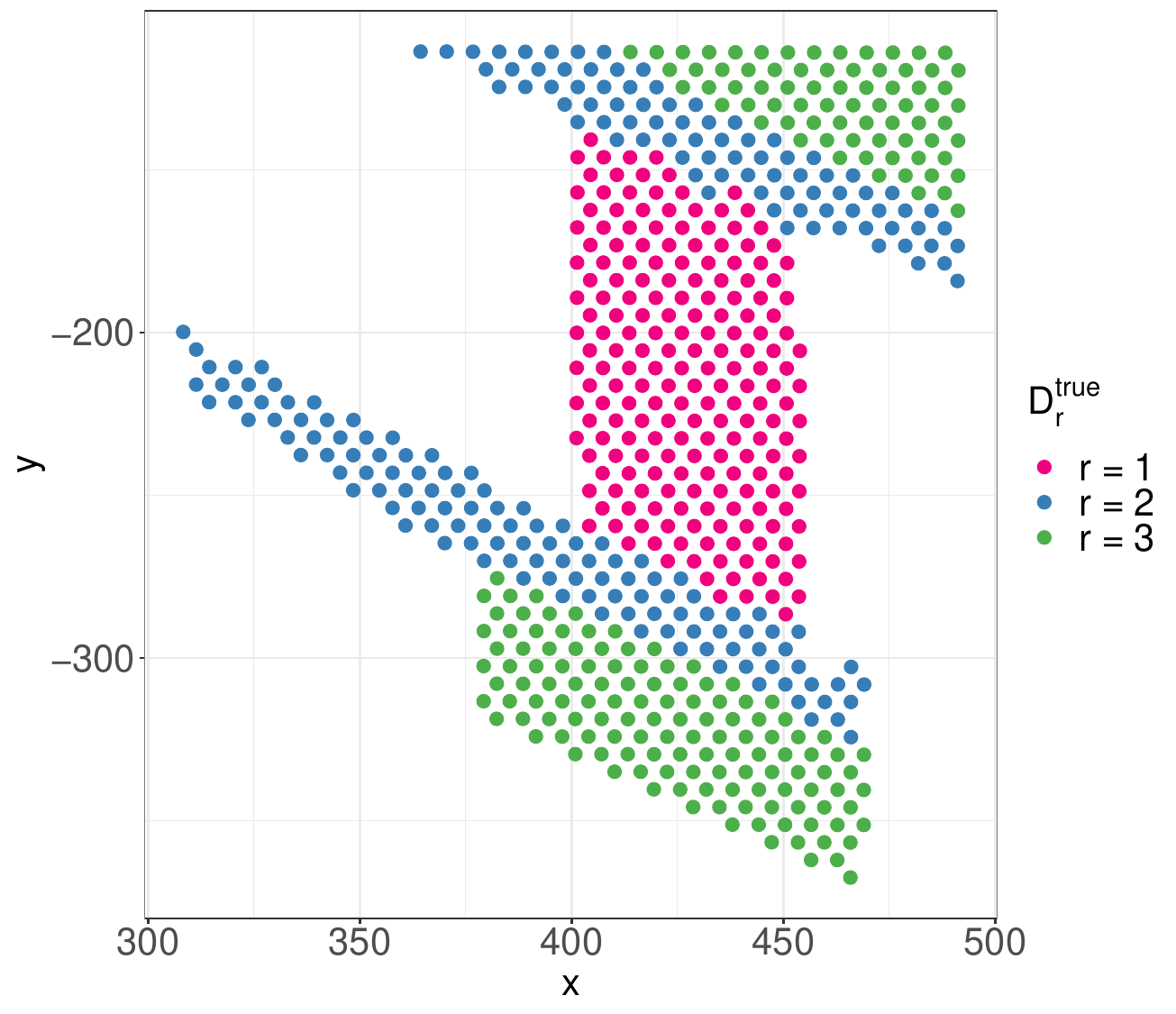}
\includegraphics[width=0.45\linewidth]{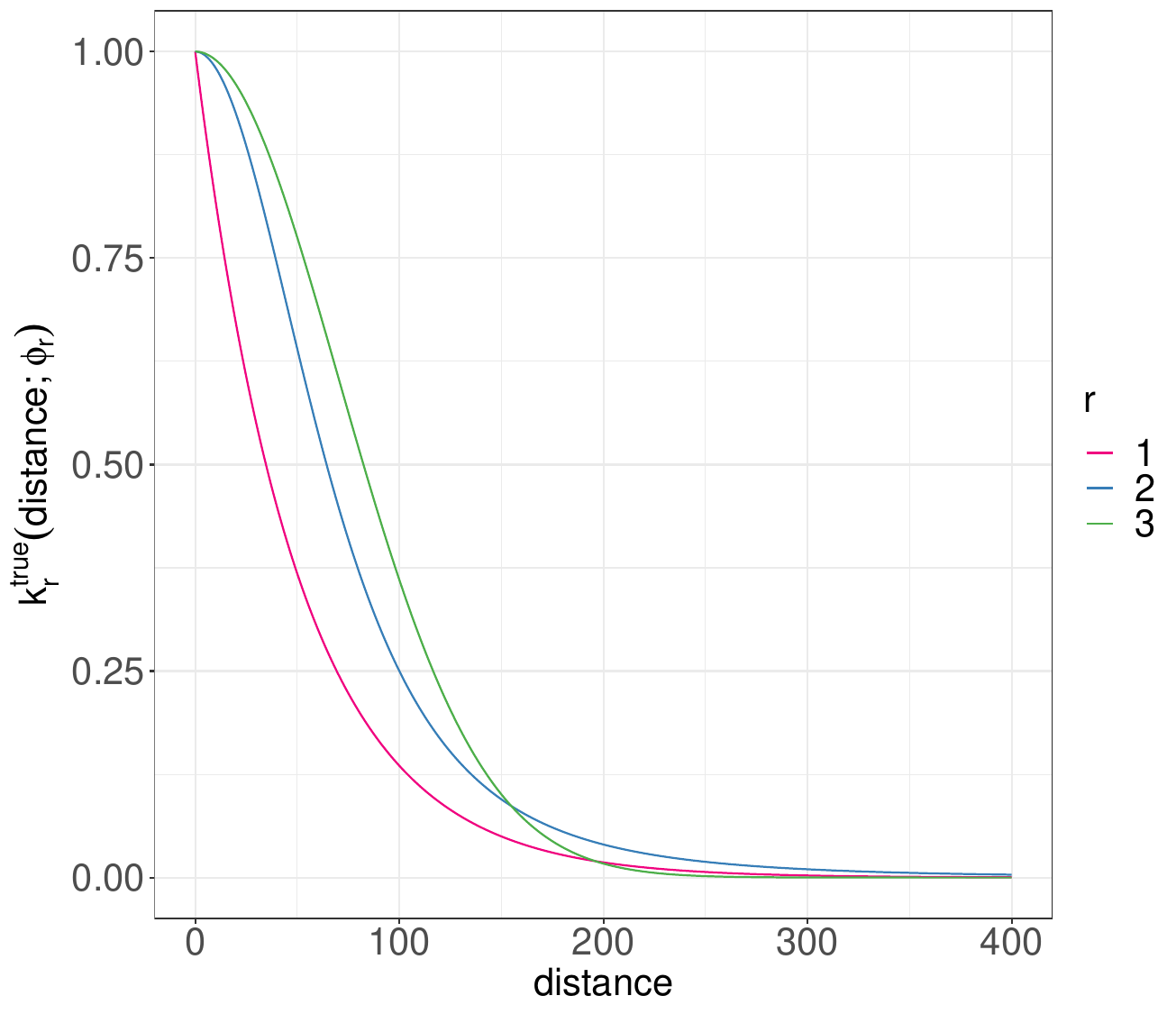}
\caption{Left: map of the spots used to generate the simulation experiments, extracted from the subject 151507 contained in the package \textnormal{\texttt{spatialLIBD}}. The clusters are of equal size, $p_1=p_2=p_3=200$. Right: comparison of the covariance functions used in the three clusters of spots. When $r = 1$, the covariance is Exponential with scale $\theta_E = 50$, when $r = 2$, it is Rational Quadratic with $\theta_R = 50$ and $\alpha_R = 2$, and when $r = 3$ it is Gaussian with scale $\theta_G = 70$.}
\label{figure:simulation_spatialcovariancefunction}
\end{figure}

We study the performance of \spartaco~with five simulated spatial experiments that recreate some possible scenarios that can be found in real data. 
We generate the latent blocks using the matrix-variate normal distribution \citep{Gupta_Nagar.2018} as follows: given the number of row and column clusters $K^{\mathrm{true}}$ and $R^{\mathrm{true}}$ (for convenience, we considered here $K^{\mathrm{true}}=R^{\mathrm{true}}=3$ in every simulation experiment),  the  clustering labels $\mathbfcal{Z}^{\mathrm{true}}$ and $\mathbfcal{W}^{\mathrm{true}}$, and the  clusters $\mathcal{C}^{\mathrm{true}}_k = \{i=1,\dots,n: \mathcal{Z}^{\mathrm{true}}_i=k \}$ and $\mathcal{D}^{\mathrm{true}}_r = \{j=1,\dots,p: \mathcal{W}^{\mathrm{true}}_j=r \}$,  the $(k,r)$-th block is drawn from 
\small{
\begin{equation}
\label{formula:simulation_model}
\X^{kr}\sim\mathcal{MVN}(\mu^{\mathrm{true}}_{kr}\mathbf{1}_{n_k\times p_r}, \bSigma^{\mathrm{true}}_{kr},\bDelta^{\mathrm{true}}_{kr}),\hspace{.6cm}\bDelta^{\mathrm{true}}_{kr} = \tau^{\mathrm{true}}_{kr}\mathbfcal{K}^{\mathrm{true}}_r(\mathbf{S}^r;\bphi^{\mathrm{true}}_r)+\xi^{\mathrm{true}}_{kr}\mathds{I}_{p_r},
\end{equation}
}
where $\mathbfcal{K}^{\mathrm{true}}_r(\mathbf{S}^r;\bphi_r) = \left( k^{\mathrm{true}}_r(||\mathbf{s}^r_j-\mathbf{s}^r_{j'}||;\bphi^{\mathrm{true}}_r)\right)_{1 \leq j,j' \leq p_r}$, and $k^{\mathrm{true}}_r(\cdot;\bphi^{\mathrm{true}}_r)$ is an isotropic spatial covariance kernel parametrized by $\bphi^{\mathrm{true}}_r$.
Note that, differently from \eqref{formula:covariancematrix}, the presence of the subscript $r$ into the kernel matrix $\mathbfcal{K}^{\mathrm{true}}_r$ denotes that the spatial covariance function can be different for any column cluster. 
In our simulations, we employed the \emph{Exponential} kernel with scale $\theta_E$ for the columns in $\mathcal{D}^{\mathrm{true}}_1$, the \emph{Rational Quadratic} kernel with parameters $(\theta_R,\alpha_R)$ for the columns in $\mathcal{D}^{\mathrm{true}}_2$, and the \emph{Gaussian} kernel (known also as \emph{Squared Exponential}) with scale $\theta_G$ for the columns in $\mathcal{D}^{\mathrm{true}}_3$. Their formulation is reported in Supplementary Section 2 and it is further discussed in Chapter 4 or \citet{Rasmussen_Williams:2006}. 
\review{The simulation model \eqref{formula:simulation_model} implies the following marginal distributions of the genes and of the spots:
\begin{equation}
\label{formula:simulation_model_row_marginal}
\mathbf{x}^{k.}_{i.}|\mathbfcal{Z}^{\mathrm{true}},\mathbfcal{W}^{\mathrm{true}}\sim \mathcal{N}_p\left\{
(\mu^{\mathrm{true}}_{k1}\mathbf{1}_{p_1},\dots,\mu^{\mathrm{true}}_{k3}\mathbf{1}_{p_3}),{\bSigma}^{\mathrm{true}}_{ii}
\mathrm{diag}(\bDelta^{\mathrm{true}}_{kr})_{r=1,2,3}
\right\},
\end{equation}
\begin{equation}
\label{formula:simulation_model_column_marginal}
\mathbf{x}^{.r}_{.j}|\mathbfcal{Z}^{\mathrm{true}},\mathbfcal{W}^{\mathrm{true}}\sim \mathcal{N}_n\left\{
(\mu^{\mathrm{true}}_{1r}\mathbf{1}_{n_1},\dots,\mu^{\mathrm{true}}_{3r}\mathbf{1}_{n_3})
,c^{\mathrm{true}}
\mathrm{diag}(\bSigma^{\mathrm{true}}_k)_{k=1,2,3}
\right\},
\end{equation}
where ${\bSigma}^{\mathrm{true}}_{ii}$ is the variance parameter of the $i$-th row and does not depend on $k$, and the notation $\mathrm{diag}(\bDelta^{\mathrm{true}}_{kr})_{r=1,2,3}$ denotes a block diagonal matrix formed by the matrices $\bDelta^{\mathrm{true}}_{1},\dots,\bDelta^{\mathrm{true}}_{3}$. Notice that, from Formula \eqref{formula:simulation_model_column_marginal}, the marginal distribution of the spots does not carry any information on the column clusters. The cross-covariance matrix of two rows $i,i'\in \mathcal{C}^{\mathrm{true}}_k$ is $\mathrm{Cov}(\mathbf{x}^{k.}_{i.},\mathbf{x}^{k.}_{i'.}) = \bSigma^{\mathrm{true}}_{k,ii'}\mathrm{diag}(\bDelta^{\mathrm{true}}_{kr})_{r=1,2,3}$, and the cross-covariance of two columns $j,j'\in \mathcal{D}^{\mathrm{true}}_r$ is $\mathrm{Cov}(\mathbf{x}^{.r}_{.j},\mathbf{x}^{.r}_{.j'}) = \mathrm{diag}\{\tau^{\mathrm{true}}_{kr} k^{\mathrm{true}}_r(||\mathbf{s}^r_j-\mathbf{s}^r_{j'}||;\bphi^{\mathrm{true}}_r)\bSigma^{\mathrm{true}}_k\}_{k=1,2,3}$.
}

We took the sets of spatial coordinates $(\mathbf{S}_1,\mathbf{S}_2,\mathbf{S}_{3})$ from the brain tissue sample of the subject with ID 151507 contained in the \texttt{R} package \texttt{spatialLIBD} and processed with Visium. As we briefly discussed in Section \ref{subsec:spatial_transcriptomics_intro}, the spots in these experiments have been manually annotated into layers.  We extracted 200 spots from each of the three layers appearing in the top-right region of the image. The resulting map of 600 spots is shown in the left plot of Figure \ref{figure:simulation_spatialcovariancefunction}; the clustering labels $\mathbfcal{W}^{\mathrm{true}}$ correspond to the  labels assigned with the manual annotation. \review{Note that, although we took the spot annotation from the real data, the image clusters in the simulation experiments have a substantially different meaning: in fact, under the simulation model \eqref{formula:simulation_model}, they denote regions of the tissue in which some genes are expressed with specific spatial variation profiles, while, in the real data, the manually annotated regions identify the morphological structure of the tissue.}
In addition, the right plot of Figure \ref{figure:simulation_spatialcovariancefunction} shows the covariance functions used for the simulations. We set the covariance parameters $(\theta_E,  \theta_R, \alpha_R, \theta_G)$ according to how much the clusters extend over the plane: the covariance function of $\mathcal{D}^{\mathrm{true}}_1$ is steeper than the one of $\mathcal{D}^{\mathrm{true}}_2$ because $\mathcal{D}^{\mathrm{true}}_1$ covers a smaller distance. Because $\mathcal{D}^{\mathrm{true}}_3$ is made of two distinct groups of spots appearing  in the top and in the bottom of Figure \ref{figure:simulation_spatialcovariancefunction} (left), we specify $k^{\mathrm{true}}_3(\cdot;\cdot)$ in such a way that only the spots within the same group are spatially correlated, while spots from different groups are poorly correlated.  
Details on the covariance parameters are given in the caption of Figure \ref{figure:simulation_spatialcovariancefunction}.

Last, we set the values of the spatial signal-to-noise ratios $\tau^{\mathrm{true}}_{kr}/\xi^{\mathrm{true}}_{kr}$. The additional identifiability constraint $ \tau^{\mathrm{true}}_{kr}+\xi^{\mathrm{true}}_{kr}=c^{\mathrm{true}}_{kr}$ leads to a unique value of the parameters  $\tau^{\mathrm{true}}_{kr}$ and $\xi^{\mathrm{true}}_{kr}$. 
Note that, due to the identifiability issue described in Section \ref{subsec:identifiability}, which holds also for the simulation model, the value assigned to $c^{\mathrm{true}}_{kr}$ is in practice irrelevant. For this reason, without loss of generality we assumed $c^{\mathrm{true}}_{kr} = c^{\mathrm{true}}=10$, for any $k$ and $r$. 
In our simulations, we considered three cases: (i) no spatial effect, $\tau^{\mathrm{true}}_{kr}/\xi^{\mathrm{true}}_{kr}=0$; (ii) the spatial effect is as much as the nugget effect, $\tau^{\mathrm{true}}_{kr}/\xi^{\mathrm{true}}_{kr}=1$; and (iii) the spatial effect is considerably larger than the nugget effect, $\tau^{\mathrm{true}}_{kr}/\xi^{\mathrm{true}}_{kr}=3$. Finally, we set $\mu^{\mathrm{true}}_{kr}=0$ to test if \spartaco~is able to recover the co-clusters using the covariance of the data without being driven by the effect of the mean.

\subsection{Competing models and evaluation criteria}
\label{subsec:competitormodels_evaluationcriteria}
We fit \spartaco~on the simulated data taking $k(\cdot;\cdot)$ in Formula \eqref{formula:covariancematrix} as the exponential  kernel, which has a lower decay than the more common Gaussian kernel considered by \citet{Svensson_etal:2018} and \citet{Sun_etal:2020}. The estimation is carried running the algorithm described in Section \ref{subsec:model_estimation} five times in parallel to avoid local maxima. The procedure is run for 5,000 iterations, and if the classification log-likelihood function is still growing, it is run until reaching 10,000 iterations. In addition to \spartaco, we consider also the following co-clustering models:
\begin{itemize}
\item two independent \textsc{k-means}, applied separately to the rows and to the columns of the data matrix, using the  \texttt{R} function \texttt{kmeans};
\item the biclustering algorithm \textsc{BC}, and its sparse version \textsc{sparseBC} with $\lambda = 1,10,20$, using the \texttt{R} package \texttt{sparseBC};
\item the matrix variate normal algorithm \textsc{MVNb} with the following setups: \emph{1)} $\lambda = 1$, $\rho_{\bSigma} = \rho_{\bDelta} = 0.25$, \emph{2)} $\lambda = 10$, $\rho_{\bSigma} = \rho_{\bDelta} = 2.5$ and \emph{3)} $\lambda = 20$, $\rho_{\bSigma} = \rho_{\bDelta}=5$. We had to implement a slight modification of the function \texttt{matrixBC} contained in the \texttt{R} package \texttt{sparseBC}, as its original form could not handle the computation of the logarithm of the determinant of some matrices.
\item \textrm{LBM}, using the \texttt{R} package \texttt{blockcluster};
\end{itemize}
\citet{Tan_Witten.2014} do not give any indication on how to select the penalization parameters $\rho_{\bSigma}$ and $\rho_{\bDelta}$ of  \textsc{MVNb}. In their simulation experiments and real data applications, they simply set $\lambda$ to be much larger than $\rho_{\bSigma}$ and $\rho_{\bDelta}$. For this reason, in our simulations we fit \textsc{MVNb} with three setups, where the $\lambda$ values are the same of \textsc{sparseBC}, and $\rho_{\bSigma}$ and $\rho_{\bDelta}$ are taken equal to a quarter of $\lambda$. \review{We measure the clustering accuracy by comparing the estimated row and column clusters with the true ones using the CER.}
In this section, we do not focus on the parameter estimates returned by \spartaco, because the principal goal is evaluating the classification accuracy of the models. We leave the interpretation of the parameter estimates to Section \ref{sec:Application}.

\begin{figure}[th!]
\centering
\includegraphics[width=0.32\linewidth]{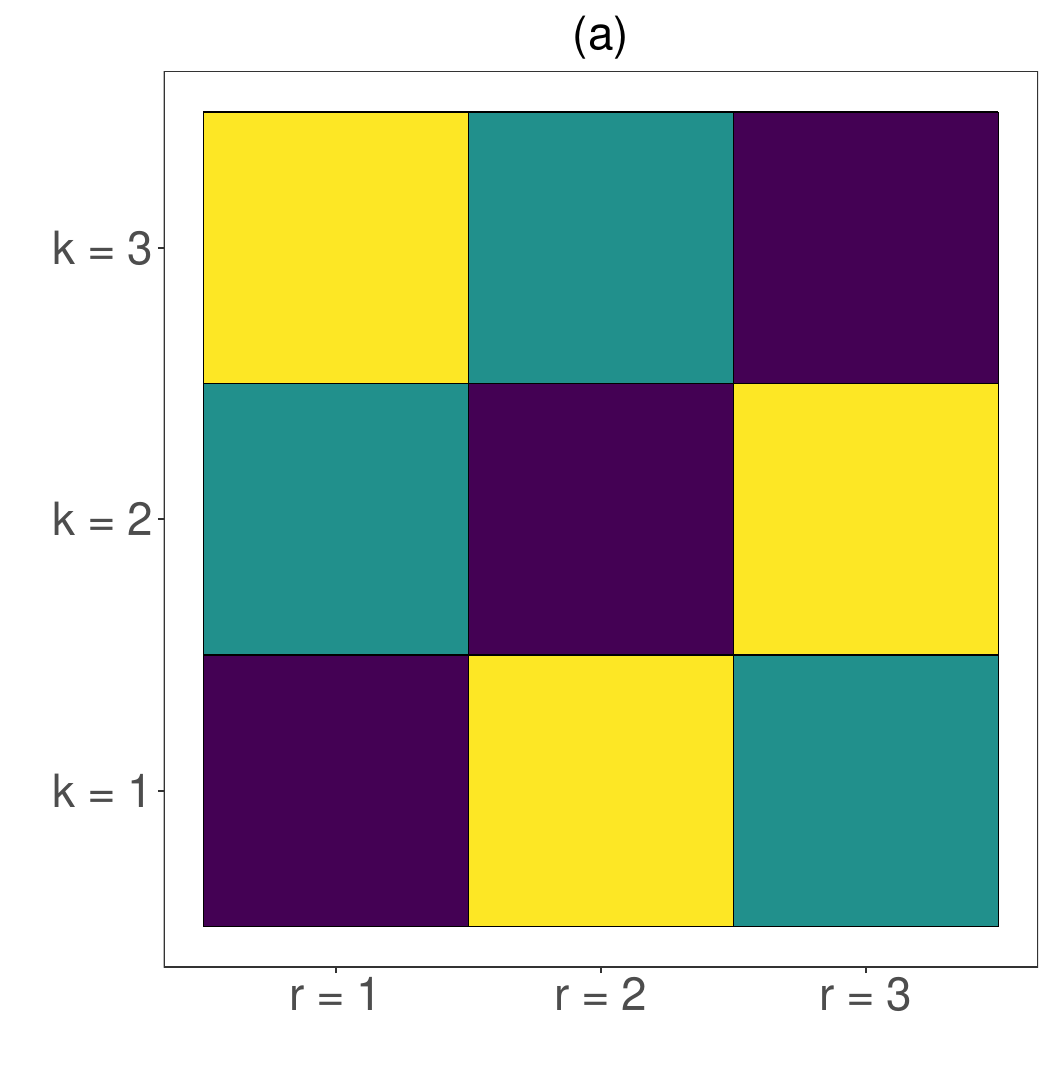}
\includegraphics[width=0.32\linewidth]{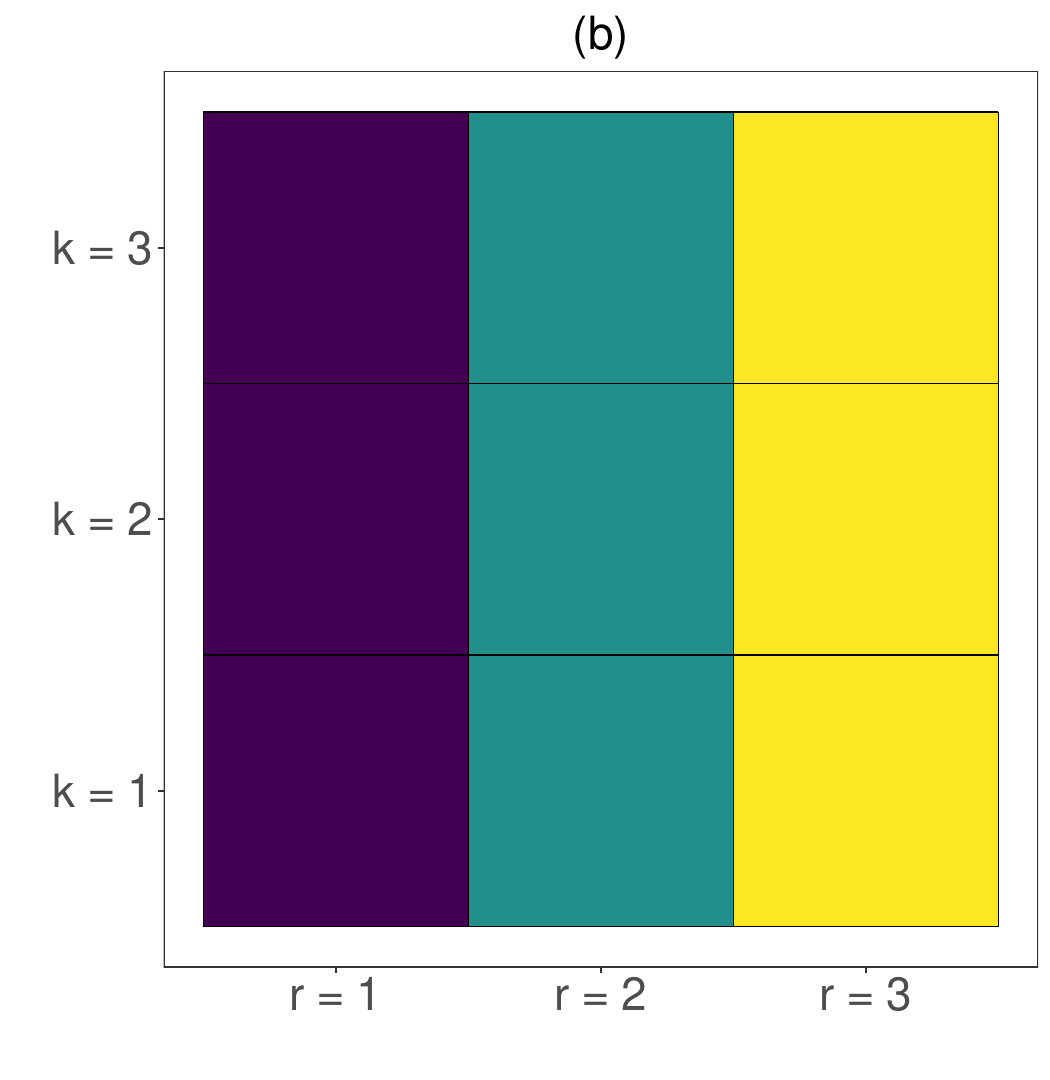}
\includegraphics[width=0.32\linewidth]{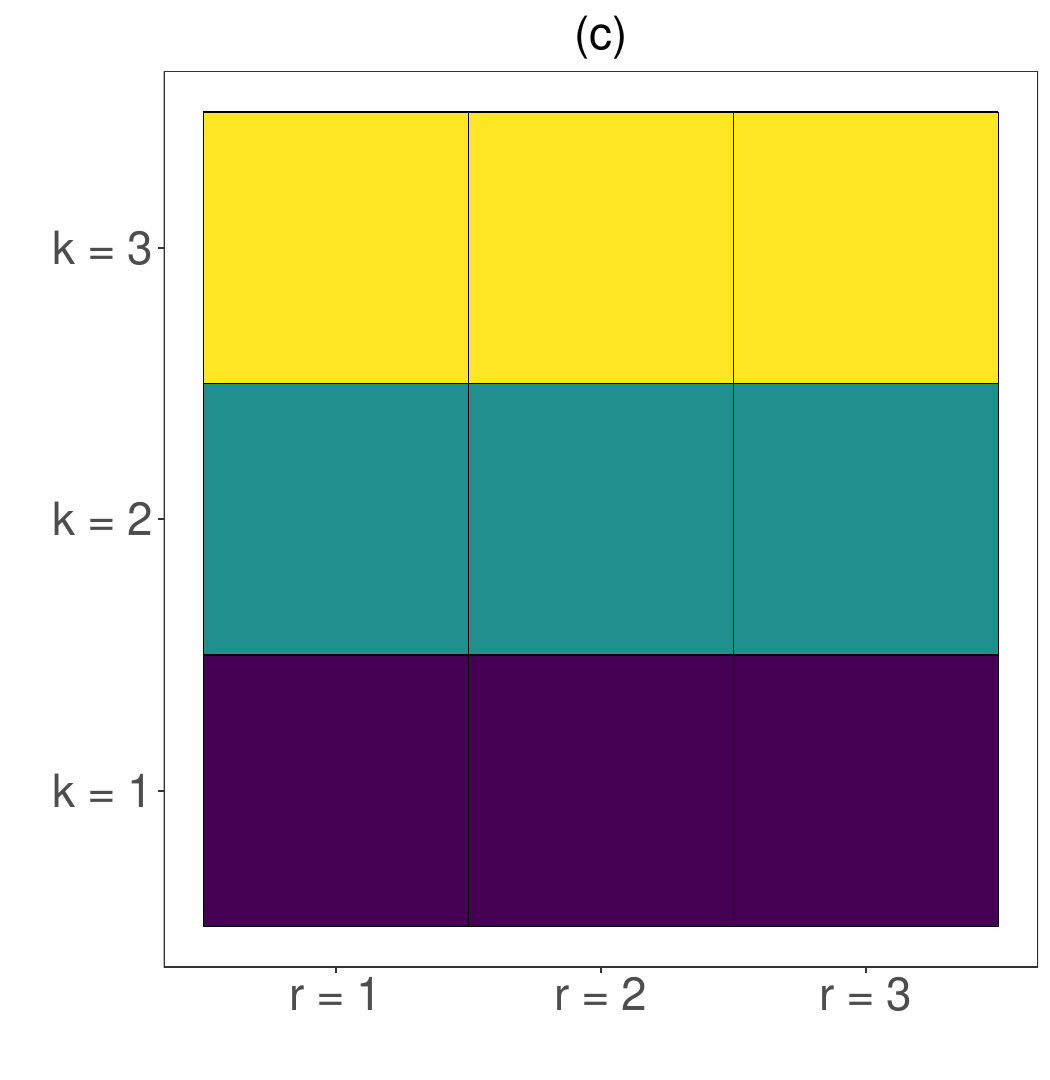}\\
\includegraphics[width=0.32\linewidth]{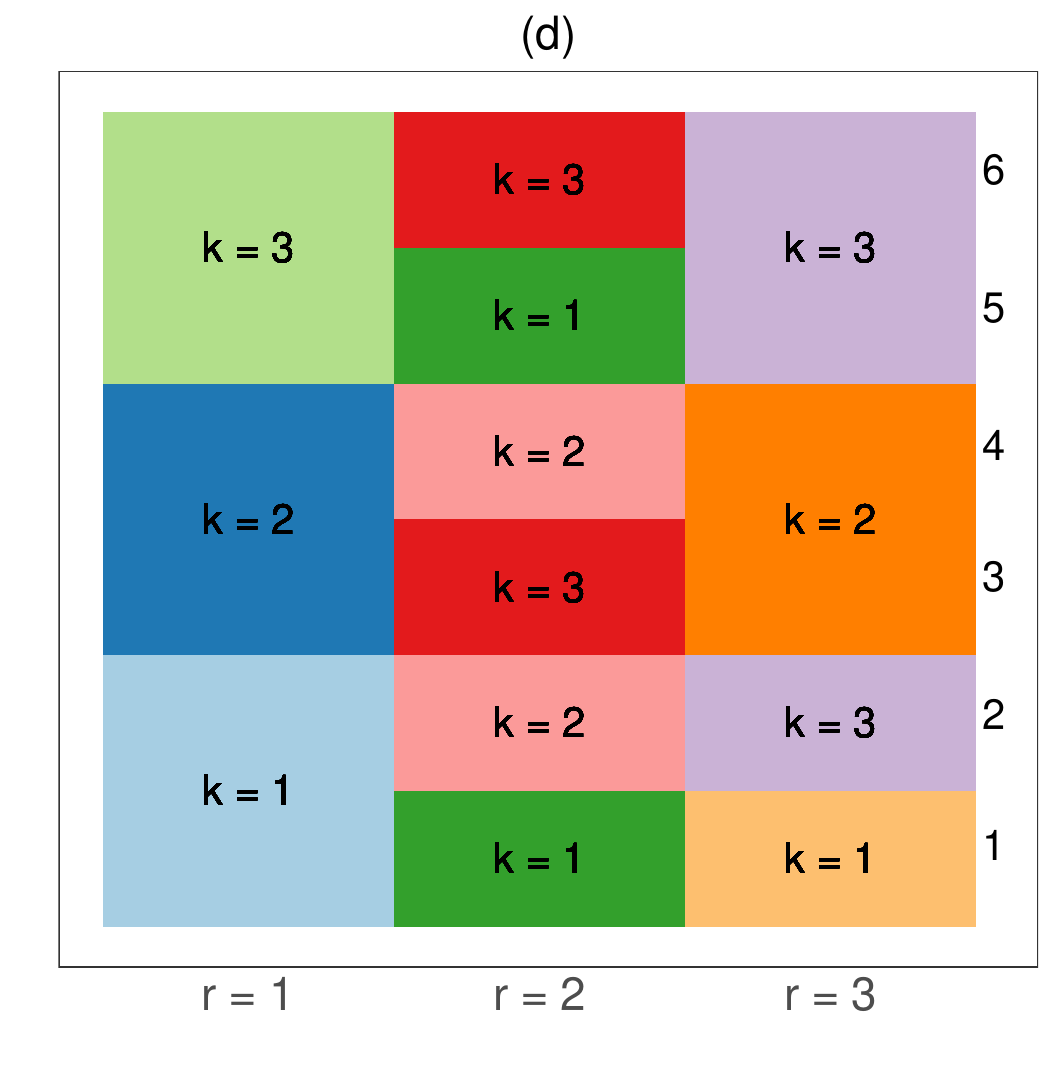}
\hspace{.5cm}
\includegraphics[width=0.32\linewidth]{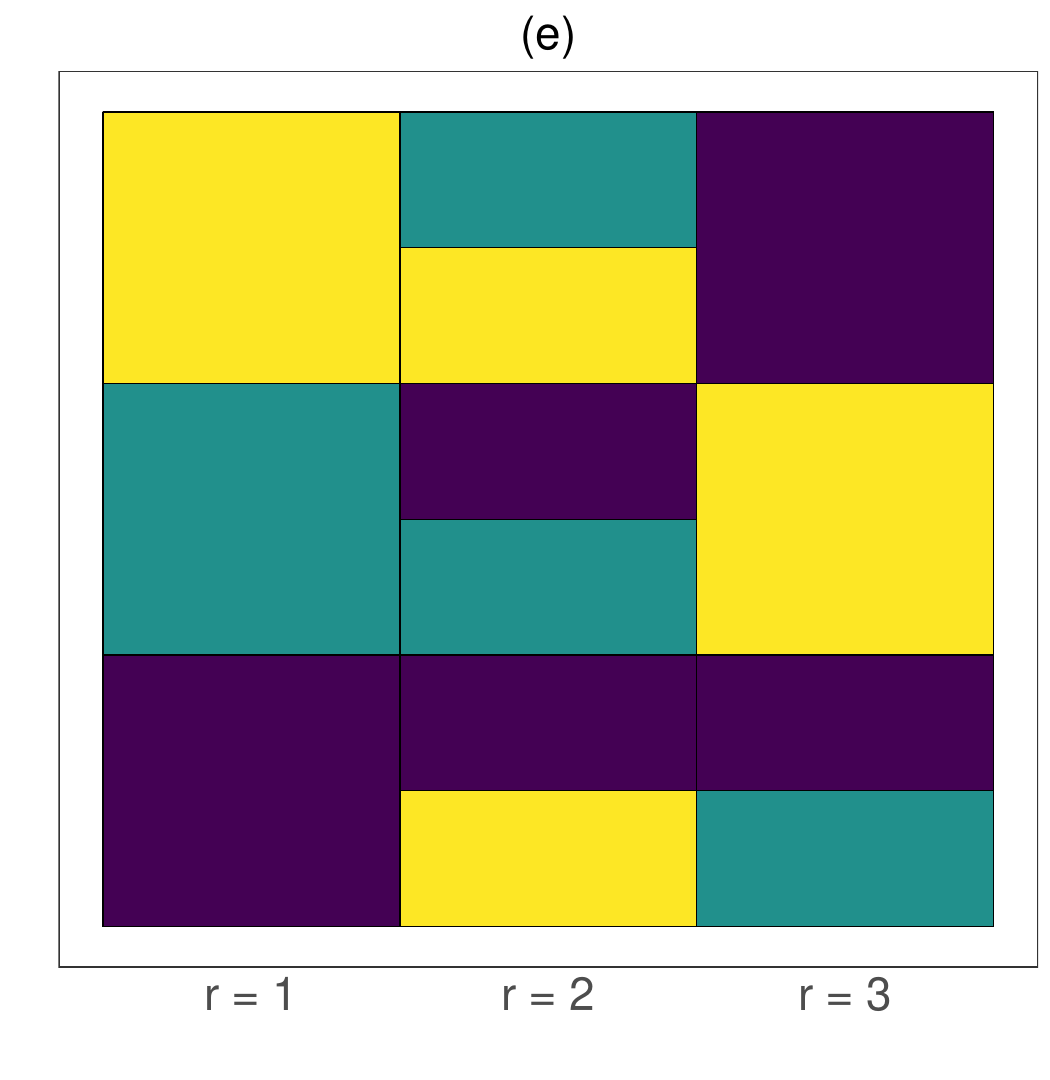}\\
\includegraphics[width=0.3\linewidth]{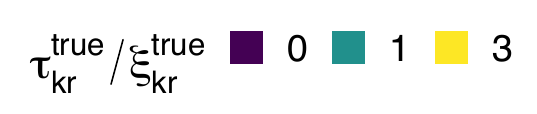}
\caption{Representation of the latent block structures used to generate the simulation experiments. All the blocks in Panels (a)-(c) have the same size and are colored according to the value of the spatial signal-to-noise ratio $\tau^{\mathrm{true}}_{kr}/\xi^{\mathrm{true}}_{kr}$. 
	The setup in Panel (a) is used in Sections \ref{subsec:simulation1} and \ref{subsec:simulation4}, Panel (b) is used in Section \ref{subsec:simulation2},  Panel (c) in Section \ref{subsec:simulation3} and Panel (e) in Section \ref{subsec:simulation5}.  Panel (d) gives the hidden block structure of Simulation \ref{subsec:simulation5}. Within the columns 1 and 2, the row clusters have the same size (200), while in the third column it is $n_{13}=100$, $n_{23} = 200$ and $n_{33}=300$. The numbers from 1 to 6 on the right denote the alternative clusters $\mathcal{C}^{*\mathrm{true}}_1,\dots,\mathcal{C}^{*\mathrm{true}}_6$.
}
\label{figure:simulation_setup_tauoverxi}
\end{figure}

\subsection{Simulation 1}
\label{subsec:simulation1}

We generated 9 blocks of  size $n_k = 200\times p_r = 200$, for every $k$ and $r$. We assume that the variances and covariances of the genes do not change with respect to the spot clusters, thus  $\bSigma^{\mathrm{true}}_{kr} = \bSigma^{\mathrm{true}}_k$ for all $r$. We draw $\bSigma^{\mathrm{true}}_k$ as follows:
\begin{equation}
{
\label{formula:simulation_gene_covariances}
\bSigma^{\mathrm{true}}_1\sim \mathcal{W}i(210, 0.03\mathds{I}_{200}),\hspace{.3cm}
\bSigma^{\mathrm{true}}_2\sim \mathcal{W}i(230, 0.05\mathds{I}_{200}),\hspace{.3cm}
\bSigma^{\mathrm{true}}_3\sim \mathcal{W}i(200,\bSigma^{\mathrm{true}}_1/150),
}
\end{equation}
where $\mathcal{W}i(a,\mathbf{b})$ denotes a Wishart distribution with degrees of freedom $a$ and scale matrix $\mathbf{b}$. Generating the covariance matrices from a Wishart distribution ensures that the draws are positive definite.
The simulation setup in Formula \eqref{formula:simulation_gene_covariances} was selected after both numerical and graphical evaluations. More details on the motivations which led to this setup are given in Supplementary Section 3.

\begin{figure}
\centering
\includegraphics[width = 0.8\linewidth]{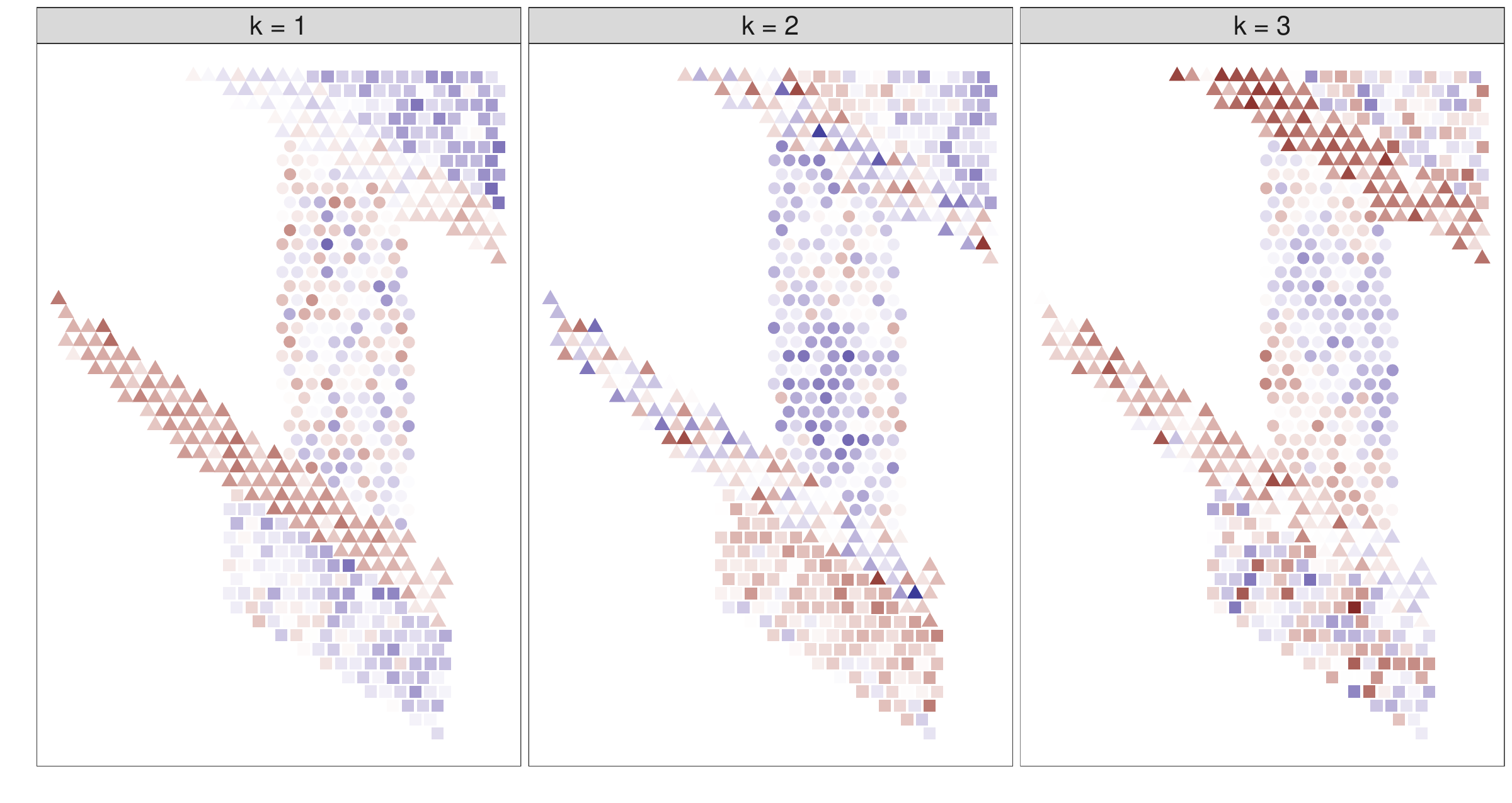}
\includegraphics[width= .4\linewidth]{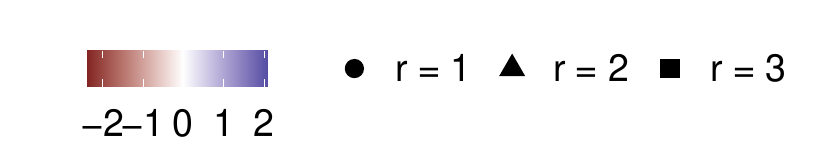}
\caption{\review{Examples of a spatial experiment generated under Simulation 1. The spots are coloured according to  $n^{-1}_k(\mathbf{X}^{k.})^T\mathbf{1}_{n_k}$, the average expression of the $k$-th gene cluster. The three spot clusters are displayed with different symbols. The co-clusters with no spatial expression are $(k=1,r=1)$, $(k=2,r=2)$ and $(k=3,r=3)$, and the co-clusters with the largest spatial signal-to-noise ratio are $(k=1,r=2)$, $(k=2,r=3)$ and $(k=3,r=1)$.}}
\label{figure:scenario1_spatialexpression}
\end{figure}

\begin{figure}[th!]
\centering
\includegraphics[width=0.72\linewidth]{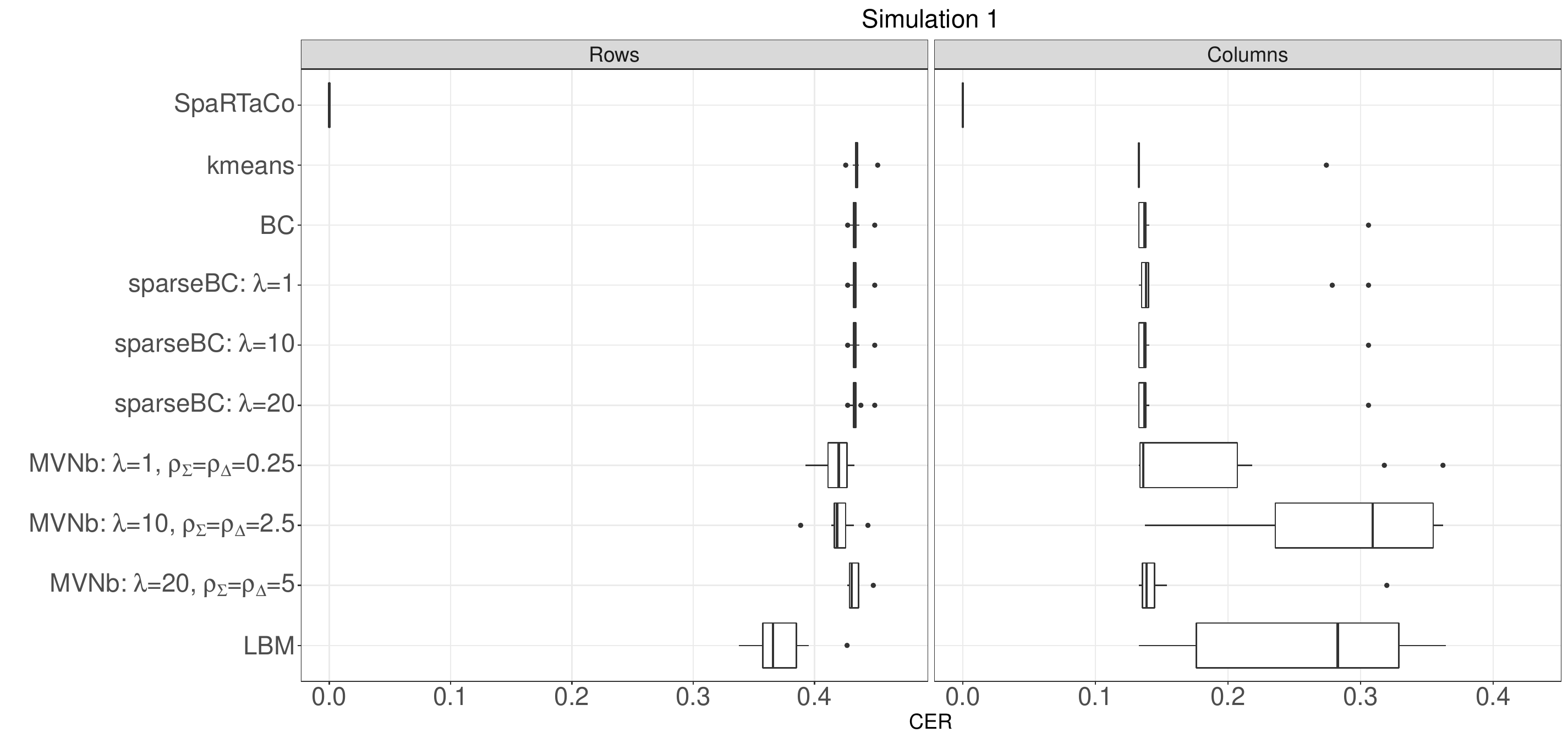}
\includegraphics[width=0.72\linewidth]{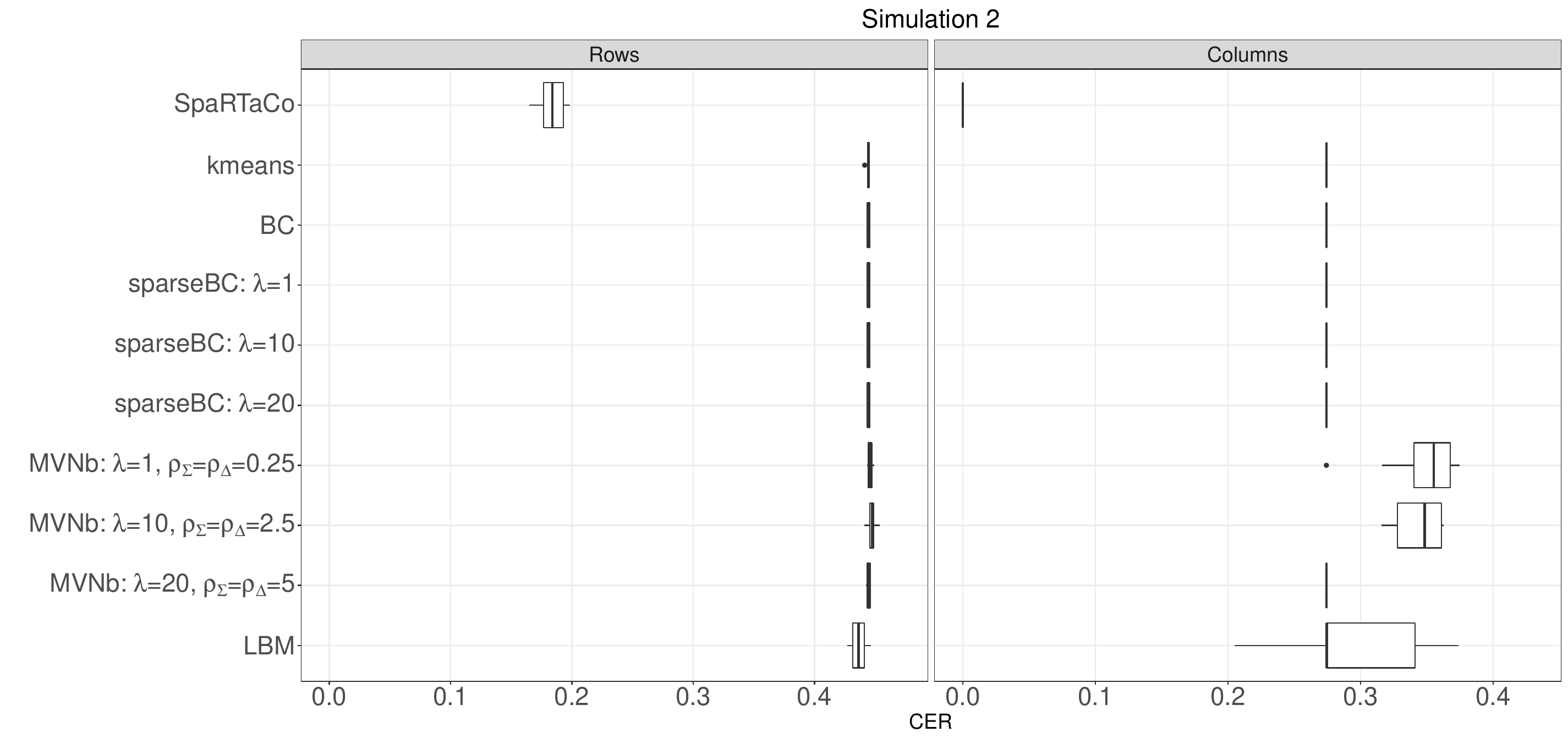}
\includegraphics[width=0.72\linewidth]{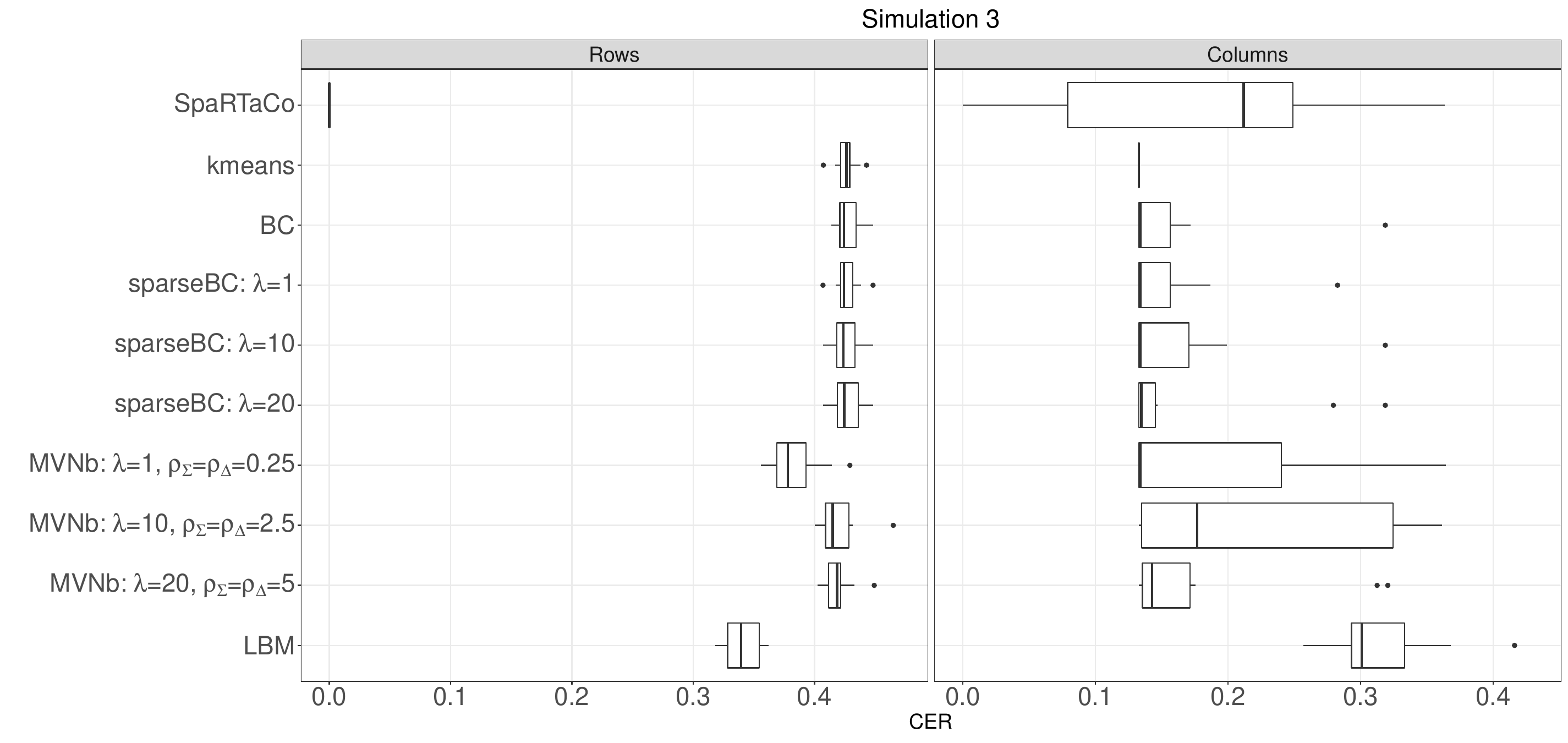}
\includegraphics[width=0.72\linewidth]{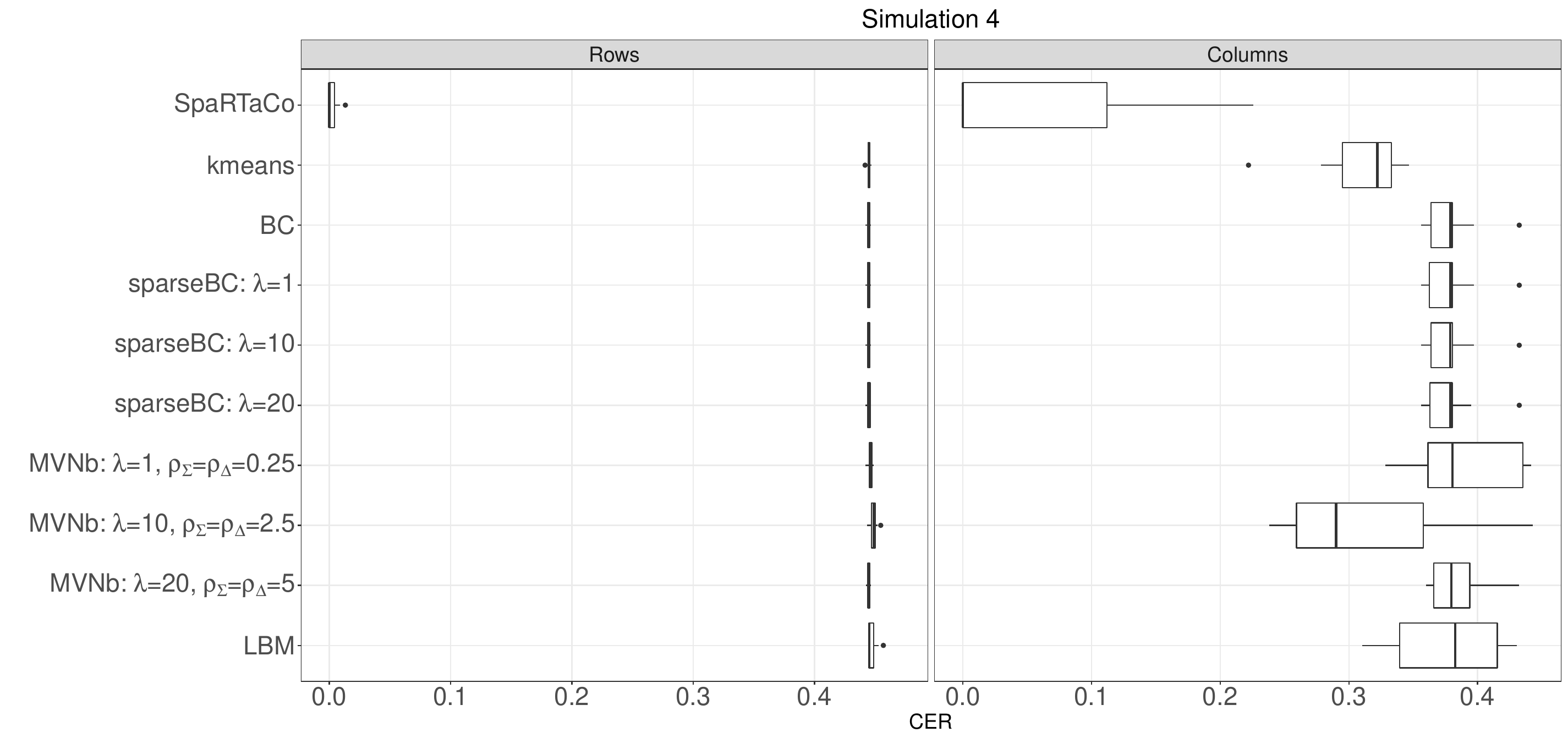}
\caption{Results from Simulations 1-4. For each scenario, we generated 10 datasets and we applied the co-clustering models listed in Section \ref{subsec:competitormodels_evaluationcriteria}. Every figure gives the boxplots of the CER obtained on the rows and on the columns.
}
\label{figure:CER_simulations}
\end{figure}

We designed a spatial experiment in which three clusters of genes have a grade of spatial expression that changes in three different areas of the tissue sample. The tessellation of the data matrix into blocks and the values of the spatial signal-to-noise ratios appear in Figure \ref{figure:simulation_setup_tauoverxi} (a). \review{Figure \ref{figure:scenario1_spatialexpression} displays a spatial experiment generated under this framework, to show how the average gene expression changes across the 9 blocks. For example, in the left panel ($k = 1$) there is an evident spatial expression across the spots from clusters $r = 2$ and $r = 3$, while the spots in $r = 1$ are randomly positive or negative due to the absence of spatial dependency. Different spatial expression profiles across the image are distinguishable also in real data, as seen in Supplementary Figure 3, which displays the expression of three genes on the subject 151507. The real and simulated experiments appear very similar, confirming that our simulations are realistic and can be used for testing methods designed for 10X Visium data.}
We simulated 10 replicates of this experiment and we fitted the co-clustering models listed in Section \ref{subsec:competitormodels_evaluationcriteria} using   $K=R=3$.  The boxplots of the row and the column CER over the 10 replicates appear in the first line of Figure \ref{figure:CER_simulations}. \spartaco~outperforms the competing models and leads to no clustering errors. Good results on the rows are achieved also by the \textsc{LBM}, while on the columns the  k-means type algorithms (\textsc{k-means}, \textsc{BC} and \textsc{sparseBC}) and the \textsc{MVNb} with $\rho_{\bSigma} = \rho_{\bDelta} = 5$ perform better than the other competitors. \review{A further confirmation of the accuracy of \spartaco~for modelling this spatial experiment comes from the value of estimated clustering uncertainties, which are $\rowepsilon < 0.001$ and $\colepsilon < 0.001$, for every $k$ and $r$. A graphical representation of these quantities across the 10 replicates is given in Supplementary Figure 4.}

This experiment has demonstrated that the presence of spatial covariance patterns, if not properly accounted for, heavily impacts on the performance of the standard co-clustering models. Since the \textsc{MVNb} is designed to flexibly estimate the covariance of the blocks, in theory it should be the best candidate for such complex experiments. \review{However, the formulation of $\hat{\bSigma}^{\textsc{MVNb}}_k$ and $\hat{\bDelta}^{\textsc{MVNb}}_r$ is too generic for capturing the spatial correlation across the spots, causing a poor clustering result. As a confirmation of this statement, we notice that the smallest classification error made by \textsc{MVNb} is reached when the penalization parameters $\rho_{\bSigma}$ and $\rho_{\bDelta}$ are large, leading the estimated matrices $\hat{\bSigma}^{\textsc{MVNb}}_k$ and $\hat{\bDelta}^{\textsc{MVNb}}_r$ to be diagonal.}

\review{As a second step of this experiment}, we tested the model selection criterion based on the ICL proposed in Section \ref{subsec:model_selection}. Using the same 10 replicates of the experiment, we ran \spartaco~with $K$ and $R$ taking values in $\{2,3,4\}$. Supplementary Figure 5 shows that the proposed ICL always selects the correct model dimension, while the classification log-likelihood favors models with a larger number of co-clusters than the truth. 

\review{
While the ICL criterion accurately selects the number of co-clusters, it is a computationally expensive procedure due to the large number of times that the estimation must be run. Hence, we compared our model selection method with two faster alternatives:  the first selects the number of row and column clusters separately by combing a dimension reduction method with  \textsc{k-means} (details are given in Supplementary Section 4), the second, proposed by \cite{Tan_Witten.2014}, performs a 10-fold cross-validation using \textsc{sparseBC}; a function that implements this last method can be found into the \texttt{R} package \texttt{sparseBC}. 
The first criterion selected 6 row clusters on the 90\% of the replicates of Simulation 1, and 5 clusters in the remaining 10\%; on the columns, it selected 3 clusters on the 33\% of the replicates, and 4 clusters on the remaining 77\%. The second criterion was applied with $K$ and $R$ taking values in $\{2,\dots,6\}$ and fixing $\lambda = 10$, but it has revealed to be inadequate for this kind of data, as it selected $K = 6$ and $R = 6$ on every replicate of the experiment.
}

\subsection{Simulation 2}
\label{subsec:simulation2}

The second simulation experiment differs from the first in the values of the spatial signal-to-noise ratios, which are now taken as in Figure \ref{figure:simulation_setup_tauoverxi} (b).
For any $r$, the signal-to-noise ratios $\{\tau^{\mathrm{true}}_{kr}/\xi^{\mathrm{true}}_{kr},k=1,\dots,K^{\mathrm{true}}\}$ have all the same value. As a consequence, $\bDelta^{\mathrm{true}}_{kr} = \bDelta^{\mathrm{true}}_{r}$ for any $k$. Under the current setup, the marginal distribution of a row $i\in\mathcal{C}^{\mathrm{true}}_k$ under the data generating model \review{given in Formula \eqref{formula:simulation_model_row_marginal} does not depend on $k$ and so it is not informative of the row clustering. The only discriminating facets are the cross-covariances of the rows and of the columns, which carry the information about the row clusters through the matrices $\bSigma^{\mathrm{true}}_k$. This framework is thus meant to evaluate the performance of \spartaco~when all the genes have the same spatial expression profiles across the tissue. A representation of a spatial experiment generated under this framework is given in the top row of Supplementary Figure 6.} 

We ran the co-clustering models using $K=R=3$ 
on 10 replicates on the proposed experiment; the results are displayed in the second line of Figure \ref{figure:CER_simulations}. Our model outperforms the competitors: on the rows, the median CER from \spartaco~is less than 0.2, while on the columns it returns a perfect classification on all replicates. \review{The estimated  row clustering uncertainty is low ($\rowepsilon < 0.15$, $\forall k$), while the column clustering uncertainty is practically null. Details are given in the second row of Supplementary Figure 4.} Both Simulations 1 and 2 have shown that \spartaco~works properly even if the spatial covariance function employed by the fitted model in Formula \eqref{formula:covariancematrix} does not match the covariance functions of the data generating process.
In particular, Simulation 2 has highlighted this remarkable result because the only cluster of columns for which the spatial covariance function is correctly specified is $r = 1$, which however is devoid of any spatial effect, as $\tau^{\mathrm{true}}_{k1}=0$ for any $k$.

The best competitor on the rows is the LBM, with a median CER of 0.44.  On the columns, the best results are from the k-means type models, or alternatively from the \textsc{MVNb} with $\lambda = 20$ and $\rho_{\bSigma} = \rho_{\bDelta} = 5$. Considerable results are obtained also with the LBM; however, its classification accuracy is more variable. This experiment hence confirms what we have already observed in Simulation 1, namely that, in the presence of spatial covariance patterns in the data,  the model of \cite{Tan_Witten.2014} tends to fail in recovering the correlation structure, at least in our simulation setup. This is demonstrated by the diagonal estimated covariance matrices  $\{\hat{\bSigma}^{\textsc{MVNb}}_k, k=1,2,3\}$ and $\{\hat{\bDelta}^{\textsc{MVNb}}_r, r=1,2,3\}$.


\subsection{Simulation 3}
\label{subsec:simulation3}

The third simulation experiment assumes that the spatial signal-to-noise ratio $\tau^{\mathrm{true}}_{kr}/\xi^{\mathrm{true}}_{kr}$ is constant across the blocks within the same row cluster $k$; as a consequence, $\tau_{kr}=\tau_k$ for any $r$. This case is illustrated in Figure \ref{figure:simulation_setup_tauoverxi} (c). Notice for example that the rows in  $\mathcal{C}^{\mathrm{true}}_1$ are not spatially expressed in any of the three column clusters. 
Under the current simulation setup, the marginal distribution of \review{a row $i\in \mathcal{C}^{\mathrm{true}}_k$ given in Formula \eqref{formula:simulation_model_row_marginal} is informative on the column clusters only through the different spatial kernels $k^{\mathrm{true}}_r(\cdot;\bphi^{\mathrm{true}}_r)$, while, as already discussed in Section \ref{subsec:simulation_model}, the marginal distribution of a column $j\in \mathcal{D}^{\mathrm{true}}_r$ is never informative on the column clusters. The cross-covariances of the rows and of the columns are informative of both rows and column clustering. Under this framework, it is challenging to determine the image areas with spatial interaction, because all the genes in a cluster $\mathcal{C}^{\mathrm{true}}_k$ are spatially expressed with the same intensity over the whole tissue. An example of a spatial experiment generated under this simulation setup is given in the bottom row of Supplementary Figure 6.}

We ran the co-clustering models on 10 replicates of the experiment using $K=R=3$; the results appear in the third line of  Figure \ref{figure:CER_simulations}. On the rows, \spartaco~outperforms the competitor models returning a  CER of zero for all replicates. On the columns, its clustering accuracy is highly variable: the median CER is 0.21, the first and the third quartiles are 0.08 and 0.25, and extremes are 0 and 0.36. The competitor models, and in particular the k-means type models, are substantially less variable  than \spartaco.  
Their median column CER is 0.13. However, none of them ever returns a perfect classification.

Even if \spartaco~ has returned unstable results on the columns, the advantages brought by our model against the competitors are many, and are particularly visible from the results on the rows. The column clustering changes considerably across the replicates because, in the current setup, our estimation algorithm  is more sensible to the starting points. \review{This aspect is highlighted also by the estimated column clustering uncertainties $\colepsilon$, whose values across the 10 replicates are now mainly between 0.3 and 0.4 (see Supplementary Figure 4)}. From our experience, if independent runs of the estimation algorithm reach distant stationary points, both the number of starting points and the number of iterations of the SE Step should be increased to favor a faster exploration of the space of the configurations.

\subsection{Simulation 4}
\label{subsec:simulation4}

Up to now, we built the simulation experiments under the framework in which \spartaco~is designed to work properly, that is the case where  the genes/spots in a cluster are correlated only with the other genes/spots of the same cluster. 
In this section, we violate this assumption and we design a spatial experiment where both the genes and the spots are correlated also with genes and spots from other clusters. This experiment aims to study the effects of an additional dependency structure across the data that is not accounted  by the fitted model.

Let $\mathbf{X}_s$ be a $600\times 600$ spatial experiment made of 9 equally sized blocks, generated as in Simulation 1, and $\mathbf{X}_b\sim \mathcal{MVN}(\mathbf{0}, \bSigma_b, \bDelta_b)$. Both $\bSigma_b$ and $\bDelta_b$ are squared matrices of size 600: the first is drawn from $\bSigma_b\sim \mathcal{W}(600, 0.015\mathds{I}_{600})$, the second is $\bDelta_b = \tau_b \mathbfcal{K}^b(\mathbf{S};\sigma_b)+\xi_b\mathds{I}_{600}$, where $\mathbfcal{K}^b(\mathbf{S};\sigma_b)=\left(k^b(||\mathbf{s}_j-\mathbf{s}_{j'}||;\sigma_b) \right)_{1\leq j,j'\leq 600}$  and $k^{b}(\cdot;\sigma_b)$ is a Gaussian kernel with scale $\sigma_b$. We set $\tau_b =\xi_b= c^{\mathrm{true}}/2$ and $\sigma_b = 50$. 
The final simulation experiment is made as follows: $\mathbf{X} = \lambda_s \mathbf{X}_s + \lambda_b \mathbf{X}_b$, where $\lambda_s,\lambda_b\geq 0$. 
We generated 10 replicates of the current experiment, each time drawing first the matrices $\mathbf{X}_s$ and $\mathbf{X}_b$, and then combining them to form $\mathbf{X}$. Supplementary Figure 7 shows a single realization of $\mathbf{X}_s$, $\mathbf{X}_b$ and $\mathbf{X}$ using $\lambda_s =\lambda_b=\sqrt{0.5}$. This value satisfies the constraint  $\lambda^2_s+\lambda^2_b = 1$ that we imposed to keep the variance of the current experiment comparable with the previous experiments proposed in this work.

We ran the co-clustering models using $K=R=3$; results appear in the last row of Figure \ref{figure:CER_simulations}. Despite the additional correlation structure in the data brought by the nuisance signal $\mathbf{X}_b$, \spartaco~outperforms its competitors on both the row and the column clustering. In the right plot, the CER boxplots are more variable than in the left plot, therefore, the nuisance component has affected more the column than the row clustering of the employed models.
Among the competitors, \textsc{k-means} and \textsc{MVNb} with $\lambda = 10$ and $\rho_{\bSigma}=\rho_{\bDelta} = 2.5$ are the least affected by the nuisance: the former because it performs the clustering on the two dimensions of the data matrix separately, the latter because it regulates the estimate of the row and column covariances with a moderate shrinkage factor. \review{The effect of the additional dependency structure is visible also on the distributions of $\rowepsilon$ and $\colepsilon$, which are displayed in the last line of Supplementary Figure 4: over the 10 replicates, the row clustering uncertainties spread between 0 and 0.17, and column uncertainties between 0 and 0.5.}

\subsection{Simulation 5}
\label{subsec:simulation5}

In the last experiment, we \review{intentionally} violate \review{two important assumptions made by \spartaco: the first states that the latent block structure of an experiment corresponds to a segmentation of the data matrix into $K$ row clusters and $R$ column clusters, the second states that the spatial covariance functions change only across the spots and not across the genes.}
For instance, we generate a spatial experiment creating first the $R^\mathrm{true}$ column clusters, and then generating the $K^\mathrm{true}$ row clusters independently for  each  column cluster. From a biological perspective, this setup simulates the case where the expression profiles of some genes are similar only in some specific areas of the tissue sample. \review{In addition, following the discoveries of \cite{Svensson_etal:2018} and \cite{Sun_etal:2020} that  different genes are s.e. according to different spatial covariance functions, we consider a data generating model where the spatial kernels change with respect to the gene cluster index $k$ and no longer with respect to the spot cluster index $r$.}

Let $\mathcal{C}^{\mathrm{true}}_{kr}$ and $\mathcal{D}^{\mathrm{true}}_r$ be the actual row and column clusters, with $k=1,\dots,K^{\mathrm{true}}$ and $r = 1,\dots,R^\mathrm{true}$, where $\mathcal{C}^{\mathrm{true}}_{kr} = \{i=1,\dots,n:\mathcal{Z}^{\mathrm{true}}_{ir} = k\}$ is the $k$-th row cluster within the $r$-th column cluster, and $|\mathcal{C}^{\mathrm{true}}_{kr}|=n_{kr}$. 
Under the current setup, \review{we draw $\X^{kr}\sim \mathcal{MVN}(\mu_{kr}\mathbf{1}_{n_k\times p_r}, \bSigma^{\mathrm{true}}_{kr},\bDelta^{\mathrm{true}}_{kr})$, where the covariance across the spots is now equal to $\bDelta^{\mathrm{true}}_{kr} = \tau^{\mathrm{true}}_{kr}\mathcal{K}^{\mathrm{true}}_k(\mathbf{S}^r;\bphi^{\mathrm{true}}_k)+\xi^{\mathrm{true}}_{kr}\mathbb{I}_{p_r}$}. 
Notice that, differently from Section \ref{subsec:simulation1}, the covariance matrices of the rows  $\bSigma^{\mathrm{true}}_{kr}$ change with respect to $r$ because the number of observations in the cluster is $n_{kr}$ (and no longer $n_k$). \review{In addition, the model assumes that the $kr$-th block has mean $\mu_{kr}$.}
The tessellation of the data matrix into blocks is shown in Figure \ref{figure:simulation_setup_tauoverxi} (d). The size of the clusters is $n_{kr}=200$ for $k=1,2,3$ and $r = 1,2$, while $n_{13}=100$, $n_{23}=200$ and $n_{33}=300$. The covariance matrices of the rows are drawn as follows:
{\small
$$
\bSigma^{\mathrm{true}}_{1r}\sim \mathcal{W}i(n_{1r}+10,0.03\mathds{I}_{n_{1r}}),\hspace{.5cm}
\bSigma^{\mathrm{true}}_{2r}\sim \mathcal{W}i(n_{2r}+30,0.05\mathds{I}_{n_{2r}}),\hspace{.5cm}
\bSigma^{\mathrm{true}}_{3r}\sim \mathcal{W}i(n_{3r},\bSigma^{*}_{3r}/150),
$$
}\noindent
where $\bSigma^{*}_{3r}\sim \mathcal{W}i(n_{3r}+10,0.03\mathds{I}_{n_{3r}})$.
Notice that this setting is nothing but a generalization of what appears in Formula \eqref{formula:simulation_gene_covariances}.  \review{Calling $\boldsymbol{\mu}^{\mathrm{true}}_{k.}=(\mu^{\mathrm{true}}_{k1},\mu^{\mathrm{true}}_{k2},\mu^{\mathrm{true}}_{k3})$, we set the mean values equal to  $\boldsymbol{\mu}^{\mathrm{true}}_{1.}=(-3,0,3)$, $\boldsymbol{\mu}^{\mathrm{true}}_{2.}=(3,-3,0)$ and $\boldsymbol{\mu}^{\mathrm{true}}_{3.}=(0,3,-3)$.
}
Finally, the employed spatial signal-to-noise ratio values $\{\tau_{kr}/\xi_{kr}\}$ are shown in Figure \ref{figure:simulation_setup_tauoverxi} (e).

To facilitate the model evaluation and the interpretation of the results, we assign to every row $i$ an alternative clustering label ${\mathcal{Z}^*_i}{^{\mathrm{true}}}$ such that ${\mathcal{Z}^*_i}{^{\mathrm{true}}}={\mathcal{Z}^*_{i'}}{^{\mathrm{true}}}$ if $i,i'\in \left(\mathcal{C}^{\mathrm{true}}_{k_1 1} \bigcap \mathcal{C}^{\mathrm{true}}_{k_2 2} \bigcap \mathcal{C}^{\mathrm{true}}_{k_3 3}\right)$, for some $k_1,k_2,k_3\in \{1,2,3\}$. In words, this means that the new clusters are formed by the rows that belong to the same cluster in all of the three column clusters. The new row clustering labels appear on the right side of Figure \ref{figure:simulation_setup_tauoverxi} (d). In our experiment, every ${\mathcal{Z}^{*}_i}^{\mathrm{true}}\in\{1,\dots,6\}$, and ${\mathcal{C}^*_b}^{\mathrm{true}}=\{i=1,\dots,n:{\mathcal{Z}^*_i}{^{\mathrm{true}}}=b\}$ is the $b$-th alternative cluster with size $|{\mathcal{C}^*_b}^{\mathrm{true}}|=100$, for $b = 1,\dots,6$. 

To reduce the computational cost spent on the simulation, we generated a single replicate of the experiment, and we fitted \spartaco~using $K=3,\dots,9$, while the number of column clusters is kept equal to its real value, $R = 3$. Supplementary Figure 8 (a) shows that the ICL criterion selects $K=8$ as the optimal model dimension; using the log-likelihood, we would have wrongly picked $K=9$, confirming the importance of using a suitable information criterion to drive the model selection. \review{In addition, one could consider also  a model with a smaller number of row clusters: for example, $K = 5$ looks also a reasonable choice, because it corresponds to a local maximum.
\spartaco~with $K=8$  returns a row CER of 0.028 and a column CER of 0. In details, the model correctly recovers the gene clusters 2, 4, 5 and 6, while the genes in ${\mathcal{C}^*_1}^{\mathrm{true}}$ and ${\mathcal{C}^*_3}^{\mathrm{true}}$ are split into two separate groups. The estimated clustering uncertainty is $\colepsilon < 0.004$, for $r=1,2,3$, while on the rows it varies between 0 and 0.19. Thus, some of the genes clusters are clearly visible, while others are unstable.
As a comparison, we give also the results using $K = 5$. The CER  on the rows is 0.056, and it is 0 on the columns; the estimated clustering uncertainties are $\rowepsilon < 0.001$, for $k=1,\dots,5$, and  $\colepsilon < 0.06$, for $r = 1,2,3$. The fact that the row clusters of the model with $K = 5$ are more stable than the ones with $K=8$ gives additional support to the idea of selecting the  model with a smaller number of blocks, but both models yield reasonably good results.
}

We finally run the competing models using $K=5$ and $R = 3$; 
\review{results are shown in Supplementary Figure 8 (b). Thanks to the difference in mean across the blocks, all the competing models can clearly distinguish the clustering structure of the spots. However, due to the spatial dependency effects, their performance in clustering the genes is poor, confirming, once again, the improvement brought by \spartaco.}

\begin{figure}[th!]
\centering
\includegraphics[width=0.45\linewidth]{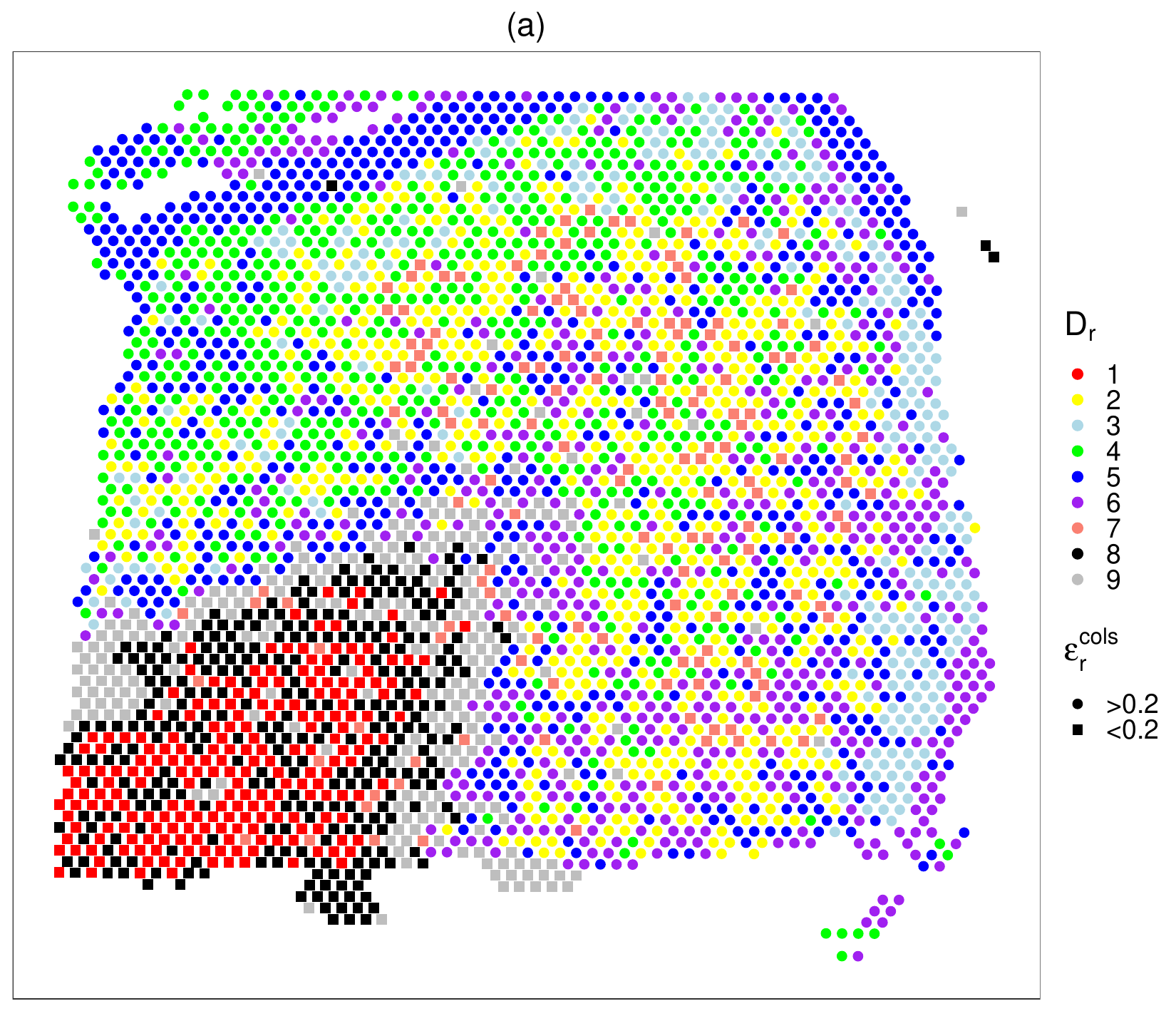}
\includegraphics[width=0.45\linewidth]{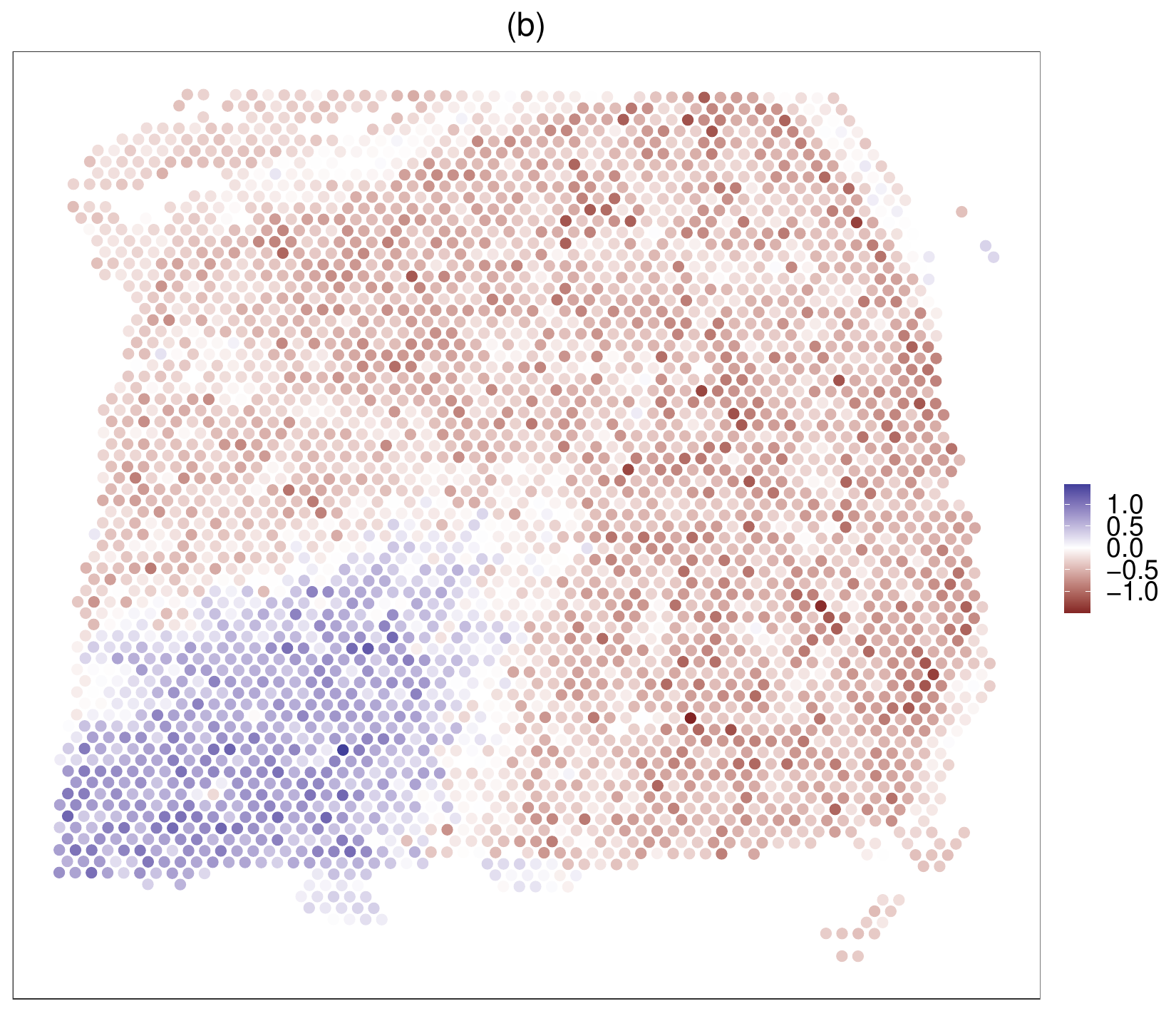}\\
\includegraphics[width=0.45\linewidth]{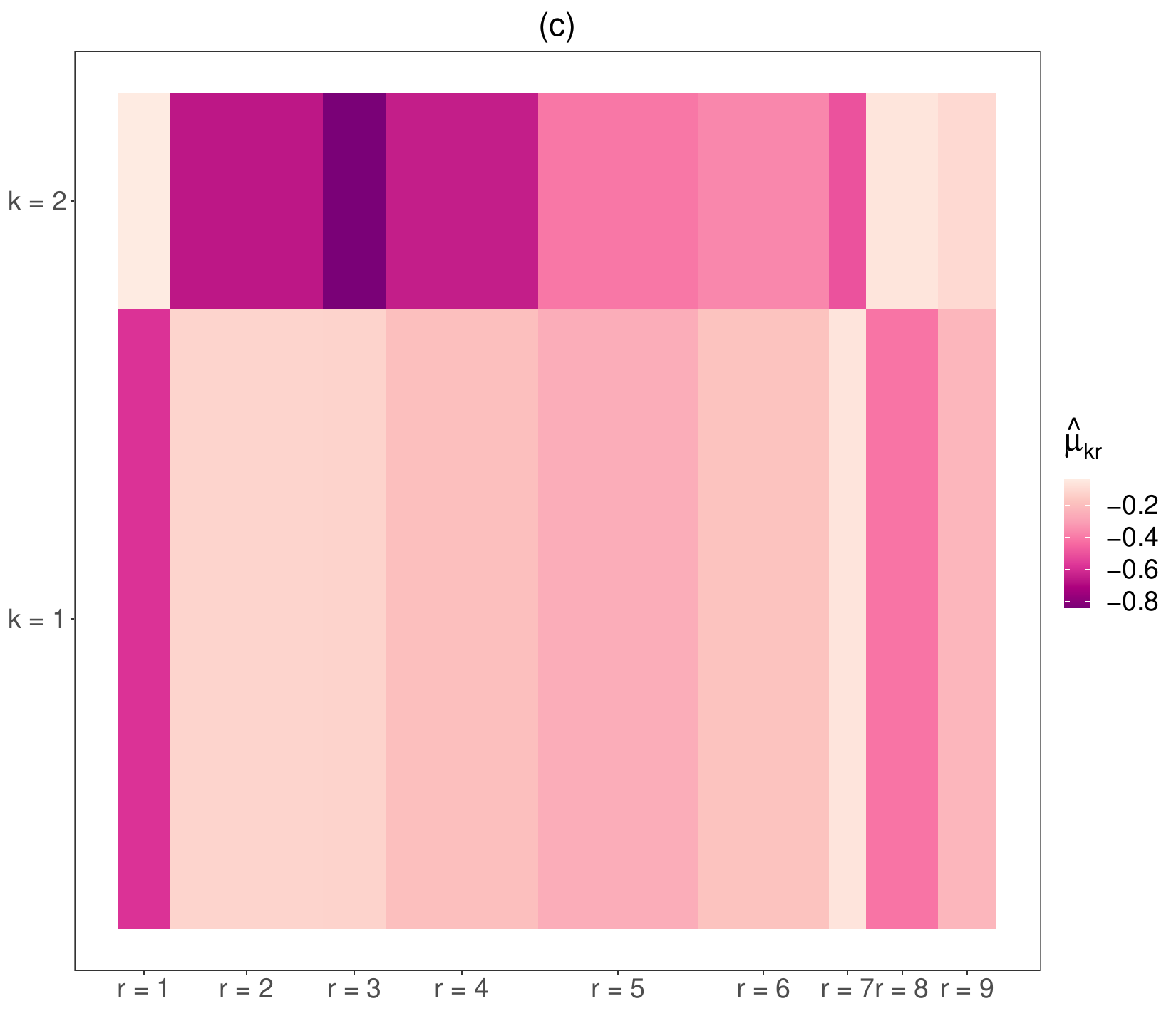}
\includegraphics[width=0.45\linewidth]{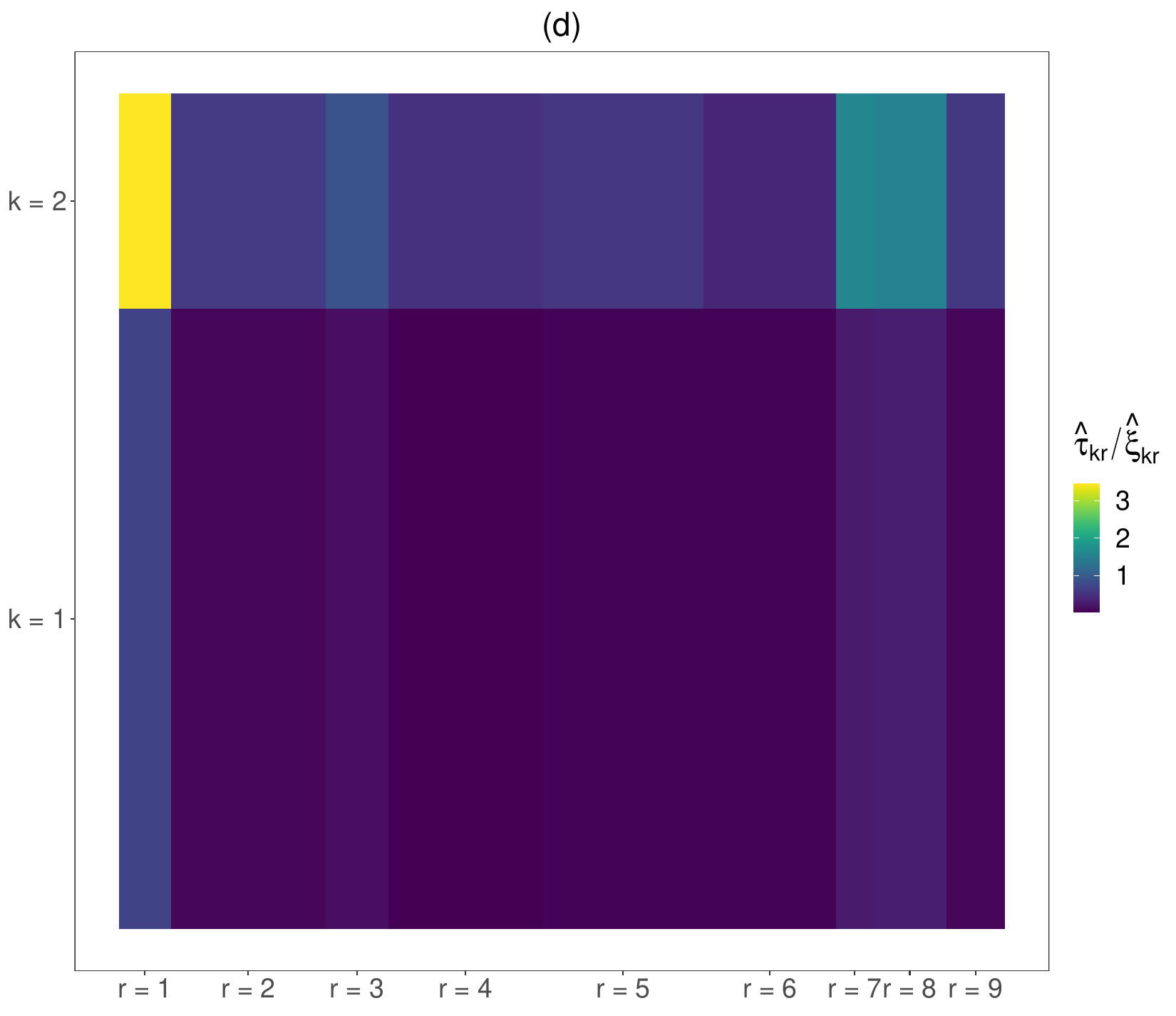}
\caption{\review{Results on the human dorsolateral prefrontal cortex data. The first row displays the 3,639 spots: in Panel (a) they are colored according to the clusters returned by \spartaco~and shaped according to the clustering uncertainty $\colepsilon$, in Panel (b) they are colored according to the average gene expression in the estimated cluster $\mathcal{C}_2$. Panels (c) and (d) represent the data matrix tessellated into the 18 discovered blocks.} Both the genes and the spots are reordered based on the estimated clusters for visualization purposes. The graphs are colored according to the estimated mean  $\hat{\mu}_{kr}$ (c) and to the estimated spatial signal-to-noise ratio $\hat{\tau}_{kr}/\hat{\xi}_{kr}$ (d).}
\label{figure:spatialLIBD_mu_tauxi}
\end{figure}

\section{Application}
\label{sec:Application}

In this section, we analyze the human dorsolateral prefrontal cortex sample from the subject 151673 studied by \cite{Maynard_etal.2021} that we briefly described in Section \ref{subsec:spatial_transcriptomics_intro} and shown in Figure \ref{figure:spatialLIBD_intro}. The dataset has 33,538 genes measured over 3,639 spots. Similarly to 10X scRNA-seq  protocols, 10X Visium yields \emph{unique molecular identifier} (UMI) counts as gene expression values. 

As a first step, we sought to exclude uninformative genes and reduce the analysis to a lower dimensional problem. We applied the gene selection procedure for UMI count data proposed by  \cite{Townes_etal.2019}, i.e., we fit a  multinomial model on every vector of gene expression and compute the deviance. Based on the criterion that large deviance values are associated to informative genes, we kept the first \review{500} genes and discarded the remaining ones. Supplementary Figure 9 shows that the deviance, which is very high for the top genes, reaches a plateau after 200 genes. To normalize the data, we computed, for each selected gene, the deviance residuals based on the binomial approximation of the multinomial distribution as done in \cite{Townes_etal.2019}. 
The result of this procedure is the expression matrix $\X$ whose entries are $x_{ij}\in\mathbb{R}$ and whose row vectors $\x_{i.}$ yield approximately symmetric histograms. \review{Boxplots of the transformed gene expression vectors are given in Supplementary Figure 10, where it is shown also that there is no practical difference between using the binomial or the Poisson for computing the the residuals.

We fitted \spartaco~with all the configurations in $\{(K,R):K = 2,\;R=7,\dots,12\}$, starting the estimation of each model from five different initial points. \review{More details about the setup of the estimation algorithm and the computational costs are given in Supplementary Section 5.}
The range of column cluster values reflects the number of biological layers that appear in Figure \ref{figure:spatialLIBD_intro}. As we already mentioned in Section \ref{subsec:spatial_transcriptomics_intro}, \spartaco~performs a substantially different image clustering than BayesSpace or GIOTTO; thus, we do not expect the clusters discovered by \spartaco~to match the cortical layers. However, we believe that their number could still be indicative of the biological diversity of this specific area. Supplementary Figure 11 (a) gives the ICL values of the  models with $K = 2$. Although our criterion selects $K=2$ and $R = 12$, we believe that the local maximum in correspondence of $(K=2, R=9)$  represents also a valid solution. In fact, a large value of $R$ would result in  too many small clusters, complicating the biological interpretation. 
Furthermore, we fixed $R = 9$ and we explored the options $K \in\{1, 3,4\}$ to investigate the absence of gene clusters ($K=1$)  and the presence of multiple clusters. However, the ICL selects $K=2$.}
Figure \ref{figure:spatialLIBD_mu_tauxi} (a) displays the tissue map colored according to the estimated clusters. \review{The White Matter spots are covered by clusters $\mathcal{D}_1$, $\mathcal{D}_7$, $\mathcal{D}_8$ and $\mathcal{D}_9$; this last one is placed at the border between the White Matter and Layer 6. The remaining clusters cover the surface within the Layers 2-6. Last, Layer 1 is covered by $\mathcal{D}_4$ and mostly by $\mathcal{D}_5$. Incidentally, we note that the spot clusters within the White Matter are the ones with the smallest grade of uncertainty (see the right plot in Supplementary Figure 11 (b)).} 

\review{As for the row clustering, 109 of the genes in Cluster $\mathcal{C}_2$ ($n_{2}= 129$) were ranked within the top 200 most informative genes by the deviance procedure of \cite{Townes_etal.2019}. Figure \ref{figure:spatialLIBD_mu_tauxi} (b) displays the spots colored according to $n_2^{-1} (\mathbf{X}^{2.})^T\mathbf{1}_{n_2}$, the average expression of the genes in $\mathcal{C}_2$, from which it emerges that the expression tends to be larger within the White Matter than in the rest of the cortical area. Panels (c) and (d) in Figure \ref{figure:spatialLIBD_mu_tauxi} display the estimated means $\hat{\mu}_{kr}$ and spatial signal-to-noise ratios $\hat{\tau}_{kr}/\hat{\xi}_{kr}$ within each block. It appears  that the spatial activity of the genes in $\mathcal{C}_2$ is largely evident within the internal area of the White Matter (${\hat{\tau}_{21}}/{\hat{\xi}_{21}}=3.45$) and progressively decreases approaching Layer 6 (${\hat{\tau}_{28}}/{\hat{\xi}_{28}}=1.55$ and ${\hat{\tau}_{29}}/{\hat{\xi}_{29}}=0.58$). These genes show also a moderate spatial expression on the rest of the cortical area (${\hat{\tau}_{2r}}/{\hat{\xi}_{2r}}\in[0.39,0.9]$, for $r = 2,\dots,6$). Last, cluster $\mathcal{D}_7$ denotes a restricted group of spots that are present both within and outside the White Matter, with a non-negligible spatial effect (${\hat{\tau}_{27}}/{\hat{\xi}_{27}}=1.60$). 
On the contrary, the genes in $\mathcal{C}_1$ ($n_1 =  371$) show a small spatial variation in every spot cluster expect in $\mathcal{D}_1$ (${\hat{\tau}_{11}}/{\hat{\xi}_{11}}= 0.71$ and  ${\hat{\tau}_{1r}}/{\hat{\xi}_{1r}}\leq 0.31$ for all $r\neq 1$), suggesting a constant variation of these genes throughout the cortical area. In fact, $\mathcal{C}_1$ is enriched for housekeeping genes with respect to $\mathcal{C}_2$ (chi-square test, $p = 2.6\times 10^{-4}$).  Housekeeping genes are maintainers of the cellular functions and their activity is not restricted to a specific cell type \citep{Eisenberg_Levanon.2003}. It is therefore expected that these genes show a small spatial variation across the tissue. We notice also from Figure \ref{figure:spatialLIBD_mu_tauxi} (c) that the estimated means $\{\hat{\mu}_{1r},r = 1,\dots,9\}$ are complementary to $\{\hat{\mu}_{2r},r = 1,\dots,9\}$: the expression level is smaller within the White Matter area than outside. To ensure that the co-clustering was not driven only by the mean effects, we run also \textsc{sparseBC} using the same number of blocks and $\lambda = 10$: the CER between the gene clusters returned by \spartaco~and \textsc{sparseBC} is 0.44, confirming that the two methods perform a substantially different grouping of the data. A further confirmation of the evidence of our gene clustering is given by the very small uncertainty displayed in the left panel of Supplementary Figure 11 (b).
}

The results discussed above allow us to answer the first two research questions listed in Section \ref{subsec:spatial_transcriptomics_intro} that motivated our work. We now turn our attention to the third research question, namely the identification of genes that exhibit high specific variation.
To do so, for every spot cluster $r$, we investigate the conditional random variables $\sigma^2_{\hat{\Z}_i r,i}|\mathbf{x}^{\hat{\Z}_ir}_{i.}$,
for $i = 1,\dots,n$, 
to determine which genes are most highly variable in each block. We display their density in Supplementary Figures 12
, highlighting in red the twenty genes with the largest $\mathbb{E}(
\sigma^2_{\hat{\Z}_i r,i}|\mathbf{x}^{\hat{\Z}_ir}_{i.}
)$, for every $r$. We expect that genes with a large gene-specific variance in some areas are likely to be informative of the biological mechanisms occurring there. 

\review{First, we notice that all the most variable genes in each of the nine spot clusters belong to $\mathcal{C}_2$.
}
Among the highly variable genes in $\mathcal{D}_1$,  $\mathcal{D}_8$ and $\mathcal{D}_9$ there are \textit{MBP} and \textit{PLP1}, which are responsible, respectively, for the production and the maintenance of myelin, the covering sheath of the nerve fibers in the White Matter. 
Conversely, among the highly variable genes in $\mathcal{D}_2$ and $\mathcal{D}_7$, we notice \textit{PCP4} and \textit{CCK}: these are markers of distinct subtypes of excitatory neurons present in Layers 5-6 \citep{Hodge_etal.2019}. \review{We display the expression of the four genes discussed here in Supplementary Figure 13, showing their pattern in the spot clusters where they appear to be highly variable.}

\review{Supplementary Figure 12 highlights some important differences between ranking genes according to the posterior distribution of our gene-specific variance $\sigma^2_i$ and the method of \cite{Townes_etal.2019} that only ranks genes based on variability without considering the spatial context. This analysis may be used to highlight important genes that would have been missed if the spatial structure of the data would not have been taken into account. Two examples are \textit{CERCAM} and \textit{SAA1}: their ranks according to Townes et al.'s method were 465 and 271, while \spartaco~places them among the most variable genes in the White Matter area (cluster $\mathcal{D}_1$) and in a region covering the Layers 3, 5 and 6 (cluster $\mathcal{D}_6$), respectively. 
We display their expression over the whole tissue in Supplementary Figure 14. \textit{CERCAM} encodes a cell adhesion protein involved in leukocyte transmigration across the blood-brain barrier \citep{starzyk2000cerebral}, while SAA1 is highly expressed in response to inflammation in mouse glial cells \citep{barbierato2017expression}.}

Taken together, these results convincingly show that our model is able to partition the tissue in coherent clusters, which exhibit cluster-specific gene expression, both spatially coordinated and otherwise\review{, and to detect highly variable genes of potential biological interest in specific areas of the tissue that would not have been found without considering their spatial variability.}

\section{Discussion}
\label{sec:discussion}

The growing demand of appropriate statistical methods to analyze spatial transcriptomic experiments has driven us to develop \spartaco, a model-based co-clustering tool that groups genes with a similar profile of spatial expression in specific areas of a tissue.
\spartaco~brings the concepts of spatial modelling into the co-clustering framework, and thus it can be applied to any dataset with entries in the real domain and whose row or column vectors are multivariate observations recorded at some fixed sites in space. 
The inference is carried out via maximization of the classification log-likelihood function. To do so, we put together two variants of the EM algorithm, the classification EM and the stochastic EM, forming what we called the classification-stochastic EM. We completed our proposal deriving the formulation of the ICL for our model to drive the model selection.

A series of simulation studies have highlighted that, in the presence of spatial covariance patterns, the major co-clustering models become inadequate to recover the hidden block structure of the data. On the contrary, \spartaco~has shown remarkable results in each simulation, managing to distinguish different spatial expression profiles in different areas of the image. It further revealed to be robust to the presence of a nuisance component into the data. \review{The model selection driven by the ICL  revealed to be precise but computationally expensive, due to the large number of times the model must be run. On the contrary, other criteria that do not exploit the spatial information of the data are computationally attractive but less accurate. We conclude that the two approaches can be used jointly, using the results given by a fast model selection criterion, such as the PCA-k-means method discussed in Section \ref{subsec:simulation1}, to restrict the range of $K$ and $R$ values to be tested with \spartaco's ICL criterion.} Lastly, we demonstrated how our proposal is capable of answering specific biological research questions using a human brain tissue sample processed with the Visium protocol. Our model has identified two clusters of genes with different spatial expression profiles in \review{nine} different areas of the tissue. A subsequent downstream analysis has allowed us to determine the highly variable genes in each of the nine pinpointed areas. \review{We additionally showed that some of the genes considered as poorly informative by the deviance method of \cite{Townes_etal.2019} are revealed by \spartaco~to be highly variable in specific areas of the tissue sample.}

Although this article has introduced a complete solution to answer some relevant questions in the analysis of spatial transcriptomics, we believe that there is space for further extensions. \review{To use \spartaco~on spatial transcriptomic experiments, the UMI counts must be transformed through a real-valuated function as discussed at the beginning of Section \ref{sec:the_statistical_model}. We performed this step using the pre-processing techniques of \cite{Townes_etal.2019}, which in our application have led to approximately symmetric distributions of the gene expression vectors $\mathbf{x}_i$. In addition, our model is theoretically robust with respect to the presence of heavy tail distributions thanks to the random parameters $\sigma^2_{kr,i}$, that allow to go beyond the normal assumption. Nevertheless, } 
\spartaco~could be extended to directly model UMI counts, similarly to how SPARK \citep{Sun_etal:2020} has extended SpatialDE \citep{Svensson_etal:2018}. Second, to overcome the limitations of the stochastic EM presented in Section \ref{subsec:simulation3}, we could explore the \emph{simulated annealing} algorithm \citep{Vanlaarhoven_Aarts.1987}, to reduce the chances of converging to local maxima.




%
%

\section*{Acknowledgments}
The authors are thankful \review{to the Editor, the Associate Editor, and the two Reviewers for their careful evaluation of our work and for their precious comments,} to Giovanna Menardi and Alessandro Casa for the precious discussions on co-clustering and to Levi Waldron and Vince Carey for help with the framing of the biological questions. We finally thank Dario Righelli for his help with the software implementation.

\section*{Fundings}
This work was supported in part by CZF2019-002443 (DR) from the Chan Zuckerberg Initiative DAF, an advised fund of Silicon Valley Community Foundation. The authors are supported by the National Cancer Institute of the National Institutes of Health (U24CA180996).

\section*{Supplementary Material}
\textbf{Supplementary to \textquotedblleft Co-clustering of Spatially Resolved Transcriptomic Data\textquotedblright}

{Contains the derivation of our information criterion, details on the spatial covariance functions and on the gene covariance matrices used in Section 4, \review{details on the PCA-k-means method for selecting the number of clusters, a discussion on the computational costs of \spartaco,} and additional figures.}
\\\\
\noindent
\textbf{Software}

{Software in the form of an \texttt{R} package that implements \spartaco~is available online at \url{https://github.com/andreasottosanti/spartaco}. All the scripts to reproduce the simulations and the real data analysis are available at \url{https://github.com/andreasottosanti/SpaRTaCo_paper}.}


\bibliographystyle{biorefs} 
\bibliography{refs.bib}       

\end{document}


\maketitle

\section{Derivation of the ICL for SpaRTaCo}
Let $m$ be the current model, and  $K$ and $R$ be the number of row and column clusters. The \emph{integrated classification likelihood} \citep{Biernacki_etal.2000} is defined as
\begin{equation}
\label{formula:jointdistribution}
\begin{aligned}
p(\mathbf{X},\mathbfcal{Z},\mathbfcal{W};m,K,R)&=p(\mathbf{X}|\mathbfcal{Z},\mathbfcal{W};m,K,R)p(\mathbfcal{Z},\mathbfcal{W};m,K,R)\\
&=p(\mathbf{X}|\mathbfcal{Z},\mathbfcal{W};m,K,R)p(\mathbfcal{Z};m,K)p(\mathbfcal{W};m,R).
\end{aligned}
\end{equation}
According to \cite{Biernacki_etal.2000},  the logarithm of the conditional distribution of $\mathbf{X}$ given the clustering labels can be approximated as
$$
\log p(\mathbf{X}|\mathbfcal{Z},\mathbfcal{W};m,K,R)\approx \max_{\boldsymbol{\Theta}} \log p(\mathbf{X}|\mathbfcal{Z},\mathbfcal{W};\boldsymbol{\Theta},m,K,R)+\dfrac{\lambda_{m,K,R}}{2}\log np,
$$
where the first component is the classification log-likelihood evaluated in its maximum, and $\lambda_{m,K,R}$ is the number of free parameters in model $m$ with $K$ and $R$ clusters. Thus, under the identifiability constraint in Section 3.1, $\lambda_{m,K,R}=4KR+\mathrm{dim}(\boldsymbol{\phi})R$. The distribution of both $\mathbfcal{Z}$ and $\mathbfcal{W}$ is Multinomial with probabilities $1/K$ and $1/R$, respectively. It follows that 
$$
\log p(\mathbfcal{Z};m,K)=-n\log K,\hspace{1cm}\log p(\mathbfcal{W};m,R)=-p\log R.
$$
Finally, taking the logarithm of \eqref{formula:jointdistribution} and replacing $\mathbfcal{Z}$ and $\mathbfcal{W}$ with their estimates $\hat{\mathbfcal{Z}}$ and $\hat{\mathbfcal{W}}$, we obtain the ICL.

\section{Spatial covariance functions}
The following isotropic spatial covariance functions have been employed to generate the spatial experiments proposed in Section 4 of the manuscript:
\small{
	$$
	k^{\mathrm{true}}_1(d;\bphi^{\mathrm{true}}_1=\{\theta_E\})=\exp\left(
	-\dfrac{d}{\theta_E},
	\right),
	\hspace{1cm}
	k^{\mathrm{true}}_2(d;\bphi^{\mathrm{true}}_2=\{\theta_R,\alpha_R\})=\left(1+\dfrac{d^2}{2\alpha_{R}\theta_{R}^2}\right)^{-\alpha_{R}},
	$$
	$$
	k^{\mathrm{true}}_3(d;\bphi^{\mathrm{true}}_3=\{\theta_G\})=\exp\left(
	-\dfrac{d^2}{2\theta^2_G},
	\right).
	$$
}
$k^{\mathrm{true}}_1(\cdot;\theta_E)$ is the \emph{Exponential} kernel with scale $\theta_E$, $k^{\mathrm{true}}_2(\cdot;\{\theta_R,\alpha_R\})$ the \emph{Rational Quadratic} kernel with non-negative parameters $(\alpha_R,\theta_R)$, and $k^{\mathrm{true}}_3(\cdot;\theta_G)$ is the \emph{Gaussian} kernel (known also as \emph{Squared Exponential}) with \emph{characteristic length-scale} $\theta_G$.

\section{Covariance matrices of the genes}

\begin{figure}[t]
	\centering
	\includegraphics[width=0.95\linewidth]{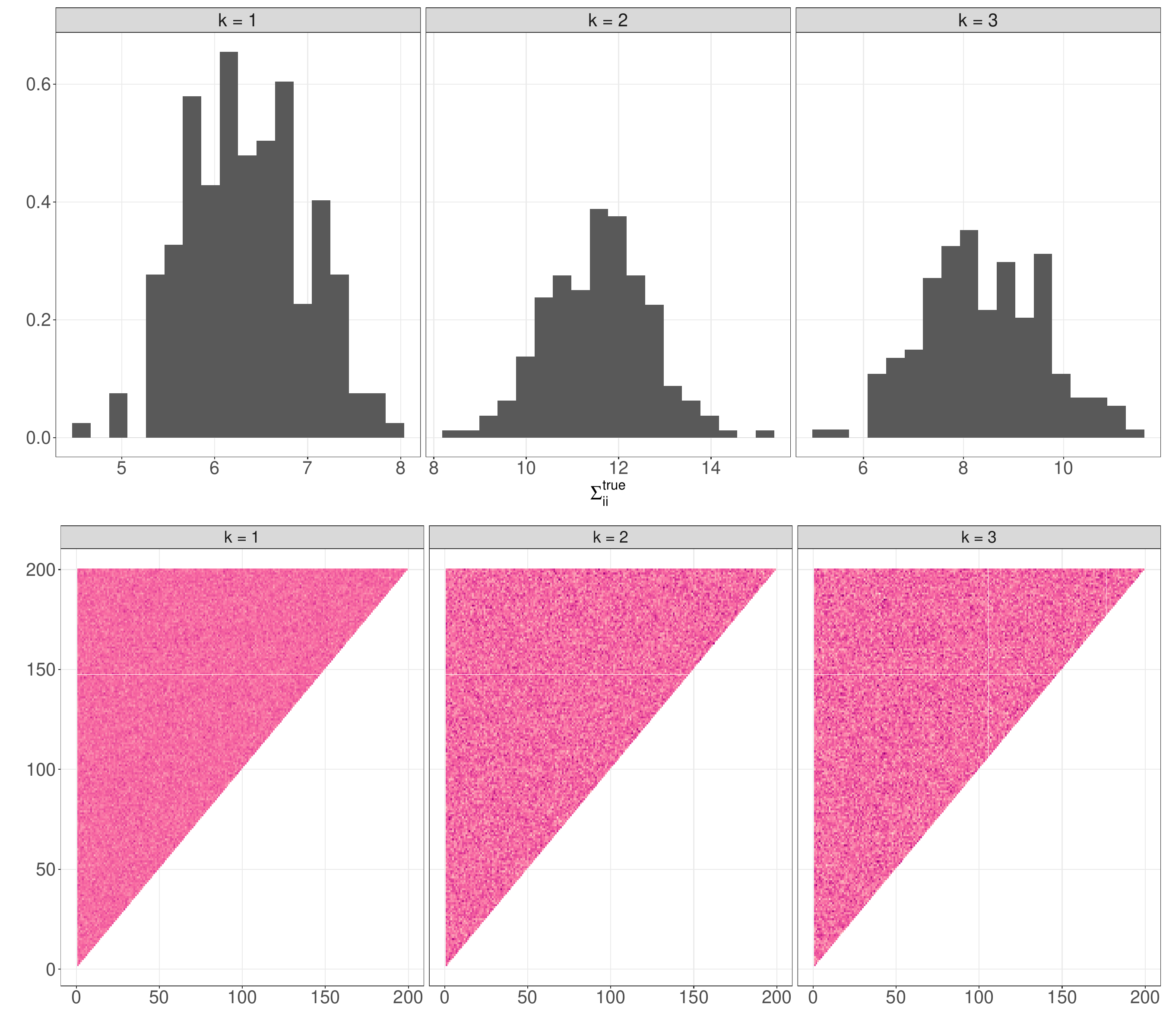}\\
	\includegraphics[width=0.2\linewidth]{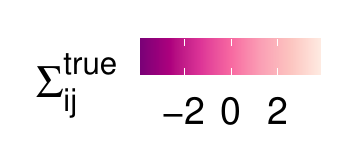}
	\caption{Plot of the row covariance matrices used in Section 2.3 of the manuscript. The top line displays the histogram of the diagonal values of $\bSigma^{\mathrm{true}}_k$, the bottom line displays the upper triangular matrix of $\bSigma^{\mathrm{true}}_k$, for $k = 1,2,3$.}
	\label{figure:scenario1_sigma}
\end{figure}

We describe here the main characteristics of the covariance matrices simulated as in Formula (4.7) of the manuscript.
The degrees of freedom of a Wishart distribution have to be at least equal to the matrix dimension, that is 200. Both the scales and the degrees of freedom are selected in such a way that the values in $\bSigma^{\mathrm{true}}_k$ have the same order of magnitude of $c^{\mathrm{true}}$. For example, using the illustrated setup, the elements on the diagonals of $\bSigma^{\mathrm{true}}_1$ and $\bSigma^{\mathrm{true}}_2$ have expected values 6.3 and 11.5, respectively. 
The top line of Figure \ref{figure:scenario1_sigma} displays the histogram of the diagonal values of a single realization of $\bSigma^{\mathrm{true}}_k$, for $k=1,2,3$.  The values are globally comparable across the three simulations.
The bottom line of Figure \ref{figure:scenario1_sigma} illustrates the elements out of the diagonal of $\bSigma^{\mathrm{true}}_k$. 
The difference between the first and the two other matrices is graphically visible: $\bSigma^{\mathrm{true}}_1$ is in fact the one with the smallest covariance values. The second and the third appear similar: in $\bSigma^{\mathrm{true}}_2$, the elements out of the diagonal are in the range $( -3.2, 3.1)$, while in $\bSigma^{\mathrm{true}}_3$ they are in the range $(-3.88, 3.81)$. 

\begin{figure}[t]
	\centering
	\includegraphics[width=0.6\linewidth]{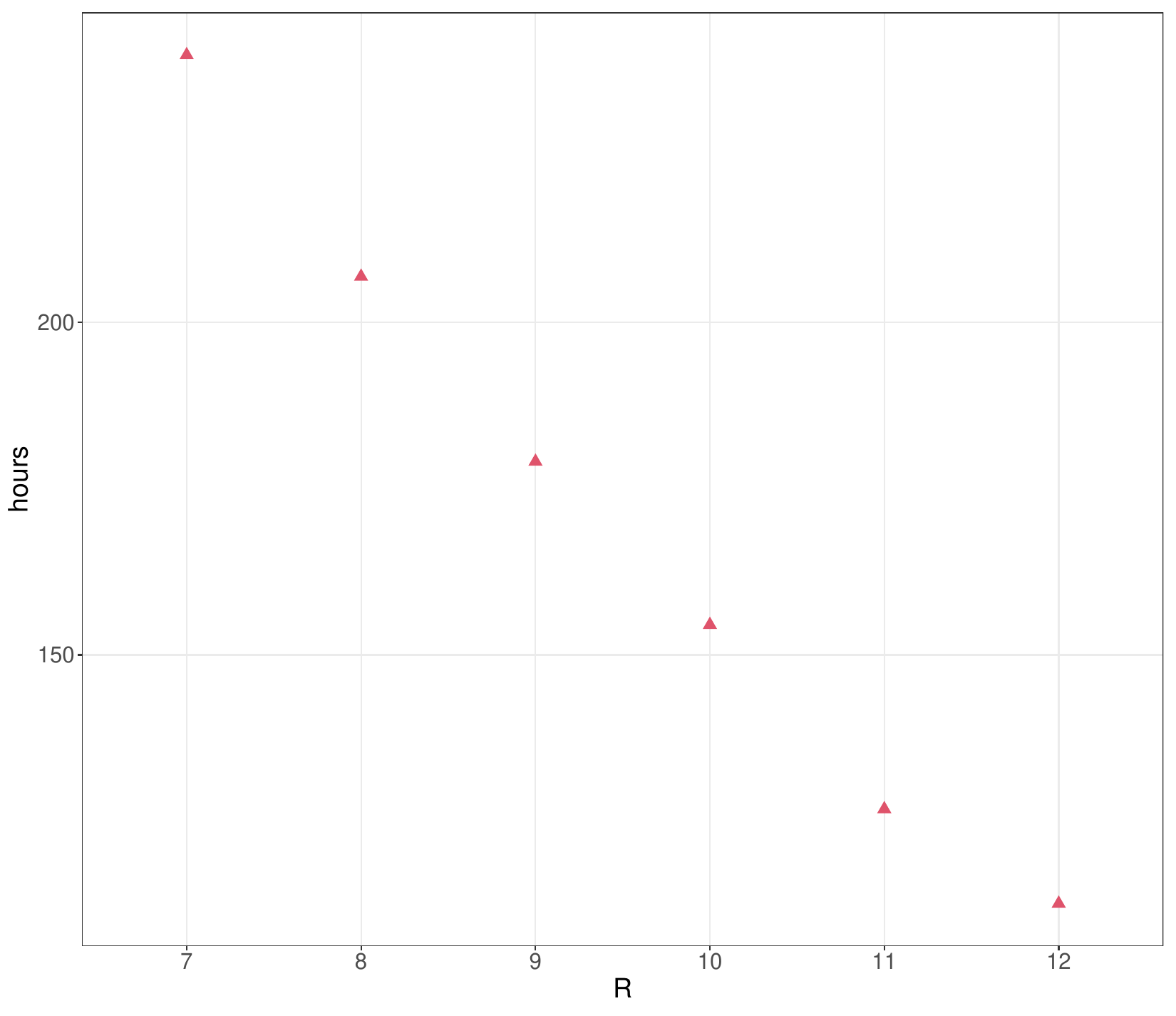}
	\caption{\review{Computation time (in hours) to perform 3,000 iterations of the CS-EM algorithm to estimate \spartaco~with $K=2$ and $R\in\{7,\dots,12\}$ on the spatial experiment studied in Section 5 of the manuscript.}}
	\label{figure:computational_burden}
\end{figure}

\review{
	\section{The PCA-k-means method for selecting the number of co-clusters}
	
	We describe a method for selecting the number of row and column clusters of a data matrix \X~separately  by combing a dimension reduction method with  \textsc{k-means}. Let $\mathbf{A}$ be the matrix obtained by rotating $\X$ with respect to its principal components. 
	The procedure fits \textsc{k-means} on the first two variables of the rotated data, i.e., the first two columns of $\mathbf{A}$, using from  1 to $m_{\mathrm{max}}$ numbers of clusters. Let $\omega^\mathbf{A}_m$ be the total within sum of squares  obtained fitting \textsc{k-means} with $m$ clusters: the integer $m^*$ that solves the following minimization problem,
	\begin{equation*}
	\min_{m^*\in\{1,\dots,m_{\mathrm{max}}\}}\min_{\beta_0,\beta_1,\beta_2} \sum_{m=1}^{m_{\mathrm{max}}} \left\{\omega^\mathbf{A}_m - \beta_0 - \beta_1 (m-m^*)\mathds{1}(m < m^*) - \beta_2\mathds{1}(m \geq m^*)\right\}^2,
	\end{equation*}
	is the selected number of row clusters. The number of column clusters can be determined by applying the same procedure on $\X^T$.
	The method can be applied also imposing $\beta_2 = 0$ to guarantee the continuity between the downward-sloping line $\beta_1 (m-m^*)\mathds{1}(m < m^*)$ and the flat line $\beta_0$.
	
	We implemented this algorithm into the function \texttt{PCA.Kmeans.KR} of the \texttt{R} package \texttt{spartaco}.
	
	\section{Computational burden}
	In this section, we illustrate the computational time spent to perform 3,000 iterations of the CS-EM algorithm on the spatial experiment described in Section 5 of the manuscript. For every iteration, we run the SE Step for 150 times consecutively to favor the exploration of the clustering configurations and speed-up convergence. The time spent (in hours) is given in Figure \ref{figure:computational_burden} for the models with $(K = 2, R \in \{7,\dots, 12\})$.
	
	The SE Step is the most computationally expensive phase because it requires the computation of the classification log-likelihood of every proposed clustering configuration $\W^*$, and thus, to invert the covariance matrices of the clusters that differ from the former configuration $\W^{(t-1)}$.  The larger is $R$, the smaller is the size of the clusters and, consequently, of the matrices to invert. For this reason, models with large $R$ are faster to be estimated.
}



\section{Additional figures}

\subsubsection*{Figures from Section 2}
\begin{figure}[t]
	\centering
	\includegraphics[width=0.5\linewidth]{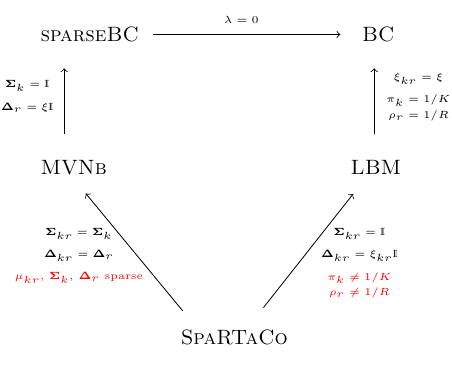}
	\caption{Map of the co-clustering models described in Section 2.2 of the manuscript. An arrow from model A to model B means that B is a special case of A. Details of how to pass from model A to model B are written in black. A red label denotes a difference between two models A an B which does not make B a special case of A.}
	\label{figure:co-clustering_models}
\end{figure}

\begin{itemize}
	\item Figure \ref{figure:co-clustering_models} gives a representation of the relations across co-clustering models described in Section 2.2 of the manuscript.
\end{itemize}

\subsubsection*{Figures from Section 4}
\begin{figure}[t]
	\centering
	\includegraphics[width=0.8\linewidth]{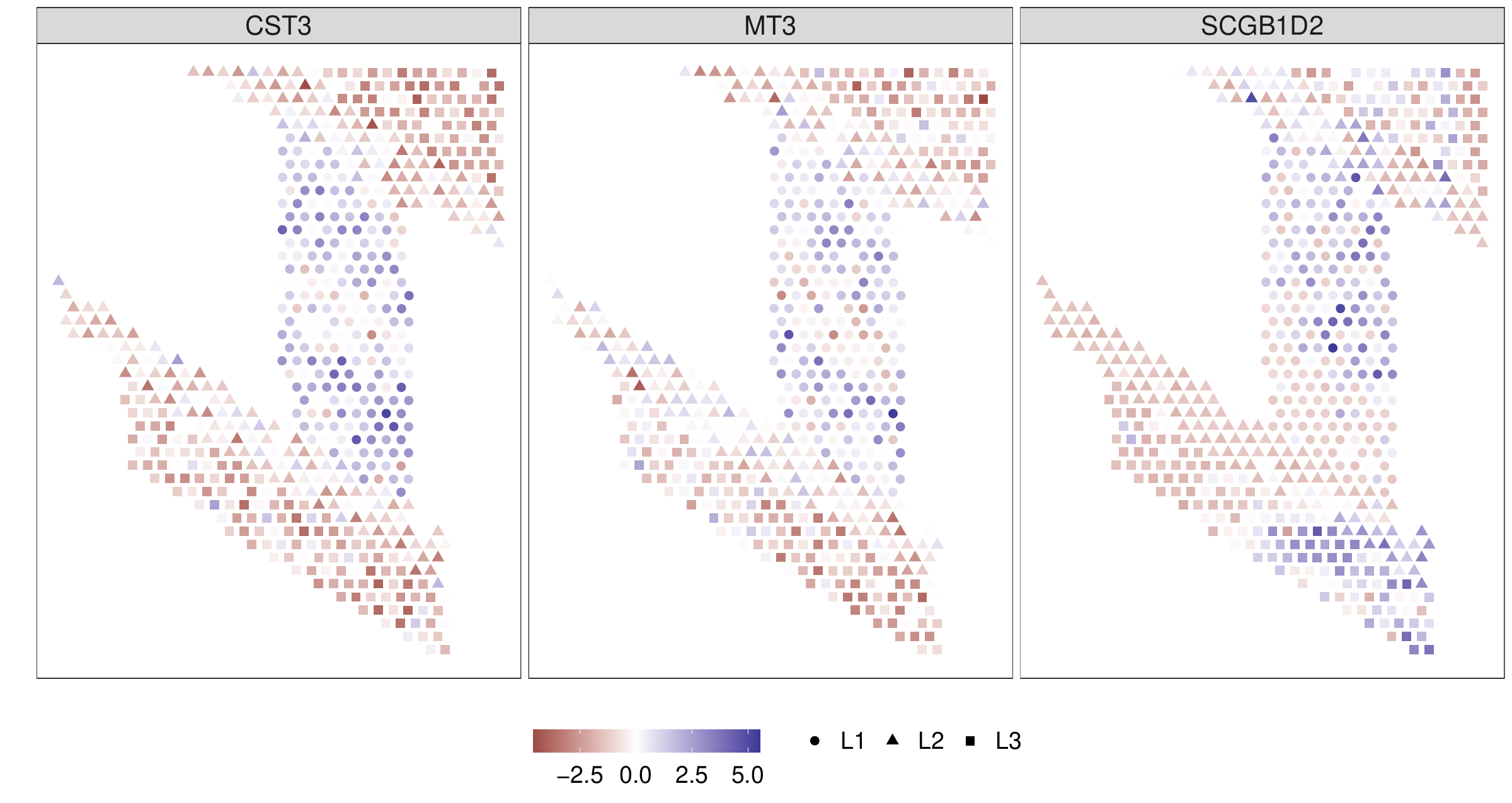}
	\caption{\review{Plot of the expression of three genes in the area used for the simulations, taken from the tissue with ID 151507. The symbols denote three different layers of the tissue. The gene expression was transformed from  counts  to a continuous measurement through the pre-processing procedure of \cite{Townes_etal.2019}. More details of this transformation are given in Section 5 of the manuscript.}}
	\label{figure:151507_genes}
\end{figure}

\begin{figure}[t]
	\centering
	\includegraphics[width=0.7\linewidth]{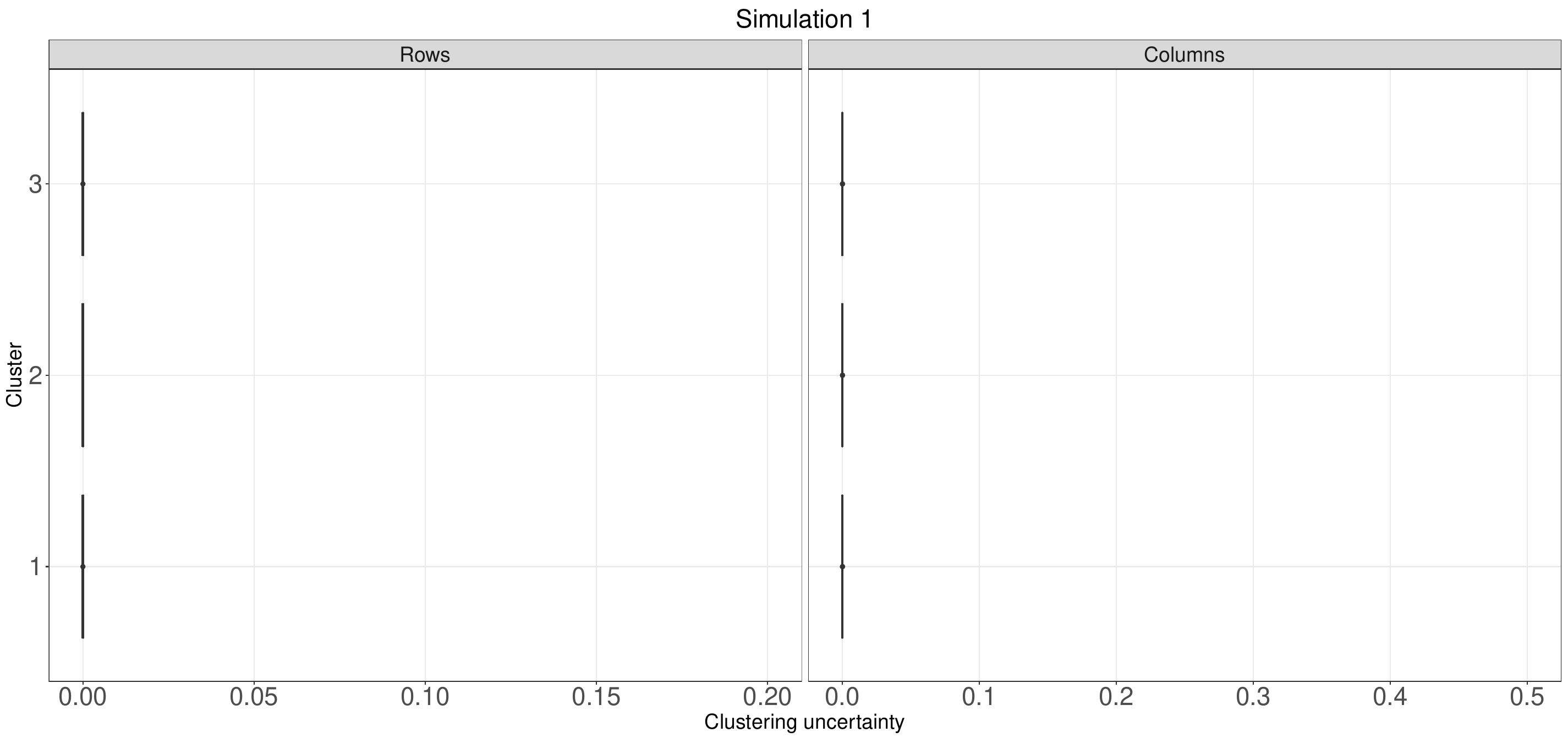}
	\includegraphics[width=0.7\linewidth]{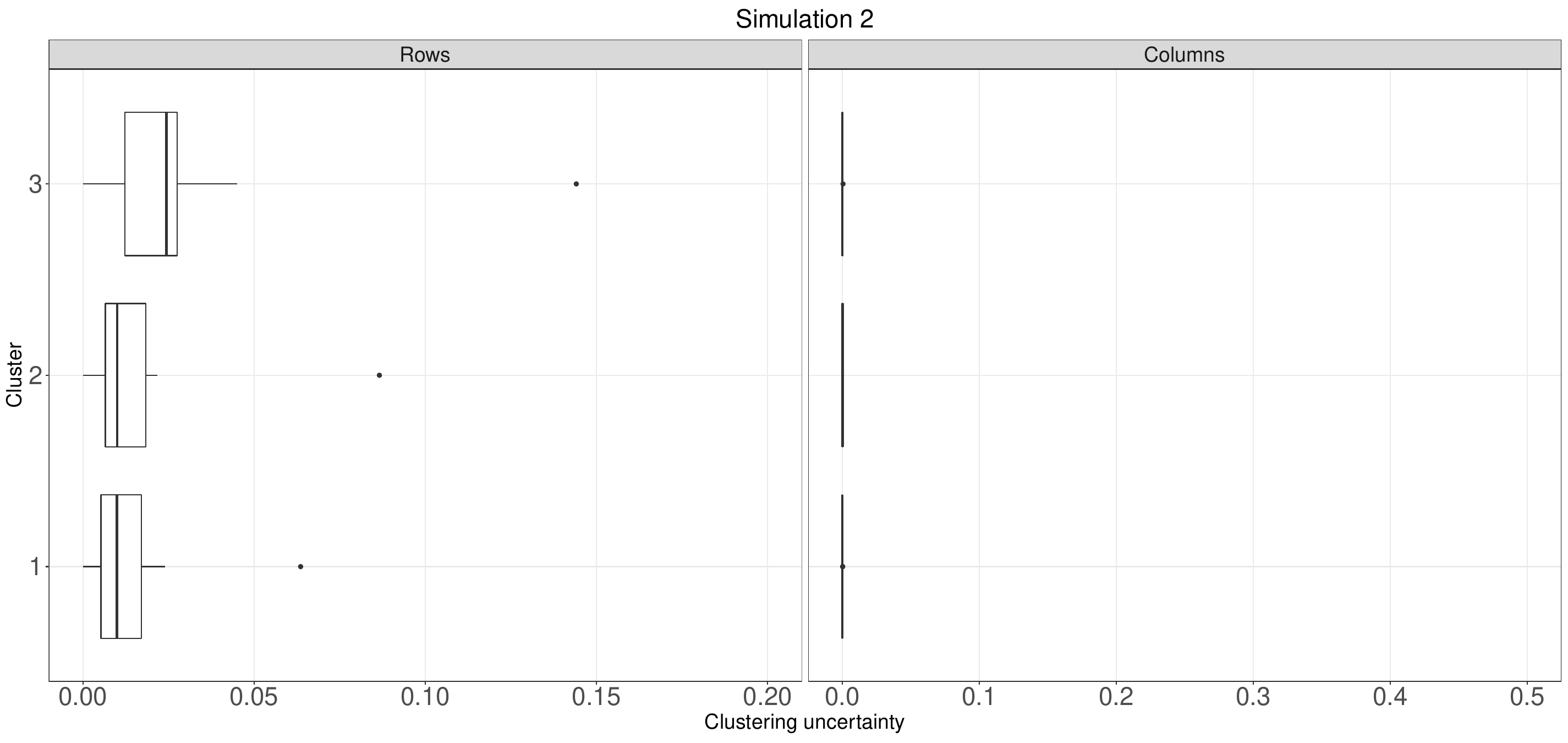}
	\includegraphics[width=0.7\linewidth]{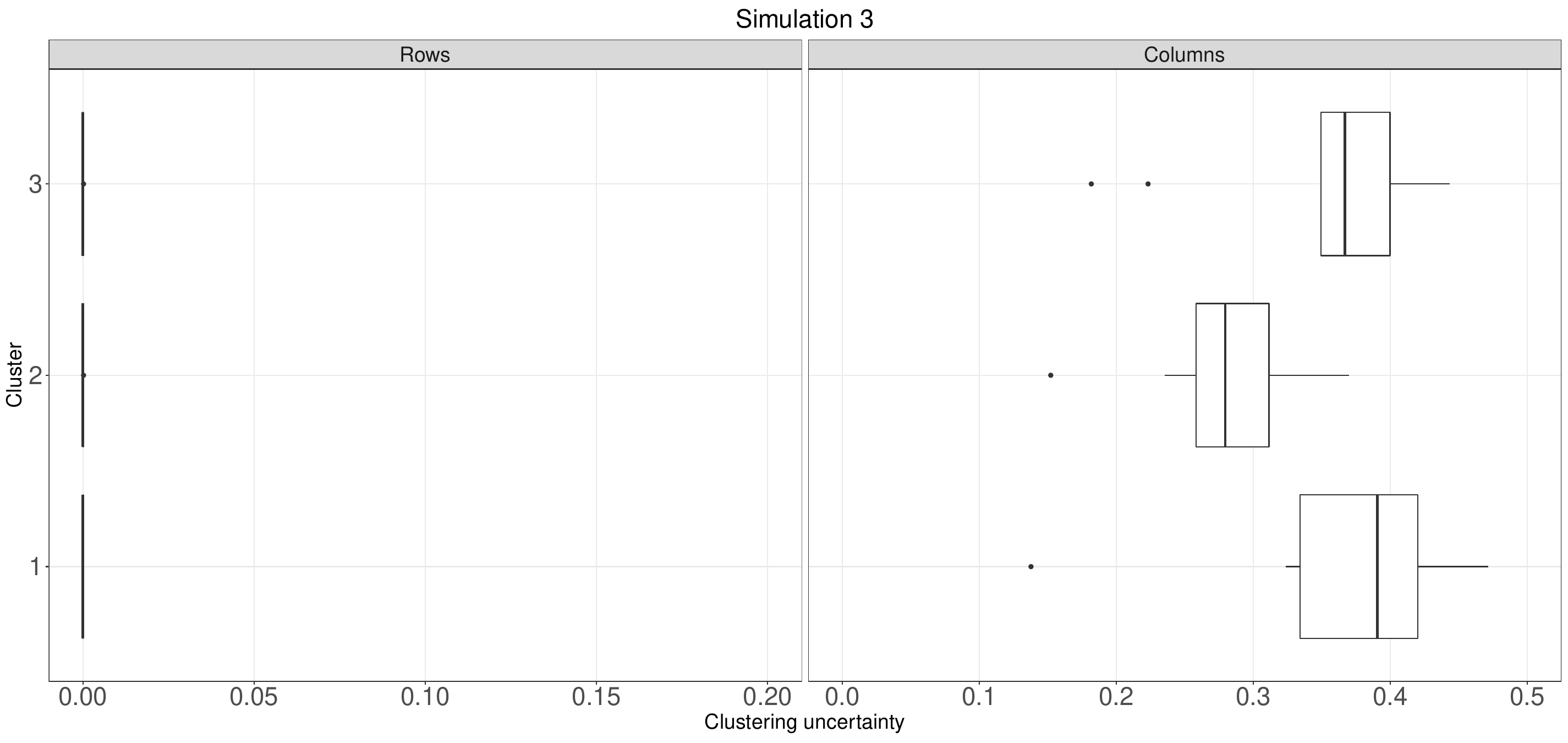}
	\includegraphics[width=0.7\linewidth]{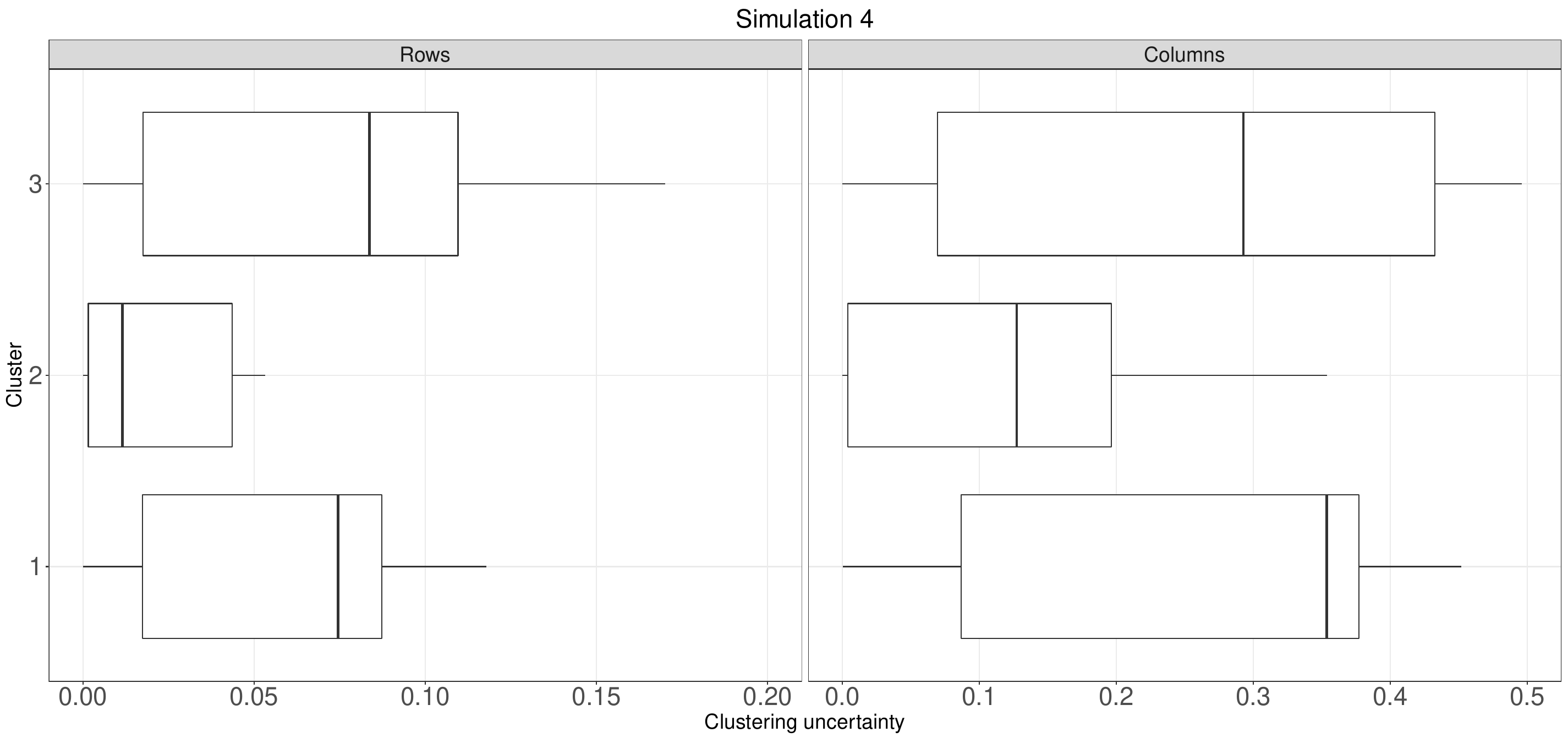}
	\caption{\review{Clustering uncertainty from Simulations 1-4. For each scenario, we fitted \spartaco~using five parallel runs and we estimated the quantities $\rowepsilon$ and $\colepsilon$ on each of the 10 replicates. Every figure gives the boxplots of $\rowepsilon$ (left panel) and $\colepsilon$ (right panel). Since the same cluster might take different labels across the replicates, we had to relabel the estimated clusters using the true clustering labels as reference.
	}}
	\label{figure:clustering_uncertainty_simulations}
\end{figure}

\begin{figure}[t]
	\centering
	\includegraphics[width=0.8\linewidth]{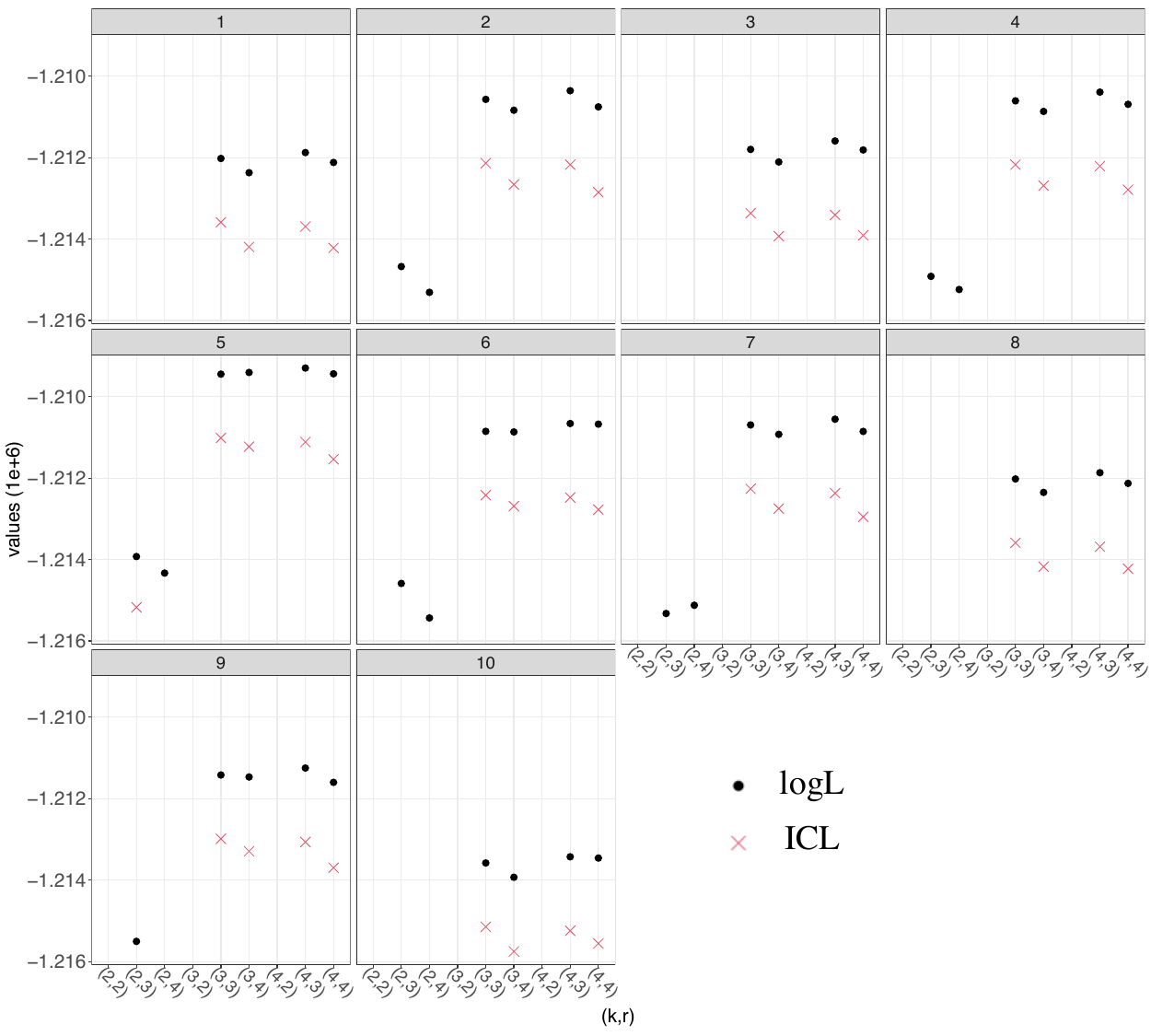}
	\caption{Detail of Simulation 1. The graphs give the log-likelihood and the ICL values on each of the 10 replicates of the experiment, using different configurations of \spartaco. We truncate on purpose the extremes of the y-axis to show only the largest log-likelihood and ICL values.
	}
	\label{figure:scenario1_multipleKR}
\end{figure}

\begin{figure}
	\centering
	\includegraphics[width= .8\linewidth]{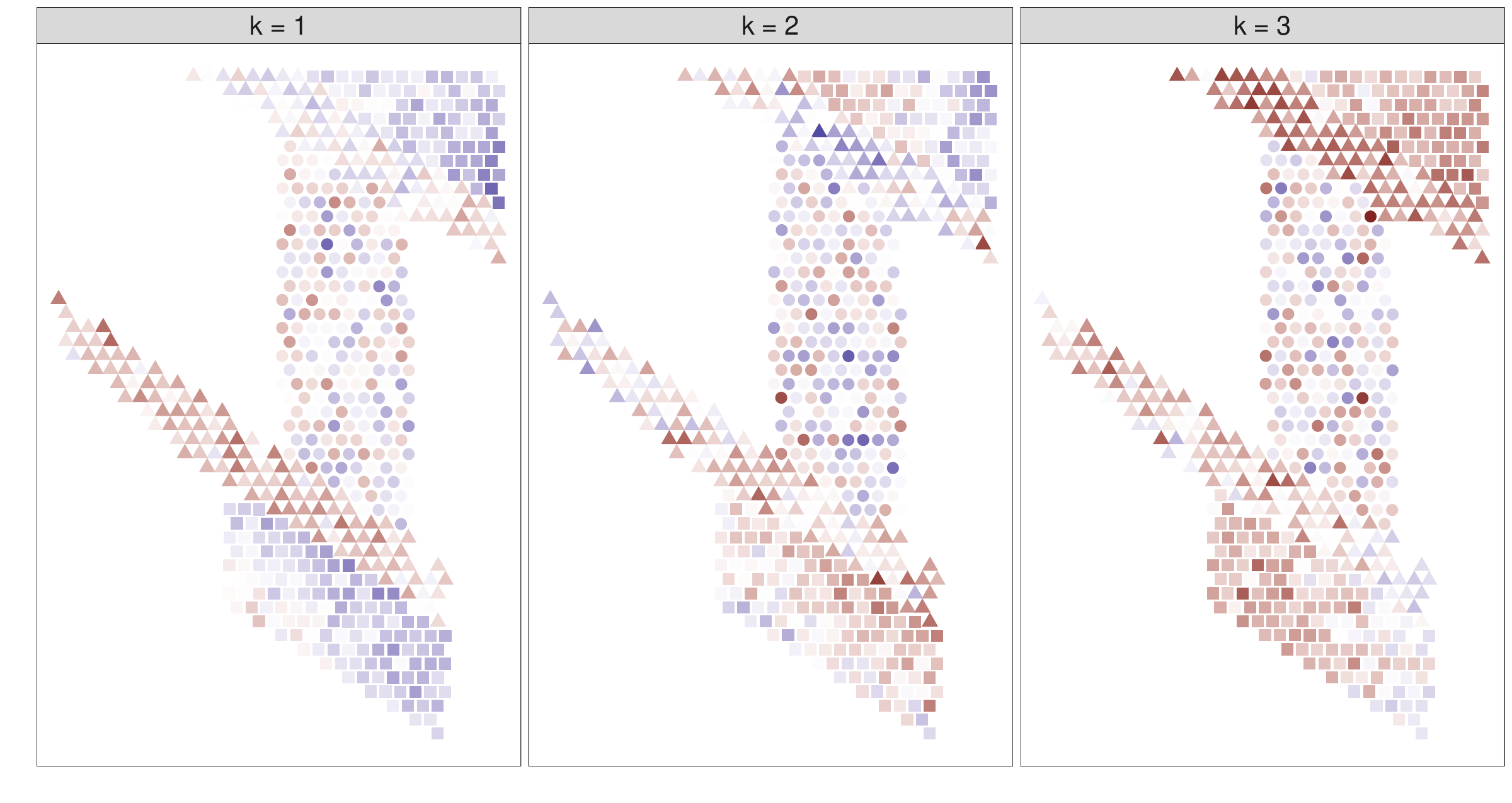}
	\includegraphics[width= .8\linewidth]{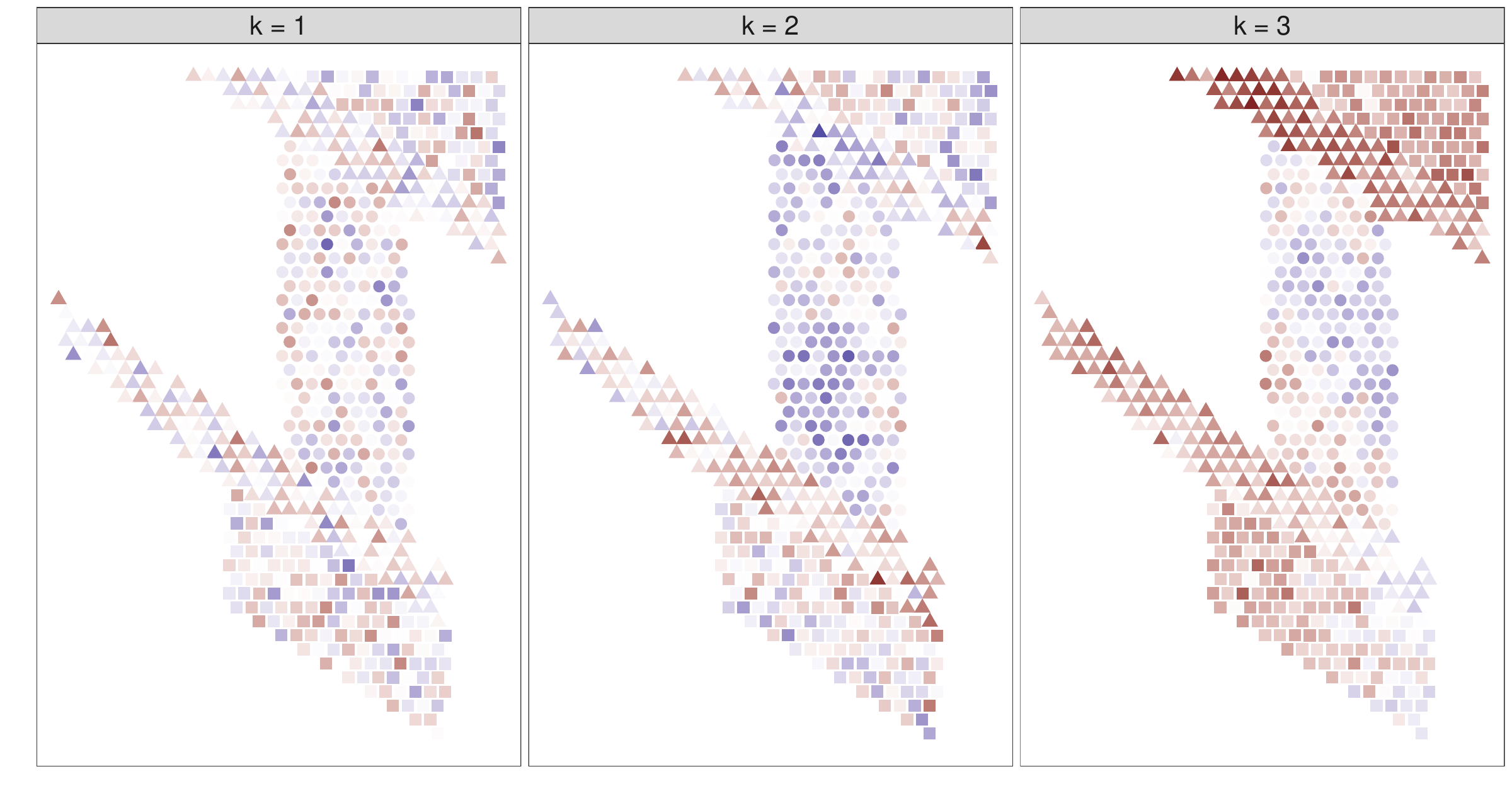}
	\includegraphics[width= .4\linewidth]{SEgenes_legend.pdf}
	\caption{\review{Examples of a spatial experiment generated under Simulation 2 (top row) and 3 (bottom row). The spots are coloured according to  $n^{-1}_k(\mathbf{X}^{k.})^T\mathbf{1}_{n_k}$, the average expression of the $k$-th gene cluster. The three spot clusters are displayed with different symbols. The co-clusters with no spatial expression are the ones associated to $r=1$ in Simulation 2, and the ones associated to $k=1$ in Simulation 3.  The co-clusters with the largest spatial signal-to-noise ratio are  the ones associated to $r=3$ in Simulation 2, and the ones associated to $k=3$ in Simulation 3.}}
	\label{figure:scenario1_geneexpression}
\end{figure}

\begin{figure}
	\centering
	\includegraphics[width=1\linewidth]{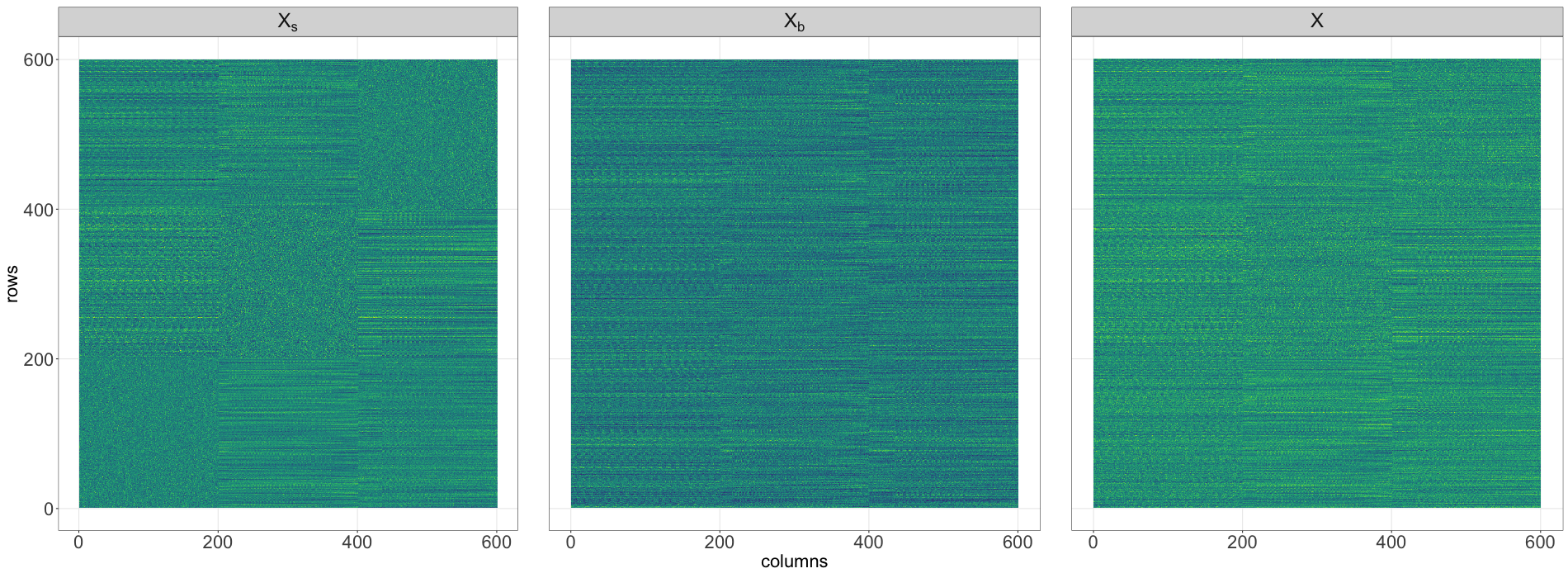}\\
	\includegraphics[width=0.15\linewidth]{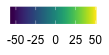}
	\caption{Simulation 4. The matrices $\X_s$, $\X_b$ and $\X$ appear from the left to the right, using $\lambda_s = \lambda_b = \sqrt{0.5}$.
	}
	\vspace{1cm}
	\label{figure:Scenario4_example}
	\includegraphics[width=0.5\linewidth]{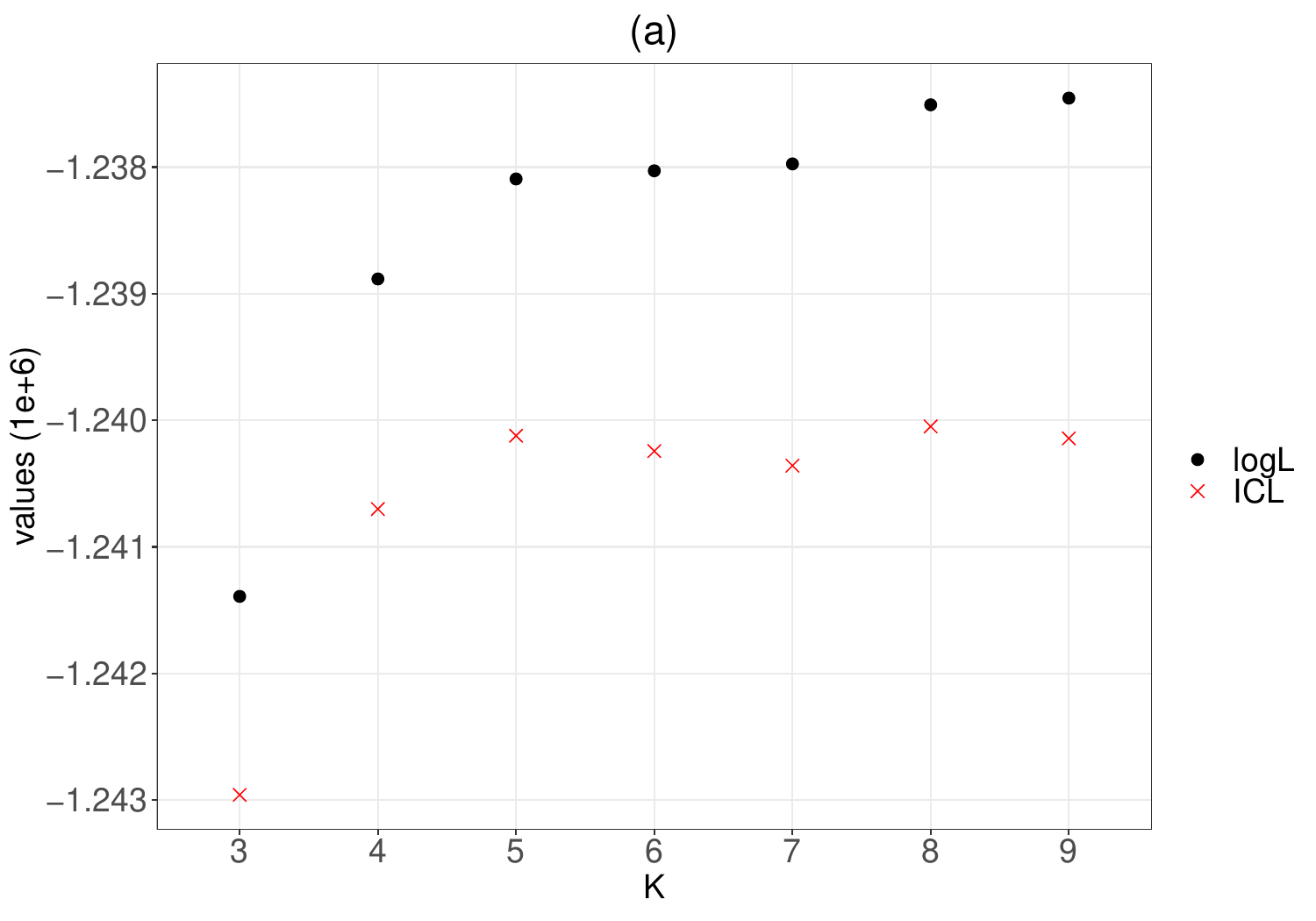}\\
	\hspace{-1.5cm}\includegraphics[width=0.8\linewidth]{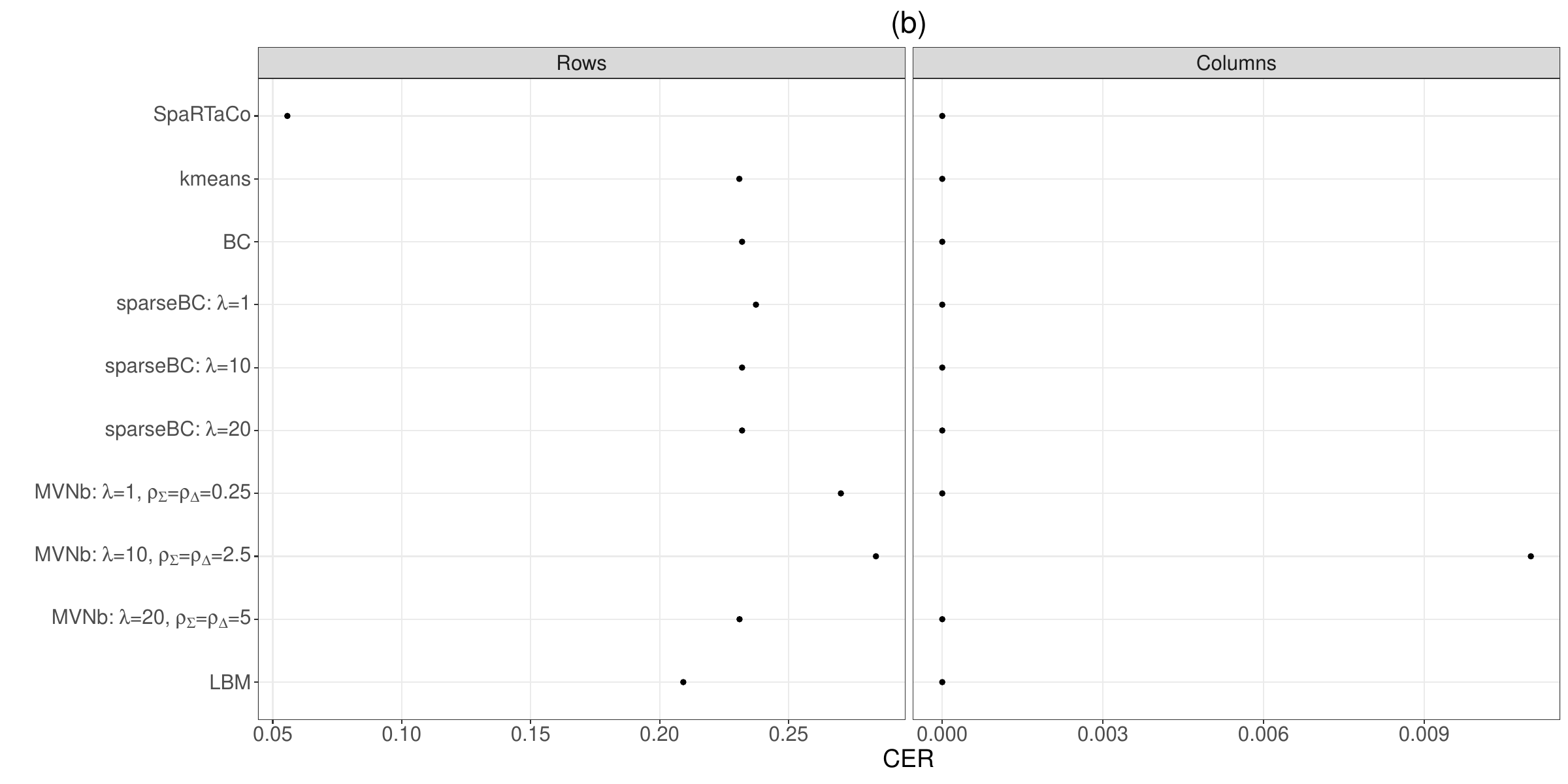}
	\caption{\review{Results from Simulation 5. Figure (a) compares the classification log-likelihood and the ICL of different \spartaco~models with $K$ varying from 3 to 8 and with $R=3$. The best model according to the ICL criterion is the one with $K=8$ row clusters. Panel (b) gives the CER obtained on the rows and on the columns using \spartaco~with and the competing models, all with $K=5$ and $R = 3$.}}
	\label{figure:Scenario5_results}
\end{figure}

\begin{itemize}
	\review{
		\item Figure \ref{figure:151507_genes} shows the expression of three genes measured on the tissue sample 151507, whose spots have been used to build our simulations.
		\item Figure \ref{figure:clustering_uncertainty_simulations} gives the boxplots of the quantities $\rowepsilon$ and $\colepsilon$, the row and column clustering uncertainties, measured over the 10 replicates of the first four simulation experiments proposed.
	}
	\item Figure \ref{figure:scenario1_multipleKR} shows the results of the model selection performed in Section 4.3 using the ICL criterion.
	\item \review{Figure \ref{figure:scenario1_geneexpression} gives an example of spatial experiments simulated under the frameworks discussed in Sections 4.4 and 4.5.}
	\item  Following the notation used in Section 4.6 of the manuscript, Figure \ref{figure:Scenario4_example} shows a single realization of $\mathbf{X}_s$, $\mathbf{X}_b$ and $\mathbf{X}$ using $\lambda_s =\lambda_b=\sqrt{0.5}$.
	\item \review{Figure \ref{figure:Scenario5_results} shows the results of the model selection performed in Section 4.7. Panel (a) compares the classification log-likelihood and the ICL, for any model dimension proposed. Panel (b) gives the CER values obtained on the unique replicate of the simulation experiment proposed, using different co-clustering models.}
	
\end{itemize}

\subsubsection*{Figures from Section 5}
\begin{figure}
	\centering
	\includegraphics[width=0.65\linewidth]{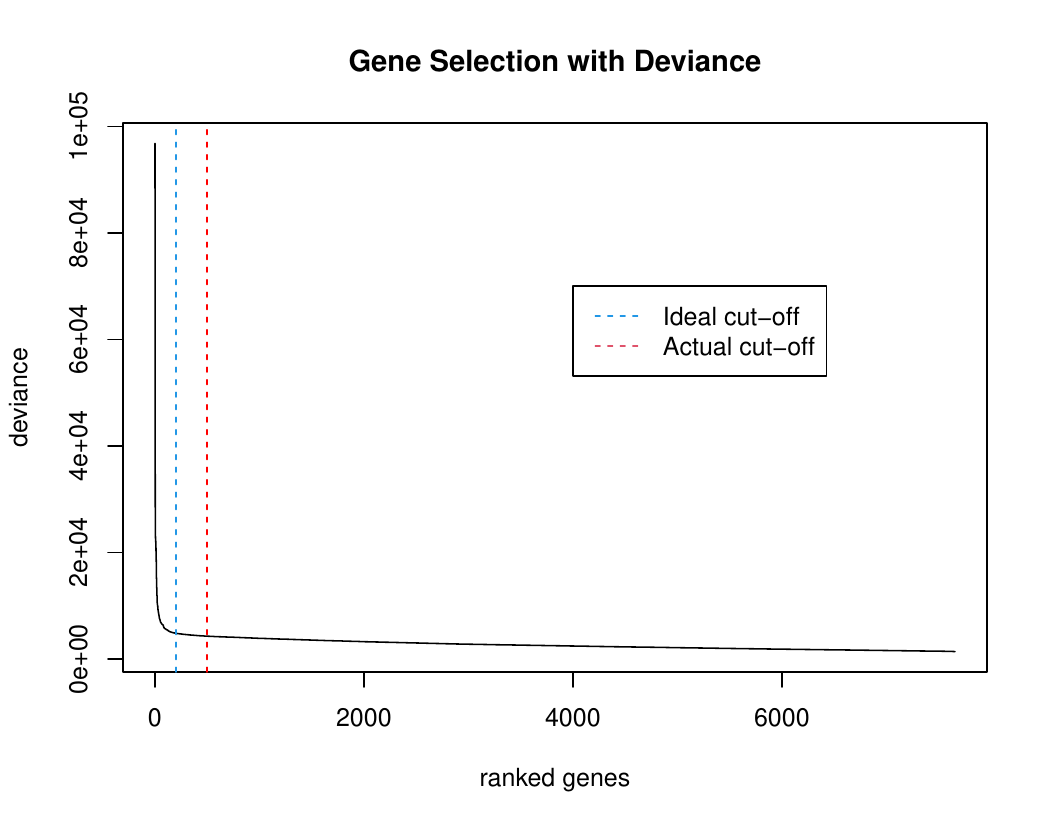}
	\caption{Graph of the genes measured on the prefrontal cortex sample analyzed in Section 5, sorted in decreasing order according to the deviance value. High deviance values are associated to informative genes. Even if from a graphical evaluation the ideal number of genes is around 200, we included in the analysis the \review{500} genes with the largest deviance.}
	\label{figure:Scenario5_geneselection}
\end{figure}

\begin{figure}[t]
	\centering
	\includegraphics[width=0.45\linewidth]{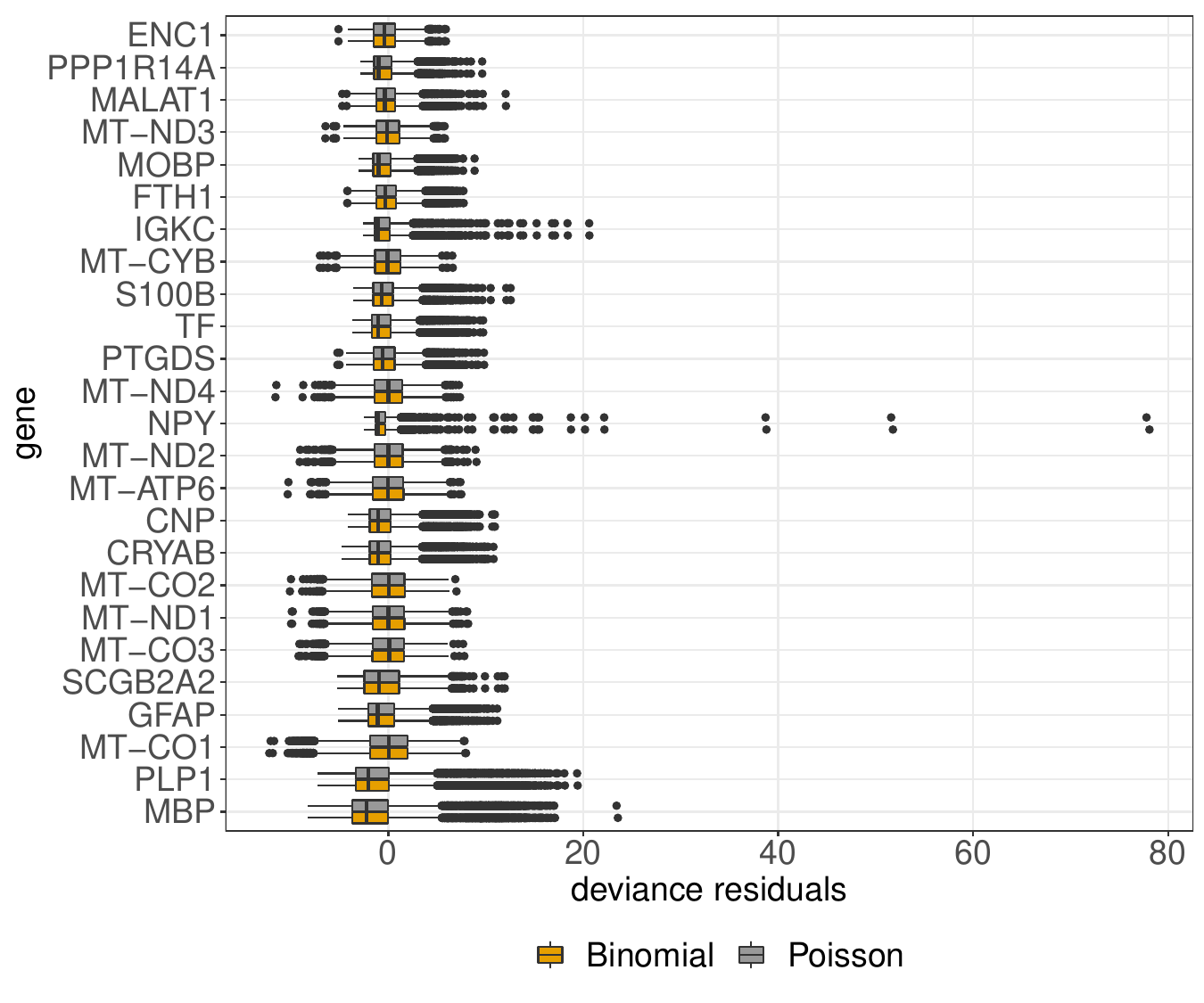}
	\includegraphics[width=0.45\linewidth]{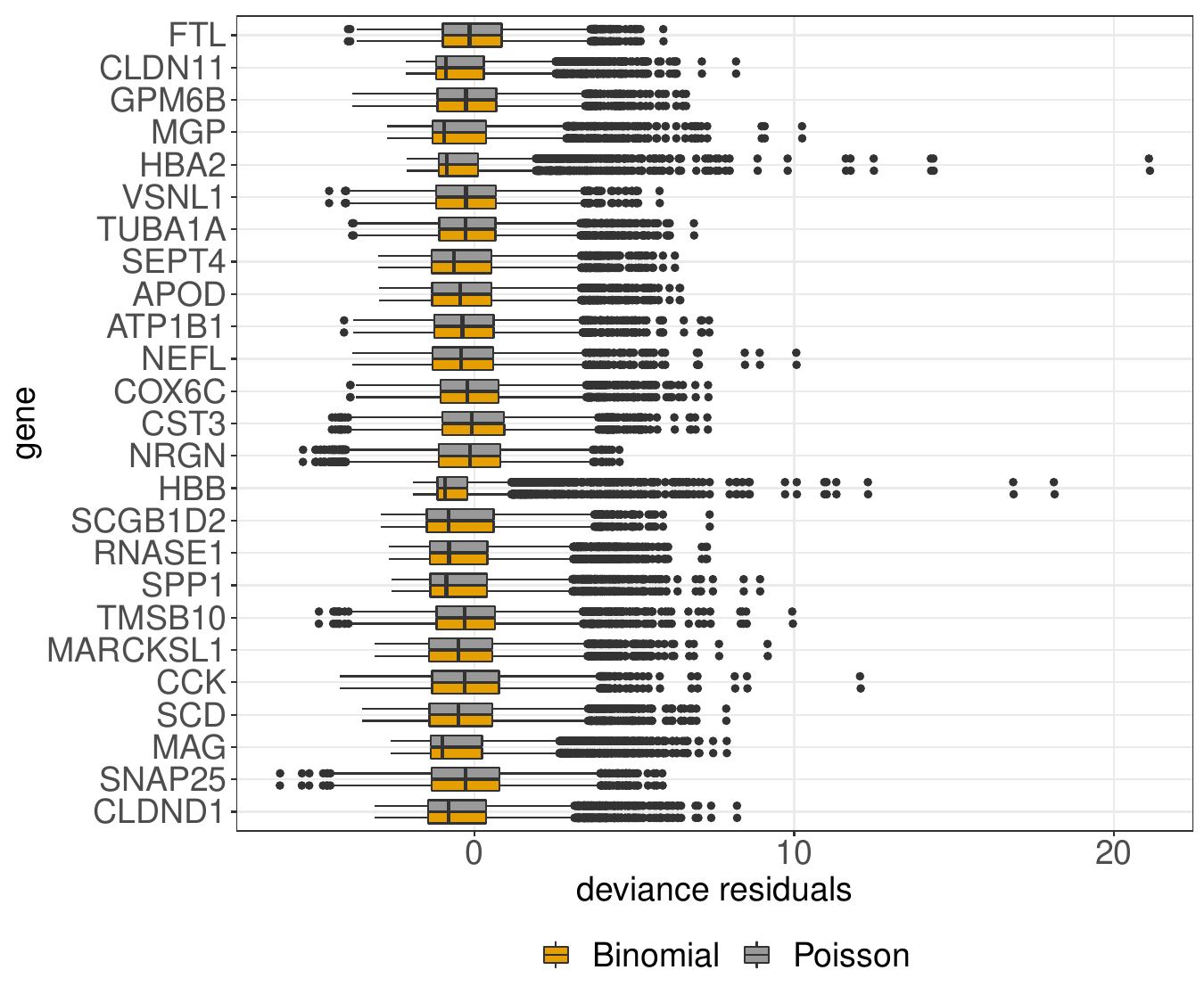}\\
	\includegraphics[width=0.45\linewidth]{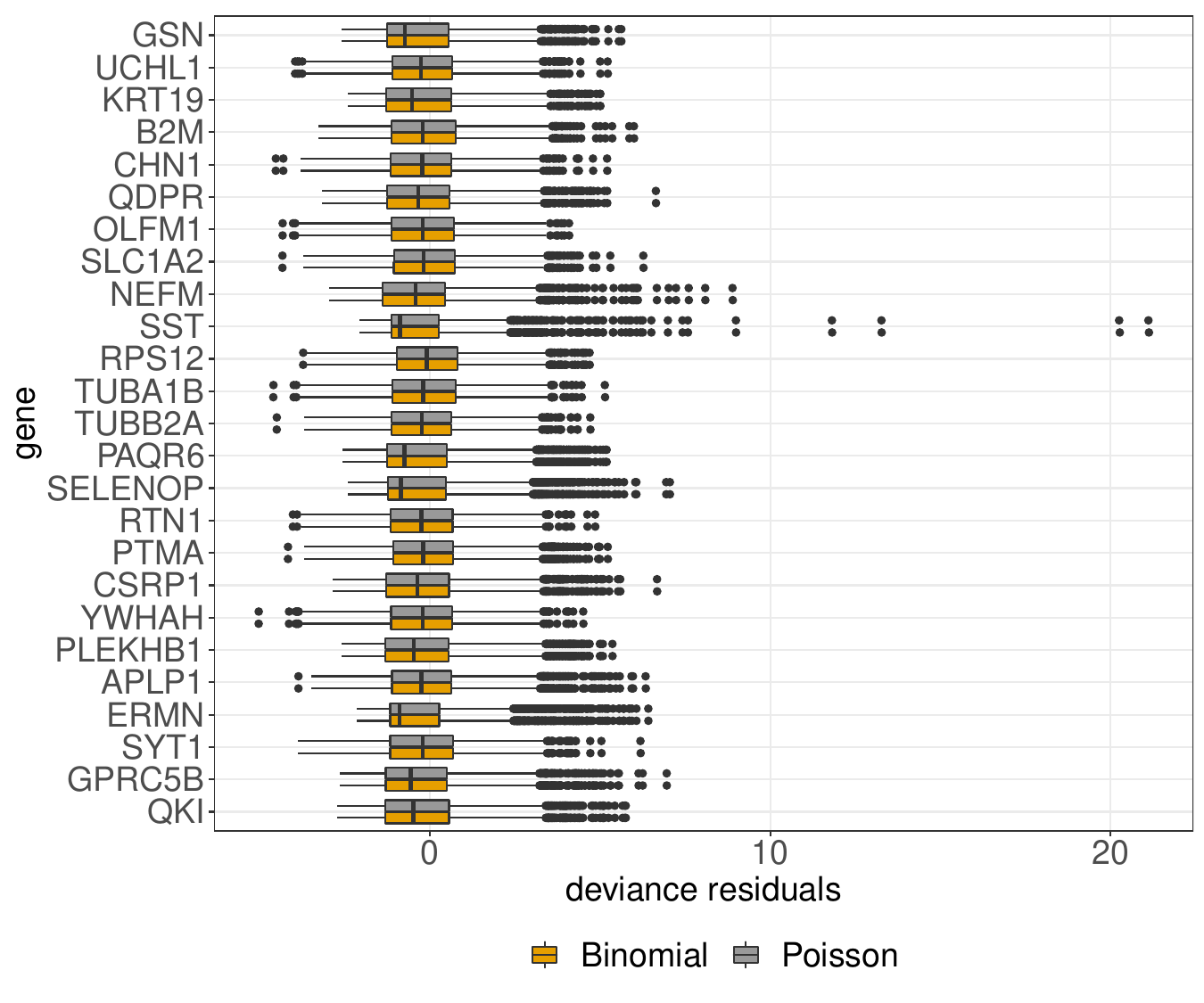}
	\includegraphics[width=0.45\linewidth]{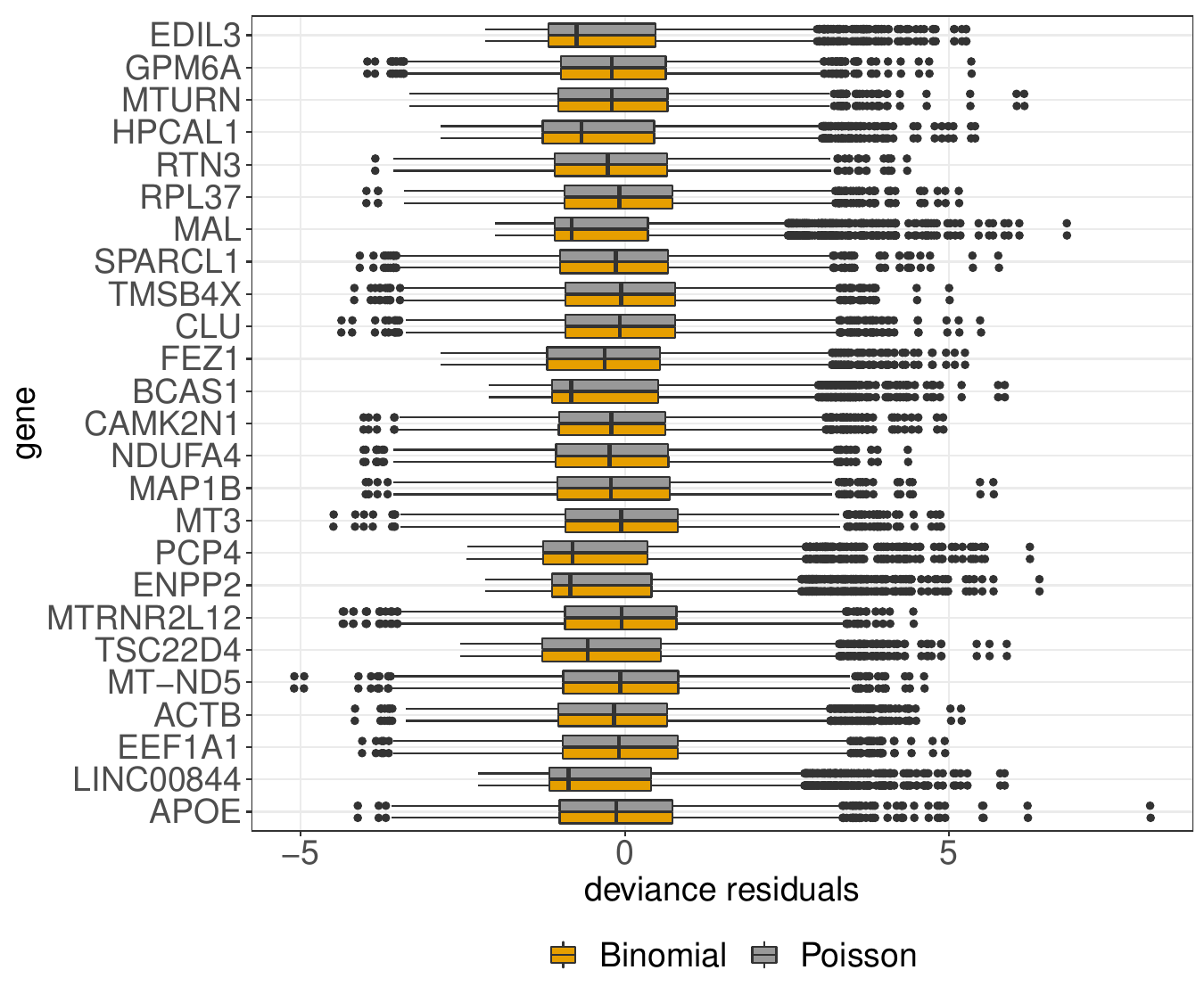}
	\caption{\review{Boxplots of the first 100 row vectors $\x_i$ from the prefrontal cortex tissue sample analyzed in Section 5. For every gene, we plot the deviance residuals using an approximation of the multinomial model based on both the binomial and the Poisson distributions, showing that the two methods are in practice equivalent on this dataset. It is worth remembering that, due to the column clustering performed by \spartaco, each boxplot must be seen as the collection of $r = 1,...,R$ different subvectors, each of length $p_r$. Therefore, it is not required to the distributions of the $x_i$ to be symmetric.}}
	\label{fig:binom_poisson_deviance}
\end{figure}

\begin{figure}[t]
	\centering
	\includegraphics[width=0.5\linewidth]{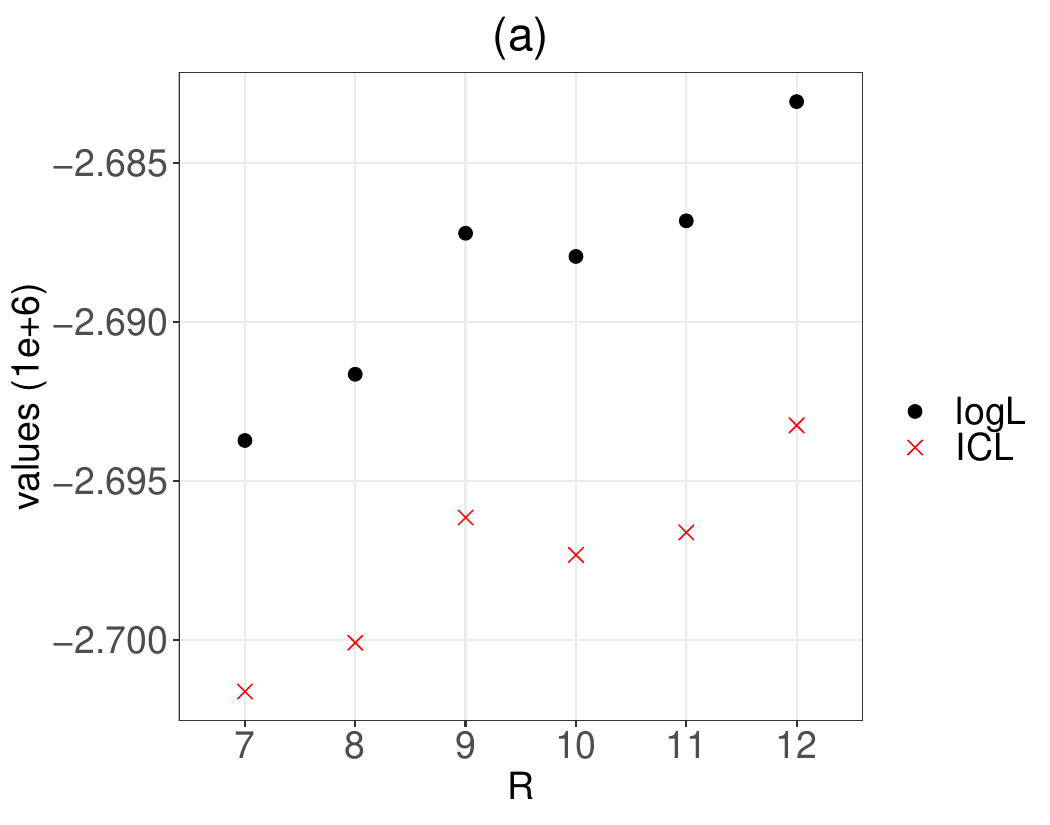}\\
	\includegraphics[width=0.9\linewidth]{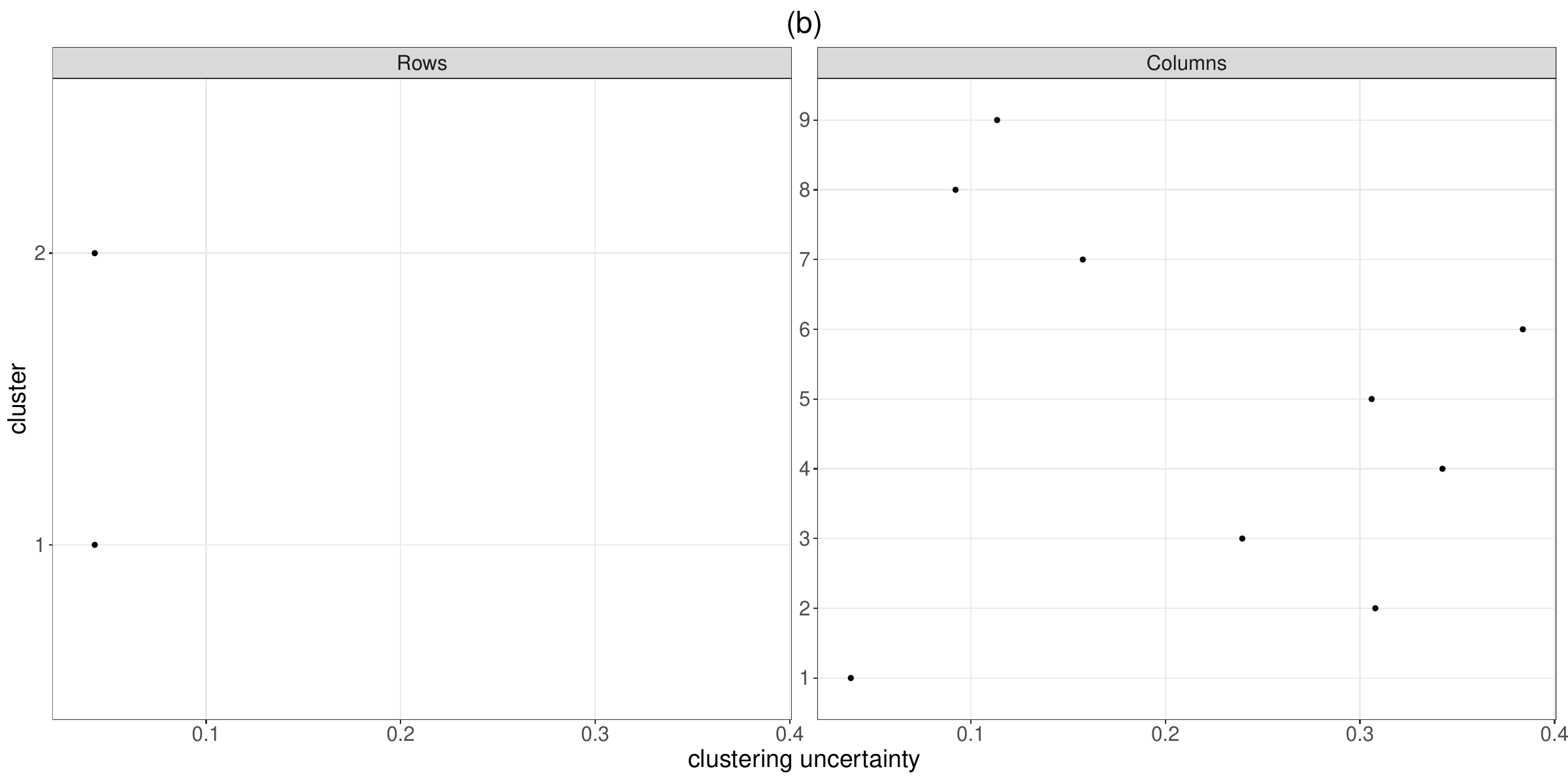}
	\caption{\review{Results from Section 5. Panel (a) compares the classification log-likelihood and the ICL of different
			\spartaco~models with $R$ varying from 7 to 12 and with $K = 2$. Panel (b) displays the clustering uncertainty measures $\rowepsilon$ and $\colepsilon$ for the selected model ($K=2$, $R = 9$).}}
	\label{fig:section5_icl_clusteringuncert}
\end{figure}

\begin{figure}[t]
	\centering
	\includegraphics[width=.43\linewidth]{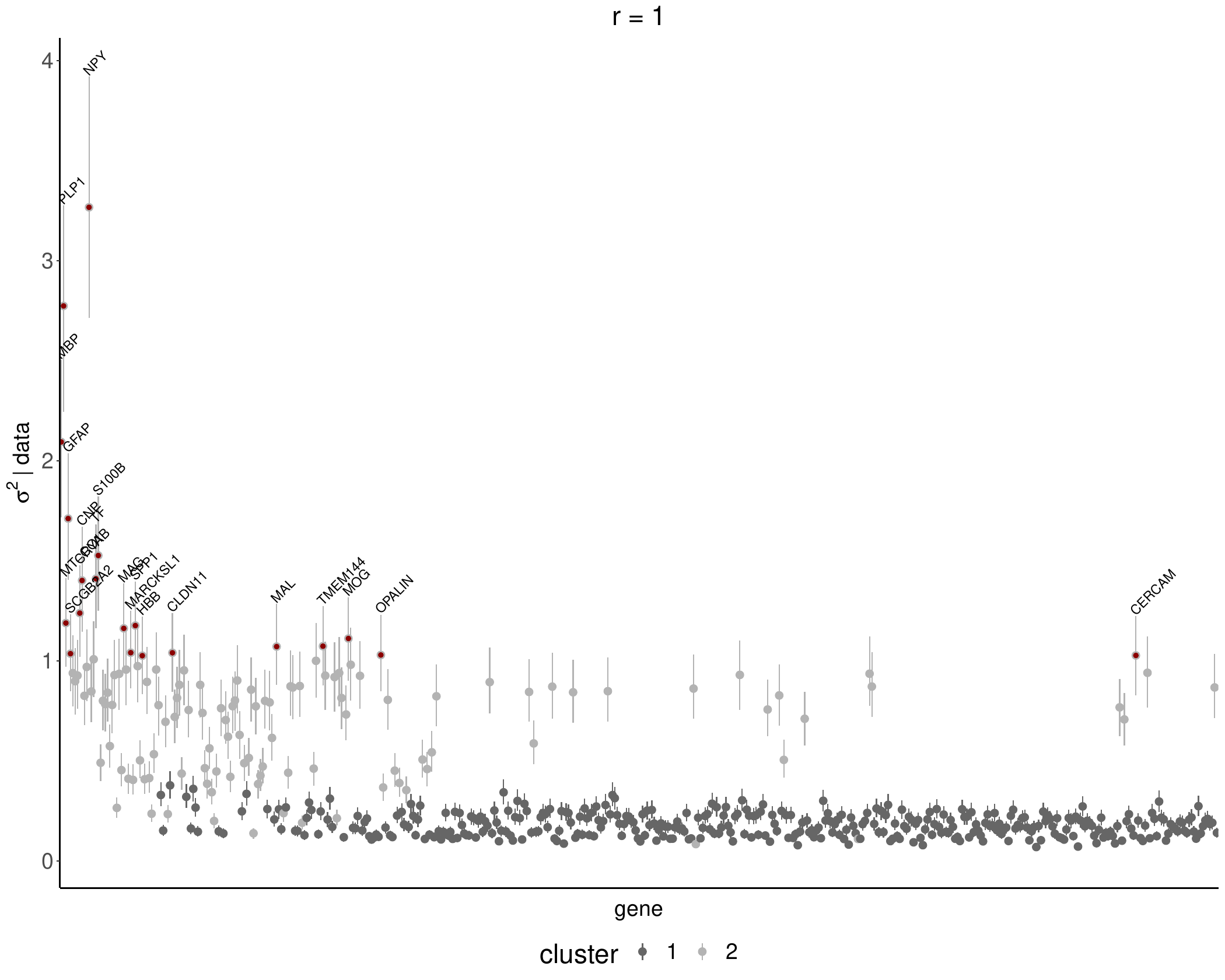}
	\includegraphics[width=.43\linewidth]{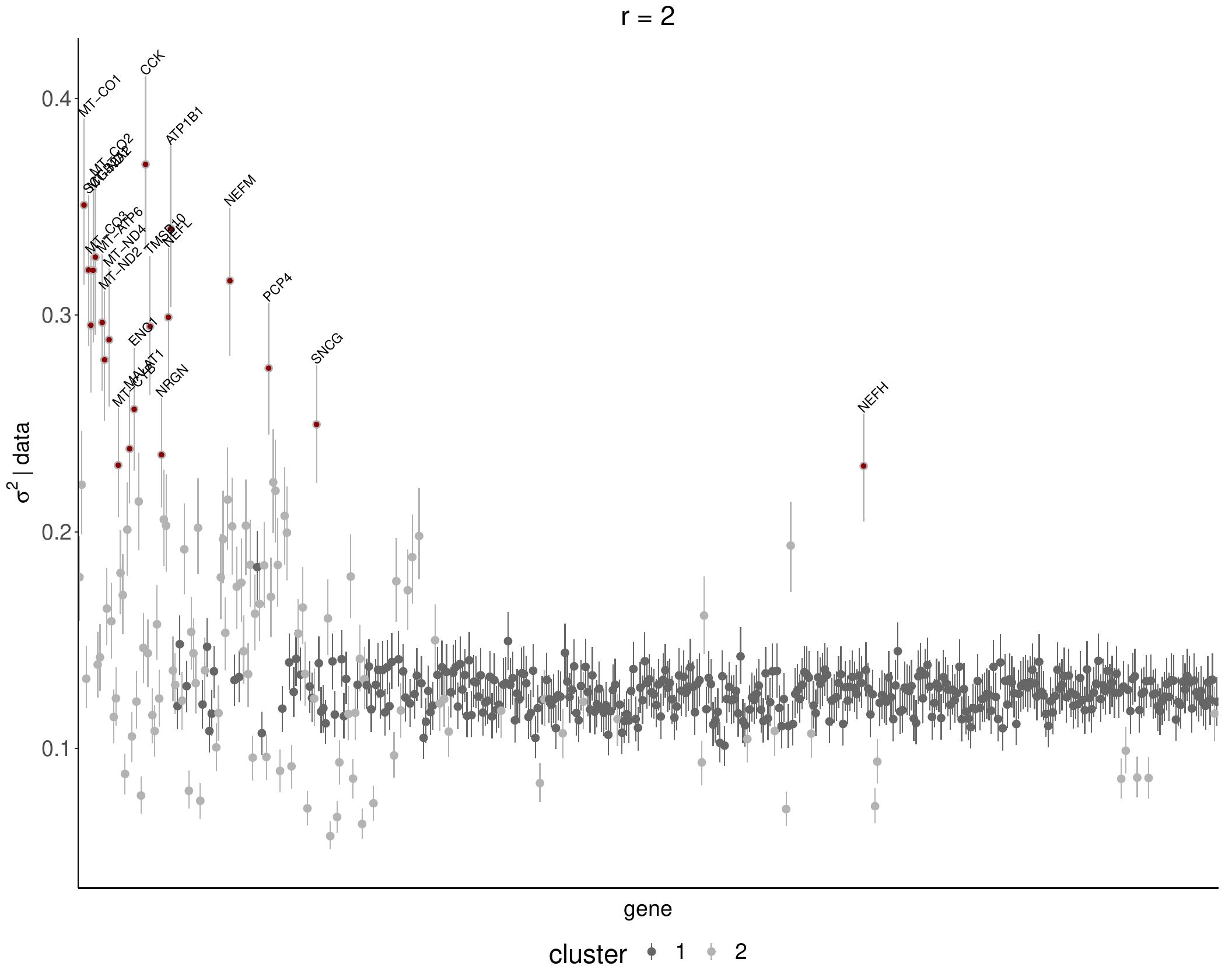}\\
	\includegraphics[width=.43\linewidth]{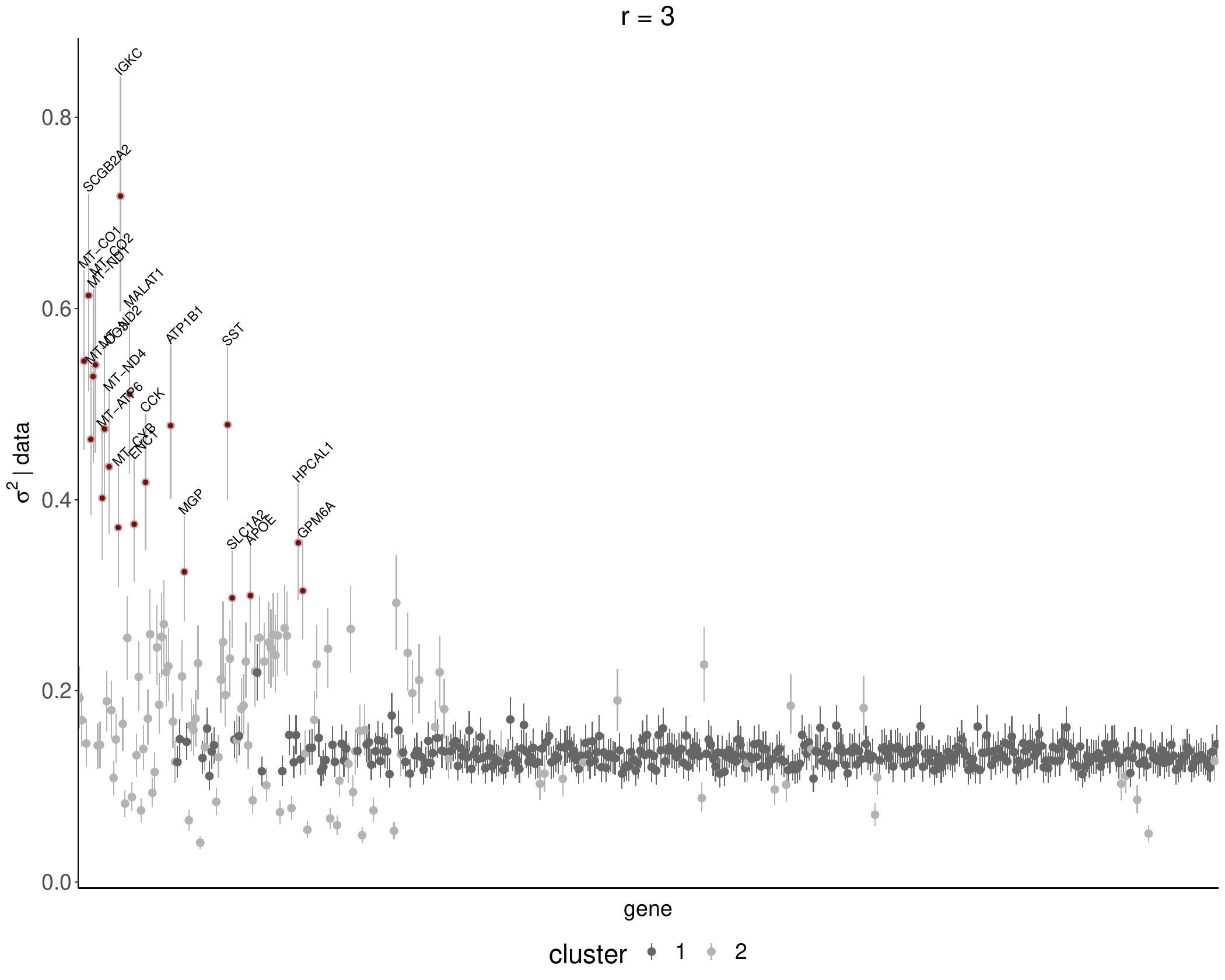}
	\includegraphics[width=.43\linewidth]{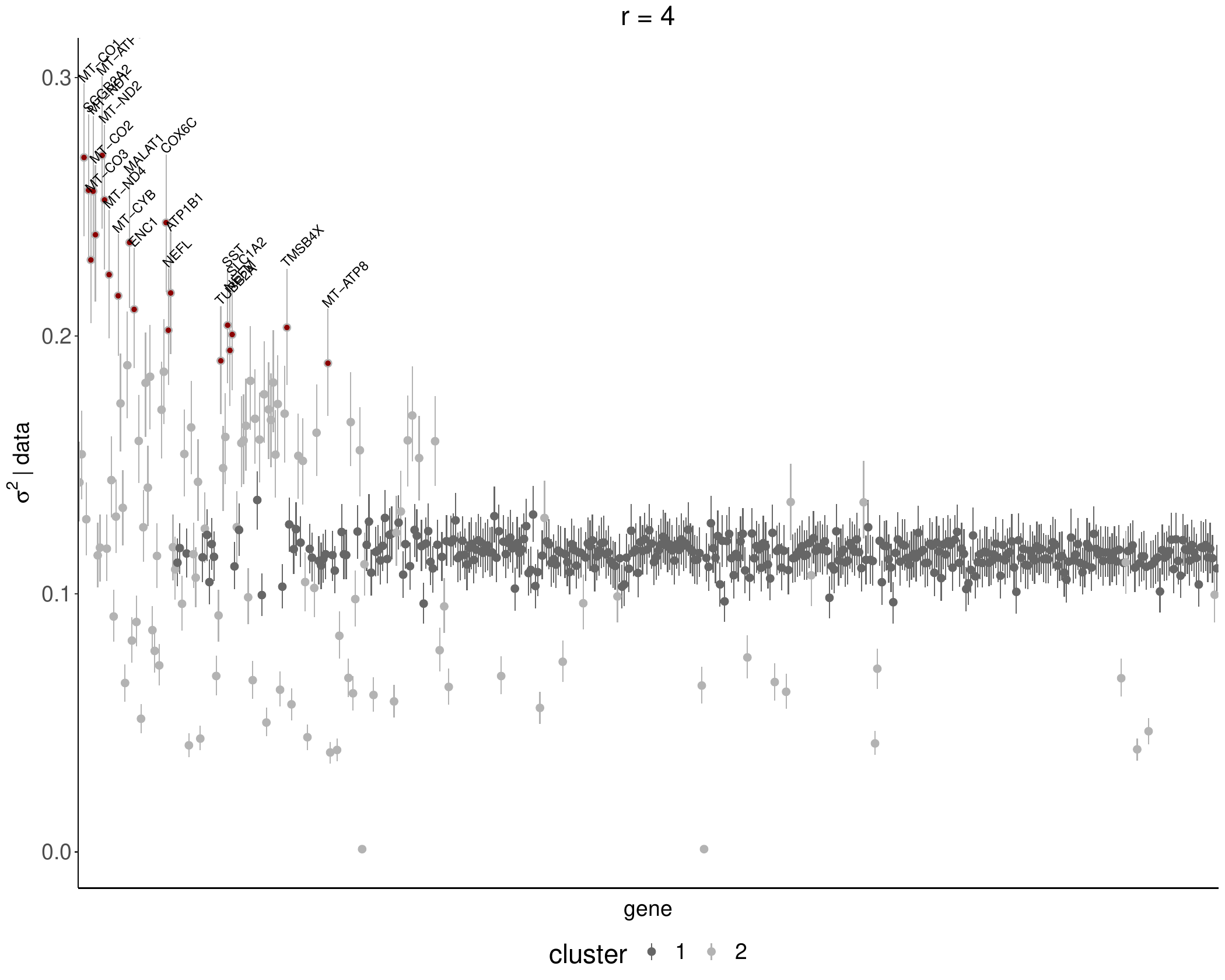}\\
	\includegraphics[width=.43\linewidth]{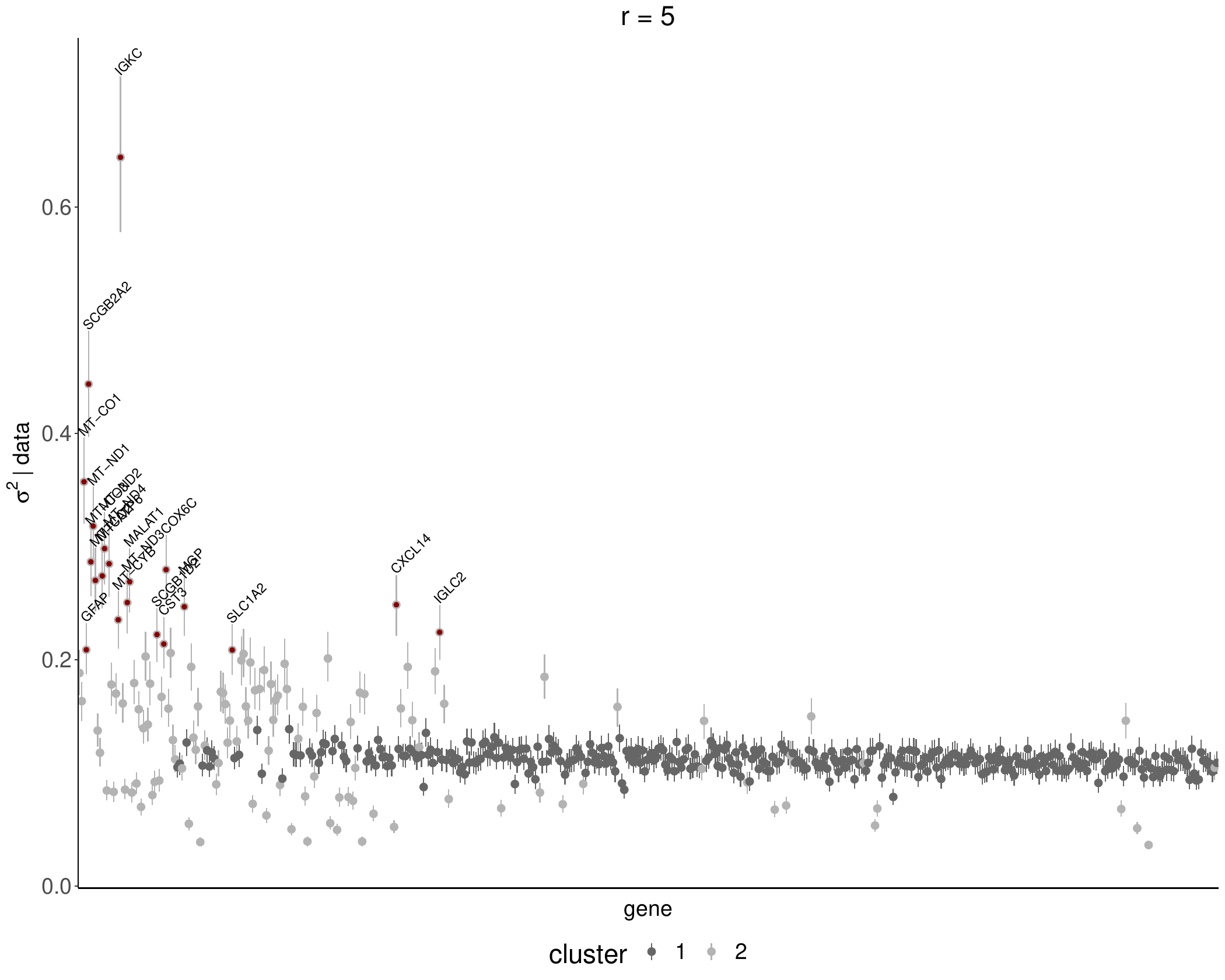}
	\includegraphics[width=.43\linewidth]{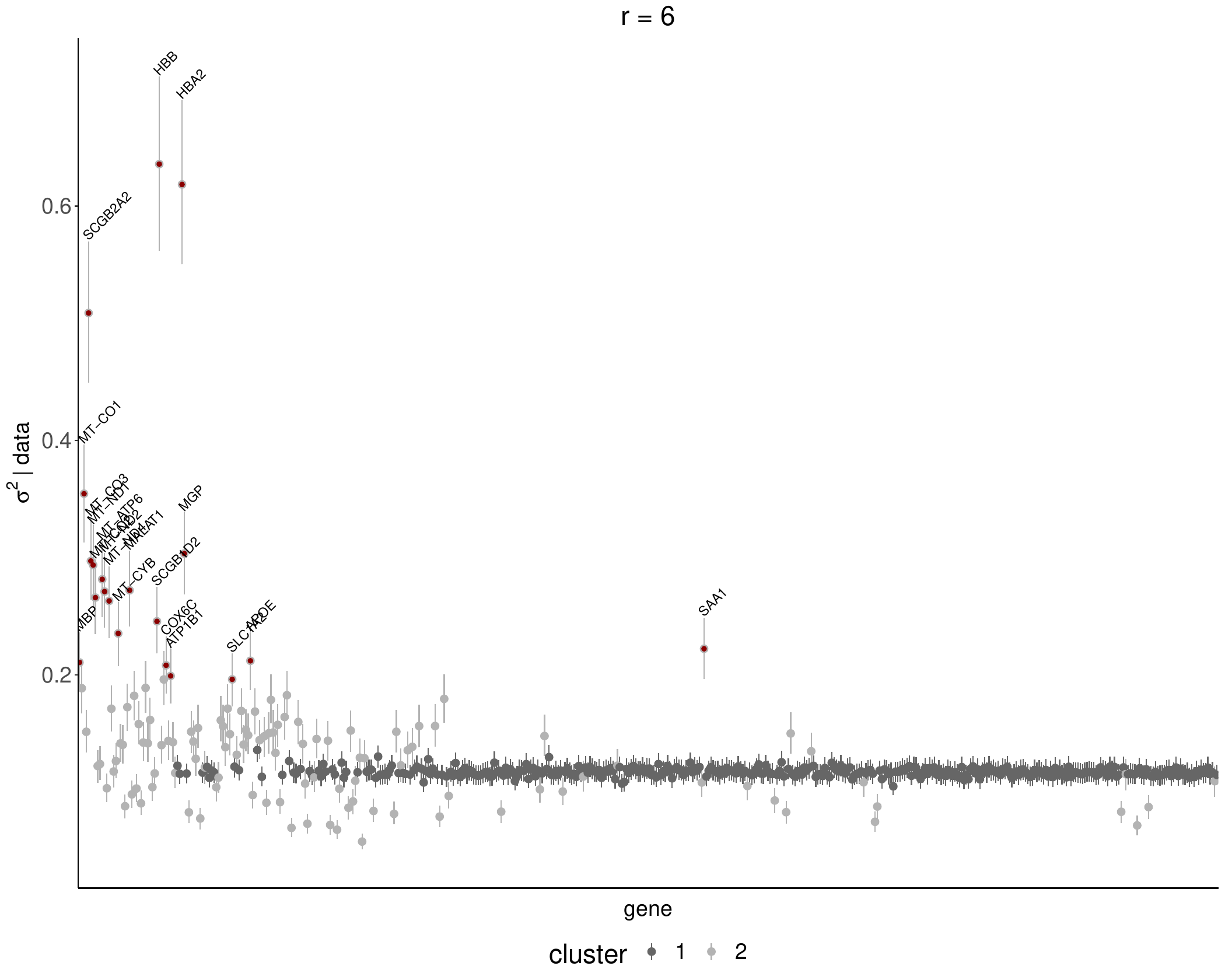}\\
	\includegraphics[width=.43\linewidth]{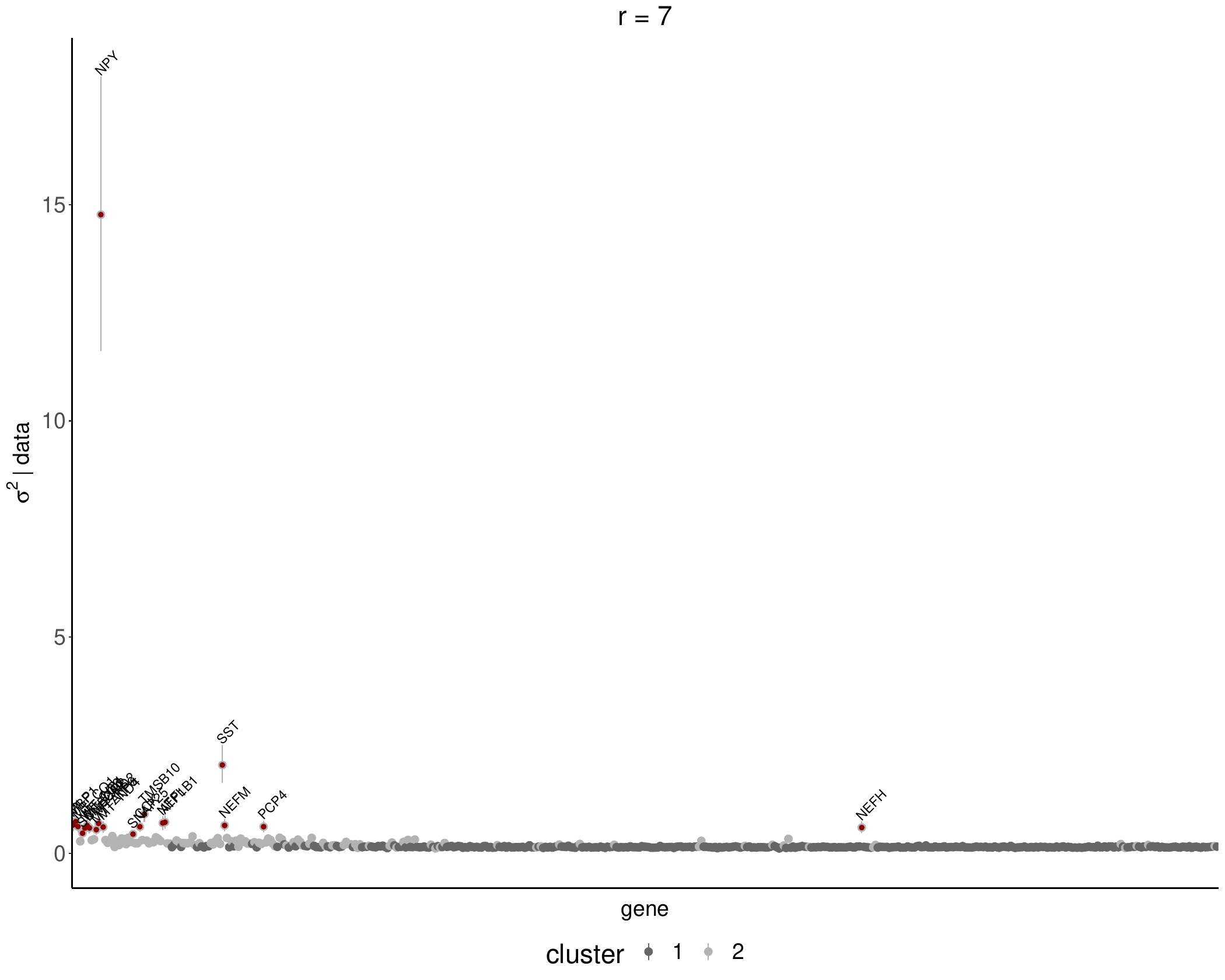}
	\includegraphics[width=.43\linewidth]{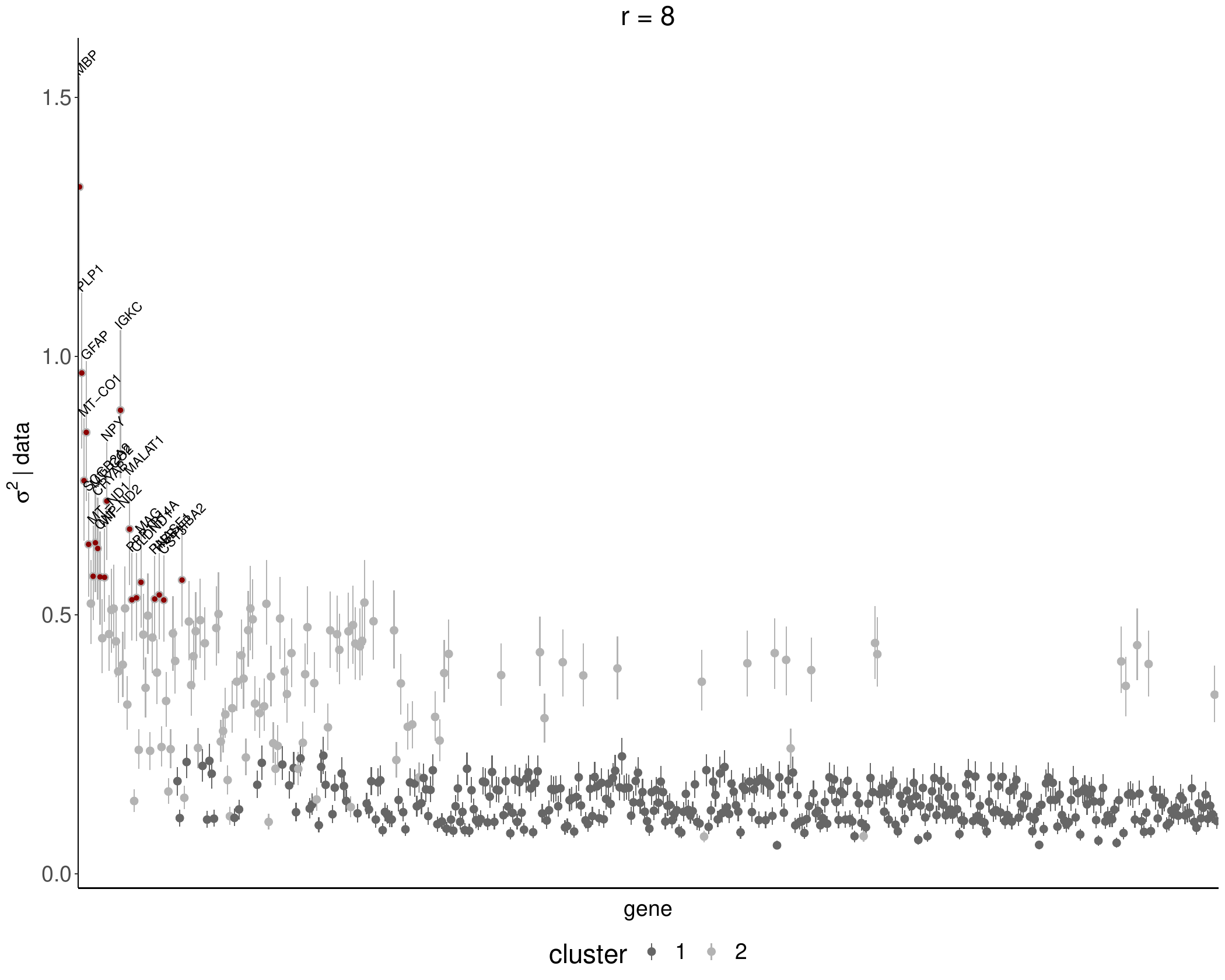}\\
	\caption{\review{Results from Section 5. Each panel gives the distribution of $\sigma^2_{.r,i}|\mathrm{data}$, where $\mathrm{data}$ denotes both the input data and the estimated quantities. The dots denote the expected values and the error bars denote the 95\% credible intervals. For each spot cluster, the twenty genes with the largest expectation are shown in red (see also Table \ref{table:high_exspressed_genes}).
	}}
	\label{figure:Sigma2_r}
\end{figure}

\begin{figure}
	\centering
	\includegraphics[width=.45\linewidth]{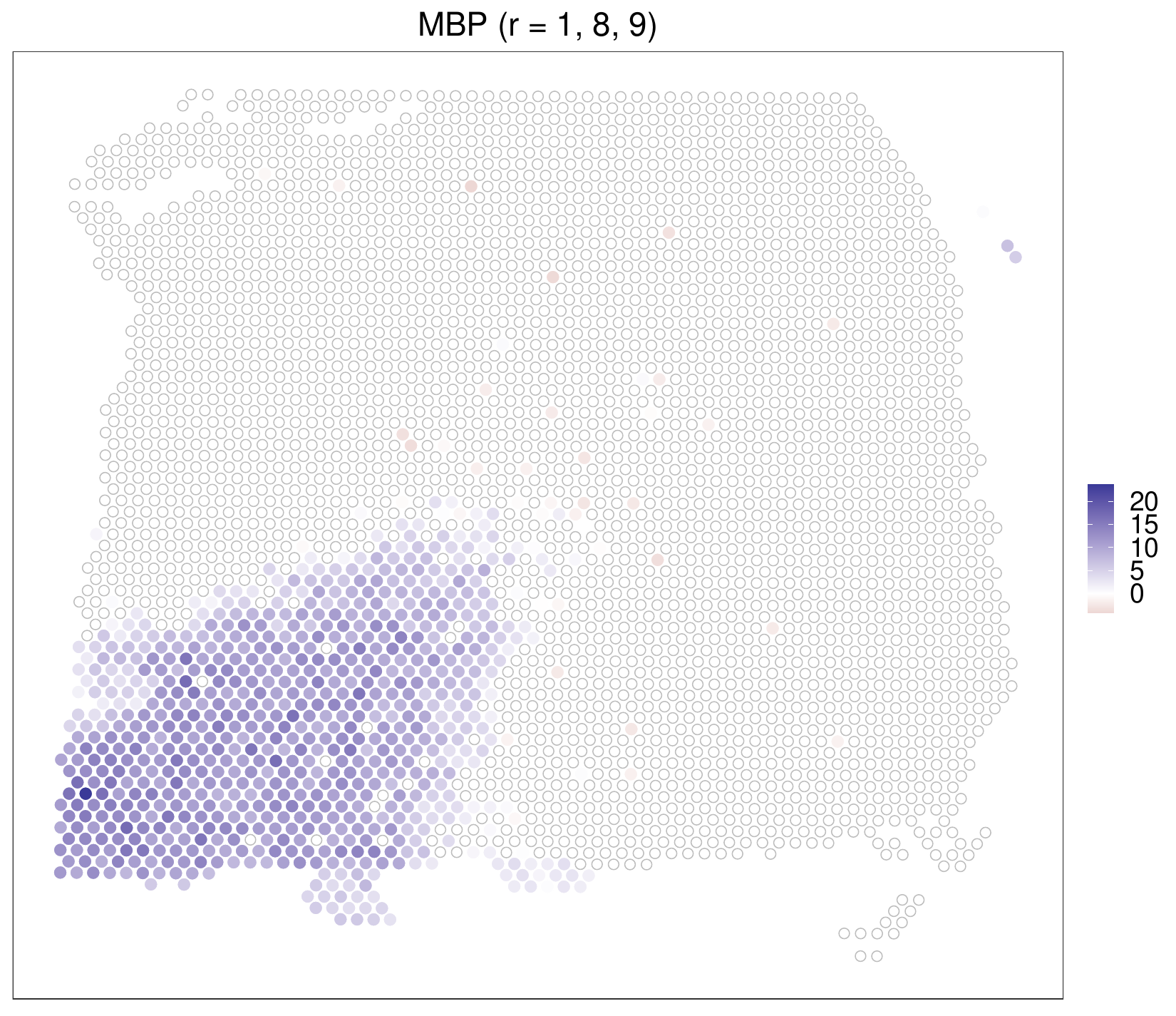}
	\includegraphics[width=.45\linewidth]{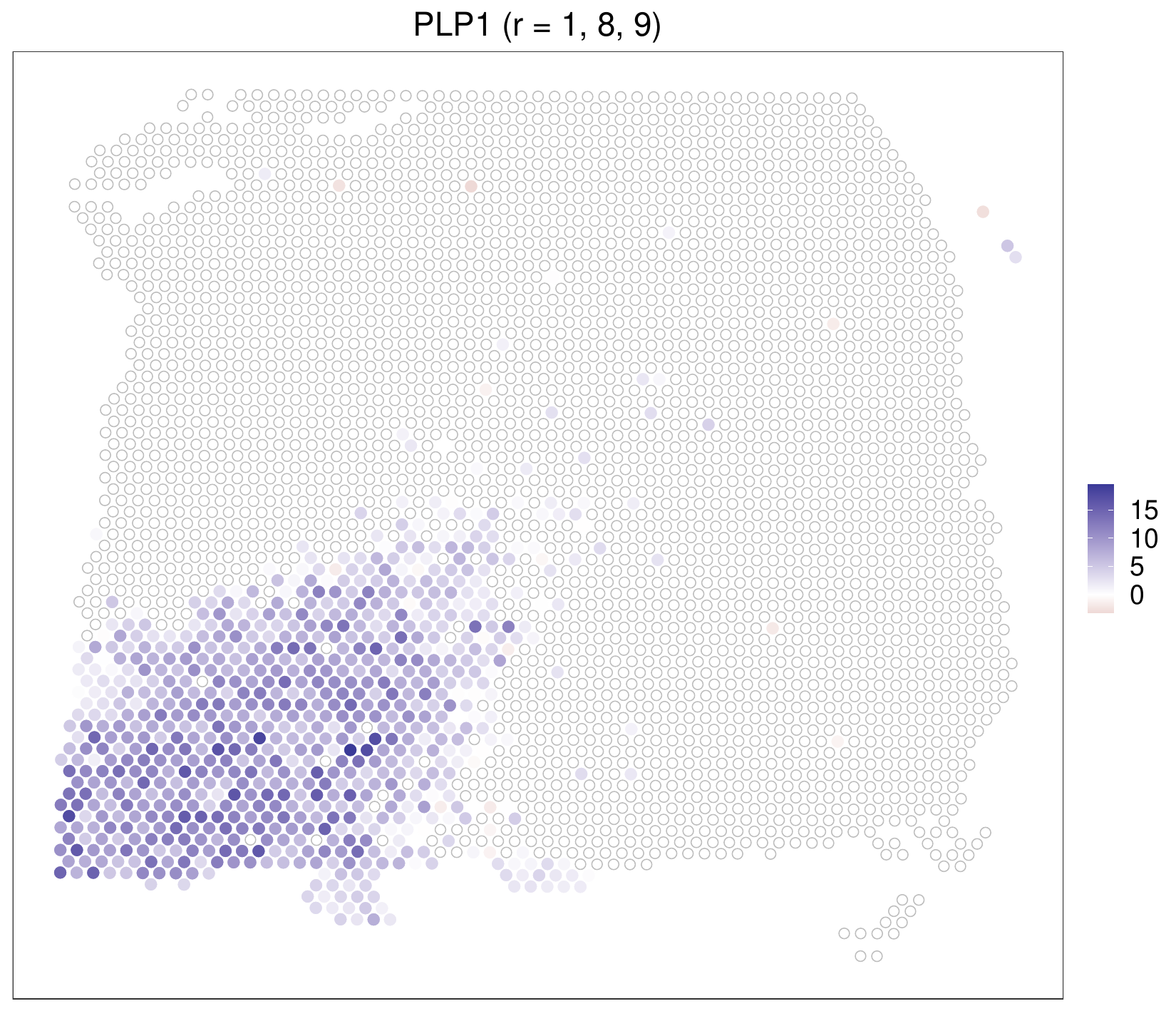}\\
	\includegraphics[width=.45\linewidth]{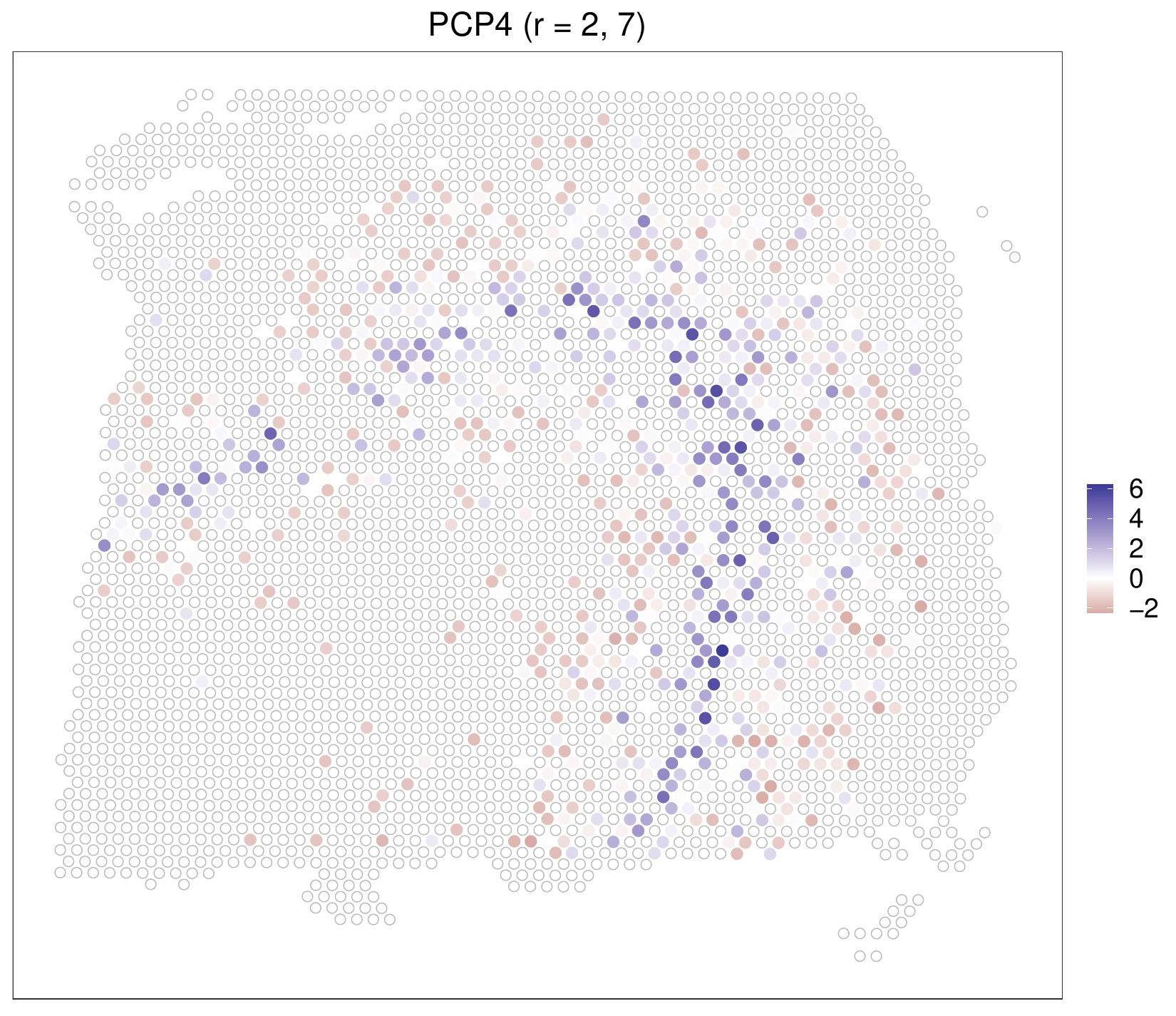}
	\includegraphics[width=.45\linewidth]{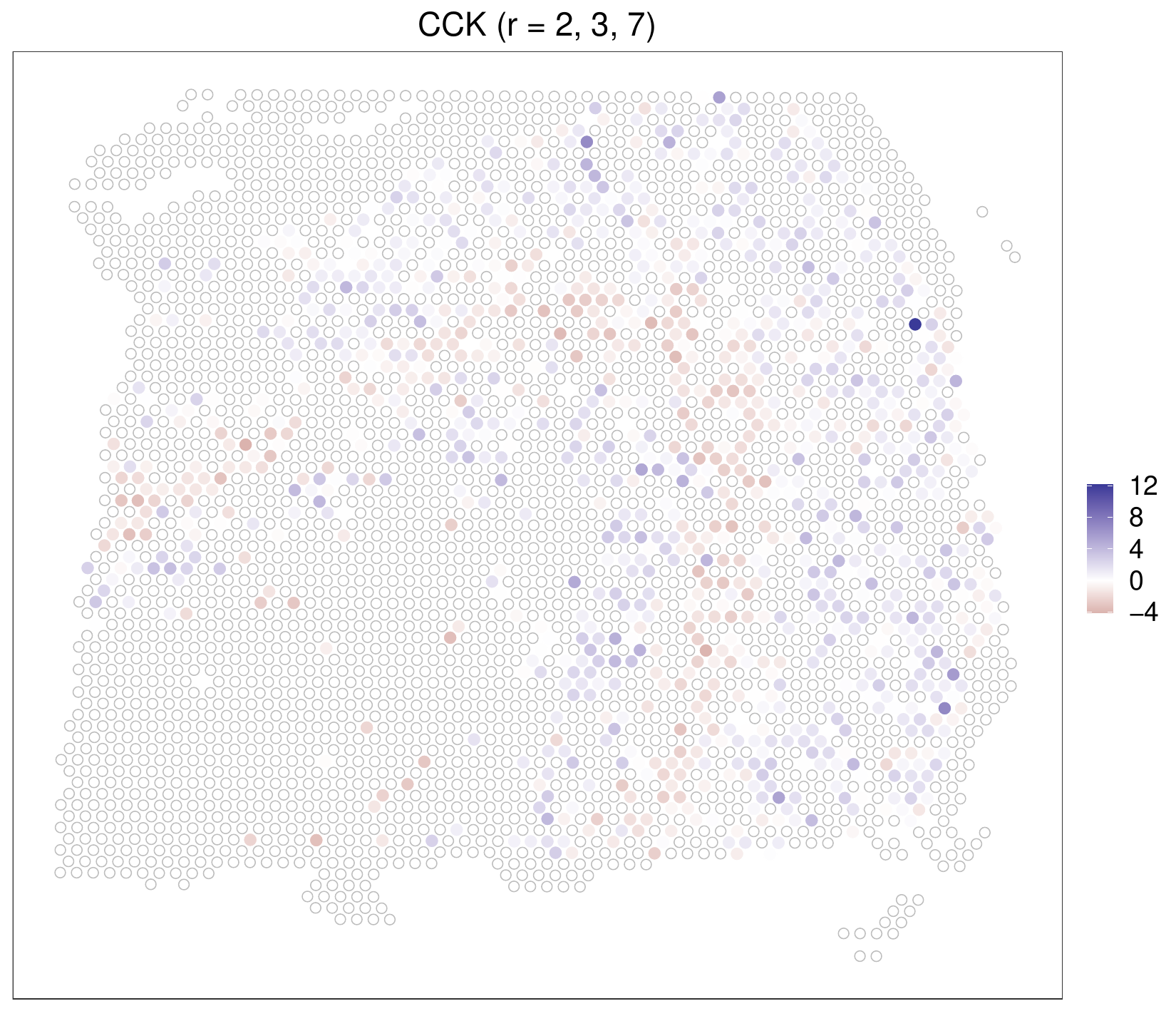}
	\caption{\review{Plot of the genes MBP, PLP1, PCP4 and CCK, discussed in Section 5 of the manuscript and selected among the most highly variable genes in specific areas of the tissue sample. The title of each figure gives both the displayed gene and the  image clusters where the expression is shown.}}
	\label{figure:variable_genes1}
\end{figure}

\begin{figure}
	\centering
	\includegraphics[width=.45\linewidth]{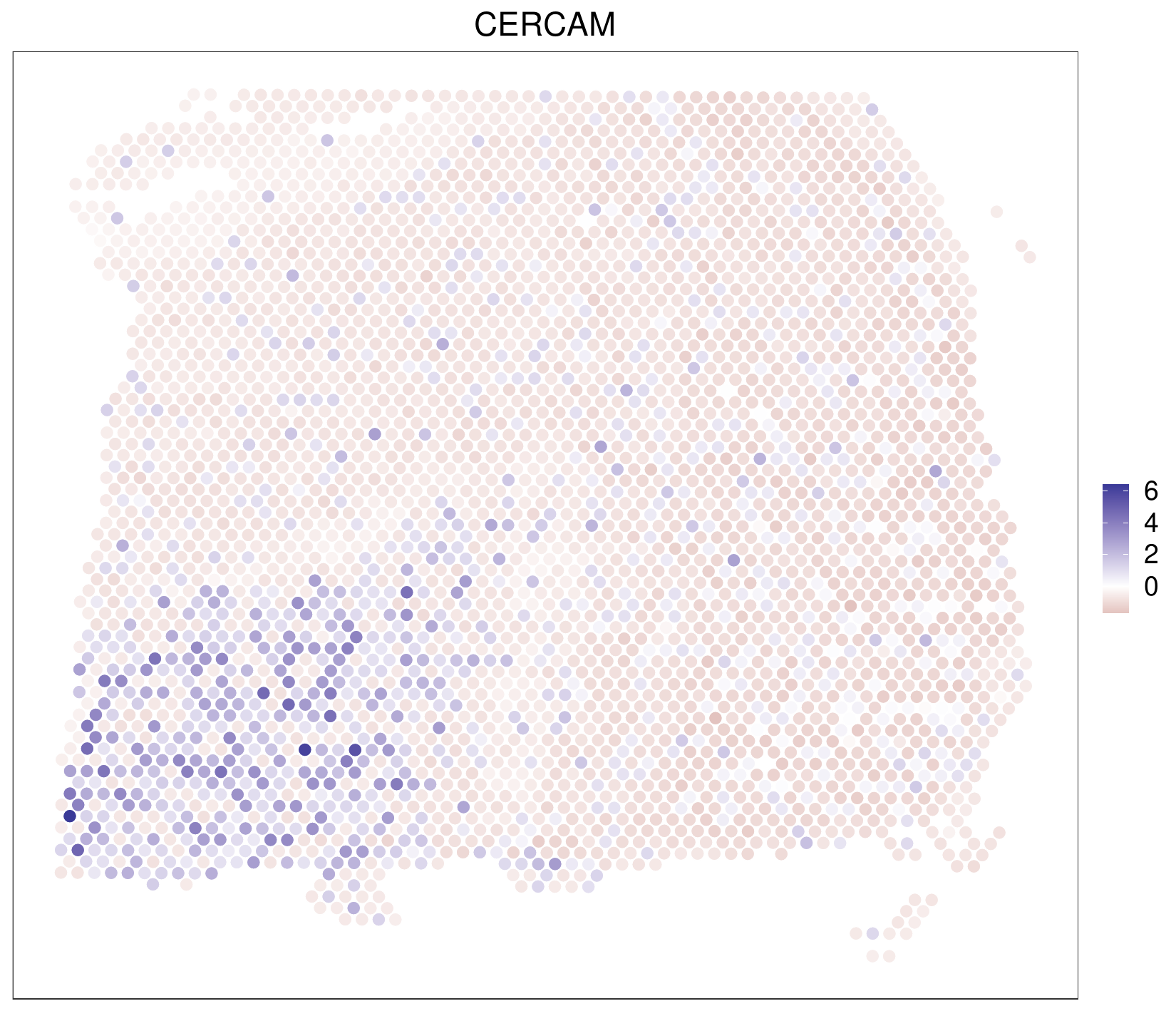}
	\includegraphics[width=.45\linewidth]{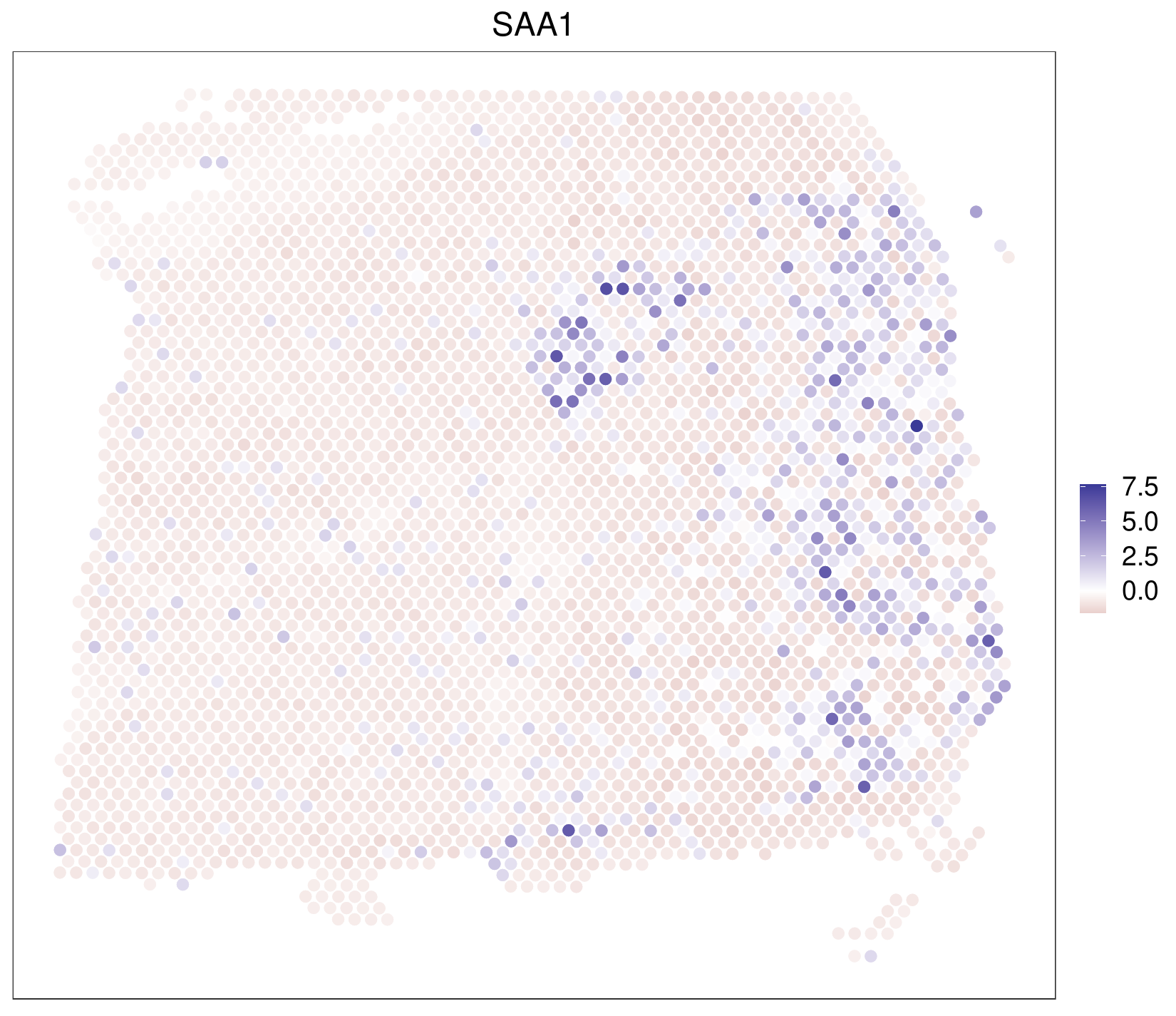}
	\caption{\review{Plot of the genes CERCAM and SAA1 over the whole tissue analyzed in Section 5.}}
	\label{figure:variable_genes2}
\end{figure}

\begin{itemize}
	\item Figure \ref{figure:Scenario5_geneselection} displays the genes ordered according to the deviance criterion proposed by \cite{Townes_etal.2019}. The red line denotes the number of genes selected for our analysis ($n = 500$), the blue line is the \textquotedblleft ideal\textquotedblright~number of genes that should be used ($n = 200$), based on where the deviance curve has a significant change in the decay. 
	\review{\item Figure  \ref{fig:binom_poisson_deviance} displays the boxplots of the first 100 row vectors of the spatial experiment matrix $\mathbf{X}$, corresponding to the gene expressions measured on the cortical tissue sample analyzed in Section 5, transformed and sorted according to the procedure of \cite{Townes_etal.2019}.
		\item Figure \ref{fig:section5_icl_clusteringuncert} illustrates some model fitting results. Panel (a) gives log-likelihood and the ICL values of the models with $K=2$ and $R\in\{7,\dots,12\}$; Panel (b) gives the clustering uncertainty measures $\rowepsilon$ and $\colepsilon$ of the model with $K=2$ and $R=9$.
		\item Figure \ref{figure:Sigma2_r} displays the conditional distributions of $\sigma^2_{.r,i}$, for $i=1,\dots,n$, given the data and the parameter estimates. In addition, Table \ref{table:high_exspressed_genes} lists the most variable genes in each spot cluster that appear also in Figure \ref{figure:Sigma2_r}.
		\item Figures \ref{figure:variable_genes1} and \ref{figure:variable_genes2} display the expression of some genes that are highly variable in specific regions of the analyzed prefrontal tissue sample.
	}
\end{itemize}

\bibliographystyle{imsart-nameyear} 
\bibliography{refs.bib}       

\newpage

\begin{sidewaystable}
	\centering
	{
		\begin{tabular}{rlllllllll}
			\hline
			& $r=1$ & $r=2$ & $r=3$ & $r=4$ & $r=5$ & $r=6$ & $r=7$ & $r=8$ & $r=9$ \\ 
			\hline
			1 & NPY & CCK & IGKC & MT-ATP6 & IGKC & HBB & NPY & MBP & MBP \\ 
			2 & PLP1 & MT-CO1 & SCGB2A2 & MT-CO1 & SCGB2A2 & HBA2 & SST & PLP1 & IGKC \\ 
			3 & MBP & ATP1B1 & MT-CO1 & SCGB2A2 & MT-CO1 & SCGB2A2 & TMSB10 & IGKC & MT-CO1 \\ 
			4 & GFAP & MT-CO2 & MT-CO2 & MT-ND1 & MT-ND1 & MT-CO1 & PLP1 & GFAP & PLP1 \\ 
			5 & S100B & SCGB2A2 & MT-ND1 & MT-ND2 & MT-ND2 & MGP & ATP1B1 & MT-CO1 & SCGB2A2 \\ 
			6 & TF & MT-ND1 & MALAT1 & COX6C & MT-CO3 & MT-CO3 & NEFL & NPY & MT-CO3 \\ 
			7 & CNP & NEFM & SST & MT-CO2 & MT-ND4 & MT-ND1 & MT-ND2 & MALAT1 & MT-CO2 \\ 
			8 & CRYAB & NEFL & ATP1B1 & MALAT1 & COX6C & MT-ATP6 & MBP & MT-CO2 & MALAT1 \\ 
			9 & MT-CO1 & MT-ATP6 & MT-ND2 & MT-CO3 & MT-ATP6 & MALAT1 & NEFM & SCGB2A2 & MT-ATP6 \\ 
			10 & SPP1 & MT-CO3 & MT-CO3 & MT-ND4 & MT-CO2 & MT-ND2 & MT-ND1 & CRYAB & MT-ND1 \\ 
			11 & MAG & TMSB10 & MT-ND4 & ATP1B1 & MALAT1 & MT-CO2 & MT-CO1 & MT-ND1 & MT-ND4 \\ 
			12 & MOG & MT-ND4 & CCK & MT-CYB & MT-ND3 & MT-ND4 & PCP4 & CNP & GFAP \\ 
			13 & TMEM144 & MT-ND2 & MT-ATP6 & ENC1 & CXCL14 & SCGB1D2 & CCK & MT-ND2 & MT-ND2 \\ 
			14 & MAL & PCP4 & ENC1 & SST & MGP & MT-CYB & MT-ND4 & HBA2 & IGLC2 \\ 
			15 & MARCKSL1 & ENC1 & MT-CYB & TMSB4X & MT-CYB & SAA1 & NEFH & MAG & HBB \\ 
			16 & CLDN11 & SNCG & HPCAL1 & NEFL & IGLC2 & APOE & MT-CO2 & HBB & B2M \\ 
			17 & SCGB2A2 & MALAT1 & MGP & SLC1A2 & SCGB1D2 & MBP & MT-CO3 & CLDND1 & SPP1 \\ 
			18 & OPALIN & NRGN & GPM6A & NEFM & CST3 & COX6C & MT-ATP6 & RNASE1 & CNP \\ 
			19 & CERCAM & MT-CYB & APOE & TUBB2A & GFAP & ATP1B1 & SCGB2A2 & PPP1R14A & TF \\ 
			20 & HBB & NEFH & SLC1A2 & MT-ATP8 & SLC1A2 & SLC1A2 & SNAP25 & CST3 & SCD \\ 
			\hline
	\end{tabular}}
	\caption{\review{List of the highly variable genes within each of the nine spot clusters discovered on the human dorsolateral prefrontal cortex tissue sample analyzed in Section 5. The genes listed here are the ones that appear in red in Figure \ref{figure:Sigma2_r}.}}
	\label{table:high_exspressed_genes}
\end{sidewaystable}

